\documentclass[preprint2]{aastex}

\newcommand{\bzcat}{Roma-BZCAT}

\newcommand{\chn}{{\it Chandra}}

\newcommand{\fer}{{\it Fermi}}

\newcommand{\swf}{{\it Swift}}

\newcommand{\xmm}{{\it XMM-Newton}}
\newcommand{\wse}{{\it WISE}}

\usepackage{graphicx}
\usepackage{longtable}
\usepackage{mdwlist}

\usepackage{multirow}
\usepackage{upgreek}
\usepackage[colorlinks=true, pdfstartview=FitV, linkcolor=blue,citecolor=blue, urlcolor=blue]{hyperref}
\usepackage{wasysym}
\usepackage{txfonts}
\usepackage{gensymb}
\usepackage{morefloats}
\usepackage{threeparttable}
\usepackage{rotating}
\usepackage{graphicx}
\usepackage[english]{babel}

\shorttitle{Optical spectroscopic observations of $\gamma$-ray blazar candidates IV}
\shortauthors{F. Ricci et al. 2014}

\begin{document} 
\title{Optical spectroscopic observations of $\gamma$-ray blazar candidates IV \\ 
Results of the 2014 follow-up campaign}


\author{F. Ricci\altaffilmark{1,2}, F. Massaro\altaffilmark{3,4}, M. Landoni\altaffilmark{5}, R. D'Abrusco\altaffilmark{2}, D. Milisavljevic\altaffilmark{2}, \\ D. Stern\altaffilmark{6}, N. Masetti\altaffilmark{7}, A. Paggi\altaffilmark{2}, Howard A. Smith\altaffilmark{2}, G. Tosti\altaffilmark{8}}

\affil{riccif@fis.uniroma3.it}

\altaffiltext{1}{Dipartimento di Matematica e Fisica, Universit\`a Roma Tre, via della Vasca Navale 84, I-00146, Roma, Italy}

\altaffiltext{2}{Harvard - Smithsonian Center for Astrophysics, 60 Garden Street, Cambridge, MA 02138, USA}

\altaffiltext{3}{Dipartimento di Fisica, Universit\`a degli Studi di Torino, via Pietro Giuria 1, I-10125 Torino, Italy}

\altaffiltext{4}{Yale Center for Astronomy and Astrophysics, Physics Department, Yale University, PO Box 208120, New Haven, CT 06520-8120, USA}

\altaffiltext{5}{INAF-Osservatorio Astronomico di Brera, Via Emilio Bianchi 46, I-23807 Merate, Italy}
 
\altaffiltext{6}{Jet Propulsion Laboratory, California Institute of Technology, 4800 Oak Grove Drive, Mail Stop 169-221, Pasadena, CA 91109, USA}

\altaffiltext{7}{INAF - Istituto di Astrofisica Spaziale e Fisica Cosmica di Bologna, via Gobetti 101, 40129, Bologna, Italy}             
 
\altaffiltext{8}{Dipartimento di Fisica, Universit\`a degli Studi di Perugia, 06123 Perugia, Italy}

\date{Received October 15, 2014; accepted  ... }

 \begin{abstract}
The extragalactic $\gamma$-ray sky is dominated by the emission arising from blazars, one of the most peculiar classes of
radio-loud active galaxies. Since the launch of \fer\ several methods were developed to search for blazars as potential counterparts
of unidentified $\gamma$-ray sources (UGSs). To confirm the nature of the selected candidates, optical spectroscopic 
observations are necessary.
In 2013 we started a spectroscopic campaign to investigate $\gamma$-ray blazar candidates selected according to different procedures.
The main goals of our campaign are: 1) to confirm the nature of these candidates, and 2) whenever possible determine their redshifts.
Optical spectroscopic observations will also permit us to verify the robustness of the proposed associations 
and check for the presence of possible source class contaminants to our counterpart selection. 
This paper reports the results of observations carried out in 2014 in the Northern hemisphere with 
Kitt Peak National Observatory (KPNO) and in the Southern hemisphere 
with the Southern Astrophysical Research (SOAR) telescopes.
We also report three sources observed with the Magellan and Palomar telescopes.
Our selection of blazar-like sources that could be potential counterparts of UGSs is based on their peculiar IR colors 
and on their combination with radio observations both at high and low frequencies (i.e., above and below $\sim$ 1 GHz) 
in publicly available large radio surveys. 
We present the optical spectra of 27 objects. We confirm the blazar-like nature of 9 sources that appear to be potential low-energy counterparts of UGSs.
Then we present new spectroscopic observations of 10 active galaxies of uncertain type associated with \fer\ sources, 
classifying all of them as blazars.
In addition, we present the spectra for five known $\gamma$-ray blazars with uncertain redshift estimates
and three BL Lac candidates that were observed during our campaign. We also report the case for
\wse\ J173052.85-035247.2, candidate counterpart 
of the source 2FGL J1730.6-0353, which has no radio counterpart in the major radio surveys.
We confirm that our selection of $\gamma$-ray blazars candidates can successfully 
indentify low-energy counterparts to \fer\ unassociated sources and allow us to discover new blazars.
\end{abstract}


\maketitle

\section{Introduction}
Blazars are the largest population of active galactic nuclei (AGNs) detected in the $\gamma$ range \citep{abdo10,nolan12}. 
Their non-thermal emission extends from radio to TeV energies and it is coupled with
very rapid variability, high and variable polarization, superluminal motion, high luminosities \citep[e.g.,][]{urry95}
and peculiar infrared (IR) colors \citep{paper1}.
Since 1978, well before the establishment of the unification scenario for the AGNs,
their peculiar properties were described in terms of emission arising from particles accelerated in a relativistic jet closely 
pointed along our line of sight \citep{blandford78}.

Adopting the nomenclature of the blazar subclasses described in the
\bzcat: Multi-frequency Catalogue of Blazars\footnote{\scriptsize{{\tt{ http://www.asdc.asi.it/bzcat/}}}} 
\citep[e.g.,][]{massaro09,massaro11}, we distinguish between BL Lac objects (i.e., BZBs) 
and the flat spectrum radio quasars indicated as BZQs.
The former present optical spectra with emission and/or absorption lines of
rest frame equivalent width \(EW \leq 5\) \AA\ \citep[e.g.,][]{stickel91,stoke91,laurent99} while
the latter show typical quasar-like optical spectra with strong and broad emission lines. 
In the \bzcat\ there are also several BZBs indicated as BL Lac candidates; these sources were indicated as BL Lacs
in literature and thus reported in the catalog but, lacking their optical spectra, a correct classification is still uncertain \citep[see also][]{sdss}.

We developed several methods to search for $\gamma$-ray blazar candidates 
that could be counterparts of the unidentified $\gamma$-ray sources \citep[UGSs, 2FGL][]{nolan12} on the basis of the peculiar IR colors
of known $\gamma$-ray blazars \citep[e.g.][]{ugs1,ugs2,ugs3}, discovered thanks to all-sky survey of the Wide-Field Infrared Survey Explorer 
\citep[\wse;][]{wright10}, or using radio observations in combination with IR colors \citep{paper5} as well as  
at low frequencies in the MHz regime \citep{paper3,ugs6}.
In addition, multifrequency analysis based on X-ray follow-up observations 
\citep[e.g.,][]{mirabal09,paggi13,takeuchi13,stroh13,acero13}
and radio campaigns \citep[e.g.,][]{petrov13} were performed to search for potential UGS counterparts.

Since the positional uncertainty of \fer\ is a few tenths of a degree, all the proposed methods
and the multifrequency follow-up observations are primarily useful to decrease the number 
of potential counterparts for the UGSs. There could be possible contamination by
different source classes in these selection procedures of $\gamma$-ray blazar candidates \citep[e.g.,][]{stern13}
and such degeneracy can be only removed with optical spectroscopic observations \citep[e.g.,][]{masetti13,shaw13a,shaw13b,paggi14,sdss}.
Furthermore, a detailed knowledge of the number of UGSs is of paramount importance 
for instance to provide constraints on dark matter models  
\citep{abdo14}. Many UGSs could be blazars, but how many of them are actually blazars 
is still unknown due in part to the incompleteness of catalogs used for the associations \citep{ackermann11a}.

Thus, motivated by these arguments, we started an optical spectroscopic campaign in 2013
aiming to confirm the real nature of the proposed low-energy counterparts of UGSs selected according to our methods.

In this paper we report the results of observations carried out since January 2014 in both hemispheres.
We mainly used the Kitt Peak National Observatory (KPNO) for our targets in the Northern hemisphere in addition 
to two more observations performed at Palomar. For targets mainly visible in the Southern sky, 
we present spectra obtained with the Southern Astrophysical Research (SOAR) telescope and one more observation
carried out at Magellan.
Preliminary results for our exploratory program in the Northern hemisphere 
obtained with the Telescopio Nazionale Galileo (TNG), 
the Multiple Mirror Telescope (MMT) and the Observatorio Astron\'omico Nacional (OAN) in San Pedro M\'artir (M\'exico) 
were already presented in Paggi et al. (2014).
In addition, the results of our observations carried out in the 2013 campaign with SOAR and KPNO 
are described by Landoni et al. (2014) and Massaro et al. (2014b), respectively.

The paper is organised as follows: Section~\ref{sec:sel} contains the different methods that we employed for the sample selection, 
in Section~\ref{sec:obs} we present our observational setup
and we discuss the data reduction procedures. Then in Section~\ref{sec:results} 
we describe the results of our analysis for different types of sources in our sample.
Finally, Section~\ref{sec:conclusions} is devoted to summary and conclusions.
We use cgs units unless otherwise stated.

\section{Selecting the sample}
\label{sec:sel}
The surveys and the catalogs used to search for the counterparts of our targets are listed in the following. These symbols are 
also reported in Table \ref{tab:obslog}.
Below 1GHz we used 
the VLA Low-Frequency Sky Survey Discrete Source Catalog \citep[VLSS;][- V]{cohen07} 
and the recent revision VLLSr\footnote{\scriptsize{\tt{http://heasarc.gsfc.nasa.gov/W3Browse/all/vlssr.html}}}
\citep{lane14}, the Westerbork Northern Sky Survey \citep[WENSS;][- W]{rengelink97},
the Sydney University Molonglo Sky Survey \citep[SUMSS;][- S]{mauch03},
the Parkes-MIT-NRAO Surveys \citep[PMN;][- Pm]{wright94},
the Parkes Southern Radio Source catalog \citep[PKS;][- Pk]{wright90},
and the Low-frequency Radio Catalog of Flat-spectrum Sources \citep[LORCAT;][- L]{lorcat}.
At higher radio frequencies we also verified the counterparts in the NRAO VLA Sky Survey \footnote{\scriptsize{\tt{http://heasarc.gsfc.nasa.gov/W3Browse/radio-catalog/nvss.html}}} \citep[NVSS;][- N]{condon98}, 
the Australia Telescope 20 GHz Survey \citep[AT20G;][- A]{murphy10},
the Combined Radio All-Sky Targeted Eight-GHz Survey \citep[CRATES;][- c]{healey07}.
In the infrared, we queried the \wse\ all-sky survey in the  
AllWISE Source catalog\footnote{\scriptsize{\tt{http://wise2.ipac.caltech.edu/docs/release/allwise/}}} \citep[][- w]{wright10}
and the Two Micron All Sky Survey \citep[2MASS;][- M]{skrutskie06}
since each \wse\ source is automatically matched to the closest 2MASS potential counterpart \citep[see][for details]{cutri12}.
Then, we also searched for optical counterparts, with or without possible spectra available, 
in the Sloan Digital Sky Survey Data Release 9 \citep[SDSS DR9; e.g.][- s]{ahn12}, 
in the Six-degree-Field Galaxy Redshift Survey \citep[6dFGS;][- 6]{jones04,jones09}.
At high-energies, in the X-rays, we searched the ROSAT all-sky survey 
in both the {\it ROSAT} Bright Source Catalog \citep[RBSC;][- X]{voges99} 
and the {\it ROSAT} Faint Source Catalog \citep[RFSC;][- X]{voges00},
as well as \xmm\ Slew Survey \citep[XMMSL;][- x]{saxton08,warwick12}, 
the Deep \swf\ X-Ray Telescope Point Source Catalog \citep[1XSPS;][- x]{evans14},
the \chn\ Source Catalog \citep[CSC;][- x]{evans10} and the \swf\ X-ray survey for all \fer\ 
UGSs \citep{stroh13}.
Note that we use the same symbol for the X-ray catalogs of \xmm, \chn\ and \swf. \\

Our sample lists a total of 27 sources divided as follows:
9 are UGSs for which the analysis based on the IR colors of blazar candidates
indicated them as blazar-like sources that were observed during our campaign;
10 are indeed classified as active galaxies of uncertain type (AGUs) according to 
The Second Catalog of Active Galactic Nuclei Detected by the Fermi Large Area Telescope \citep[2LAC;][]{ackermann11a}
for which no optical spectra were available when the catalog was released and have been observed as part of our campaign. 
Some sources have been also selected based on low-frequency radio information \citep{ugs3}.
The remaining 8 sources are BL Lac candidates and BL Lacs, 
either detected or not by \fer\, listed in the \bzcat\ 
for which no optical spectroscopic information were found in literature \citep{massaro11} or with uncertain/unknown redshift estimate.

\begin{figure*}
	\begin{center}
	\includegraphics[width=0.25\paperwidth]{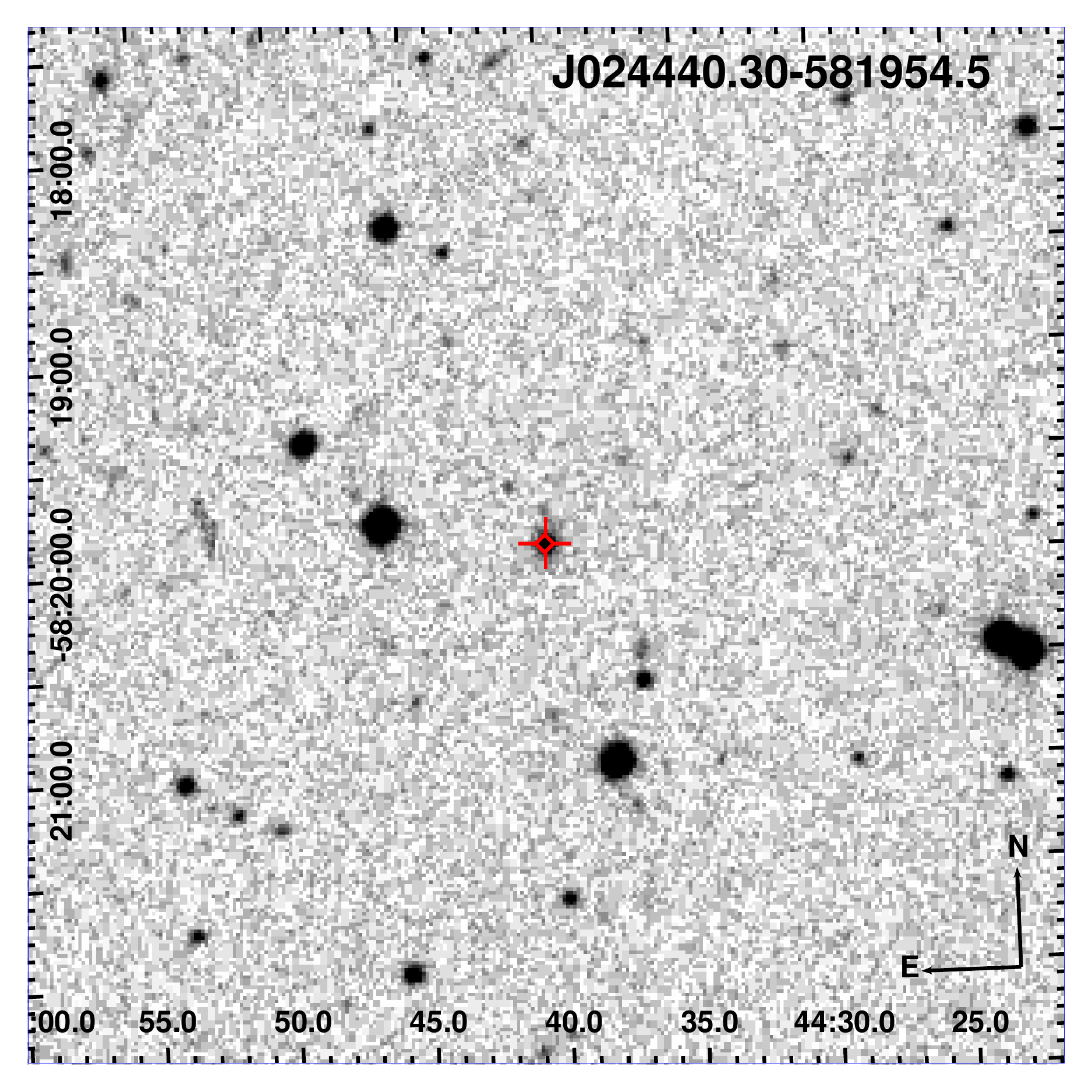}	
	\includegraphics[width=0.25\paperwidth]{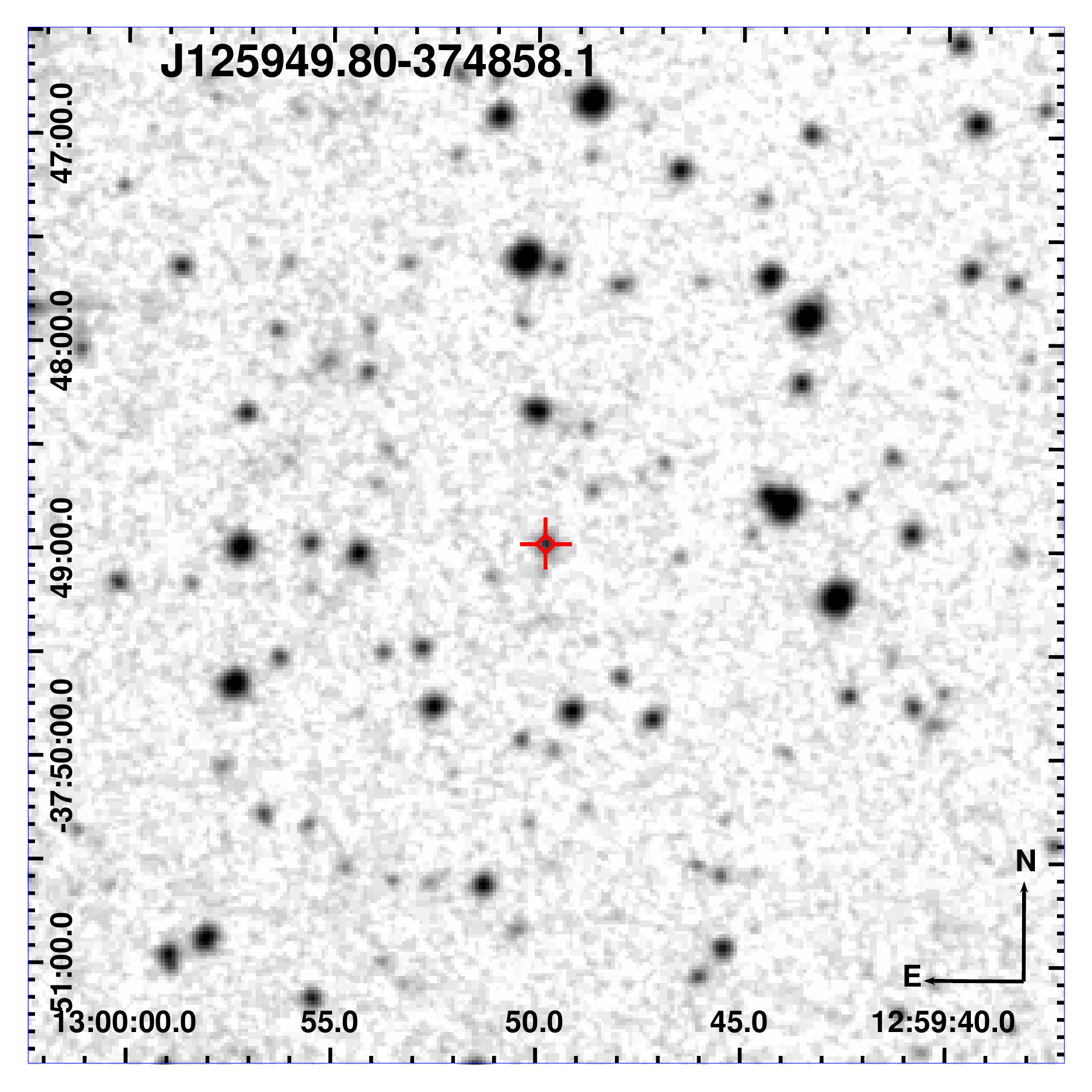}	
	\includegraphics[width=0.25\paperwidth]{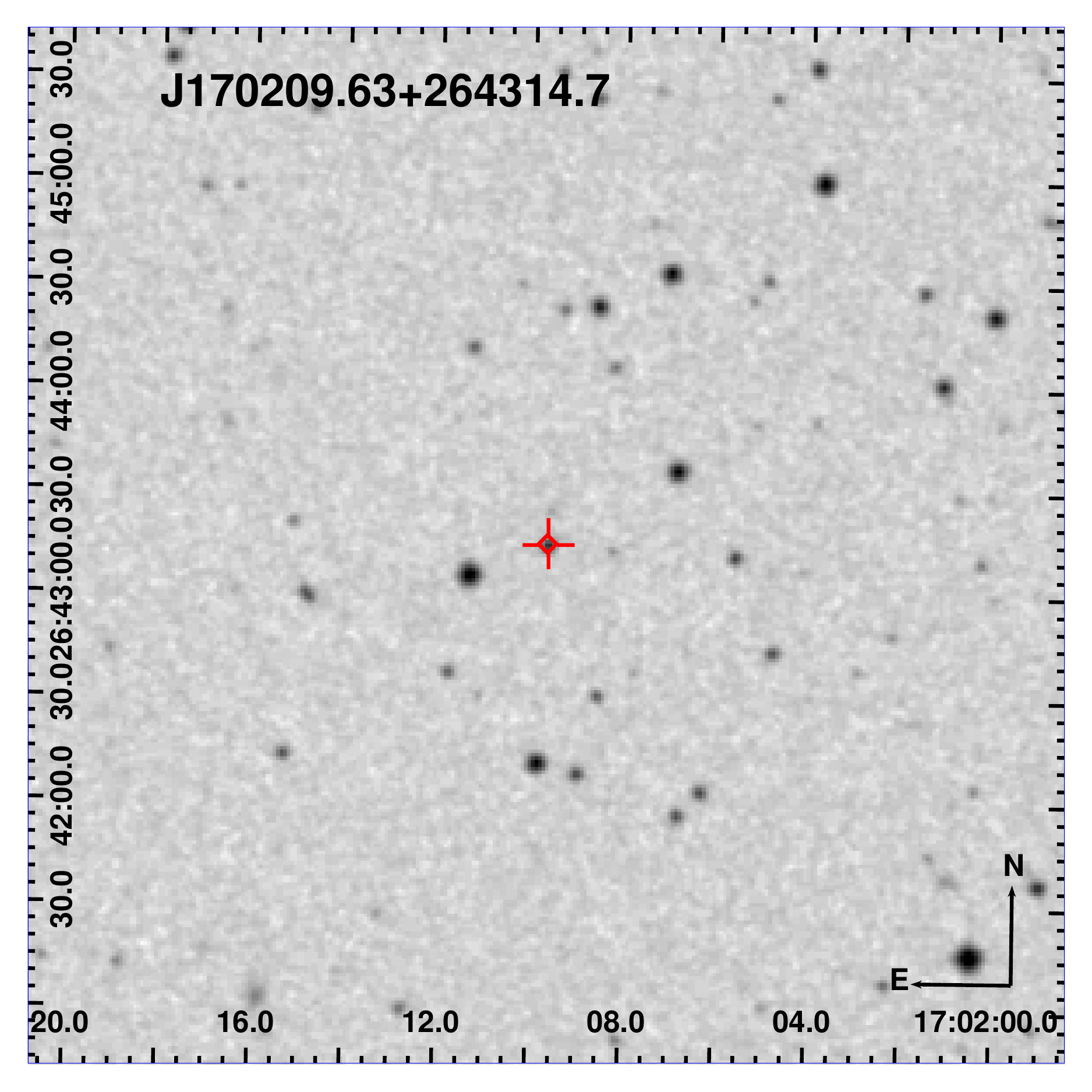}
	\includegraphics[width=0.25\paperwidth]{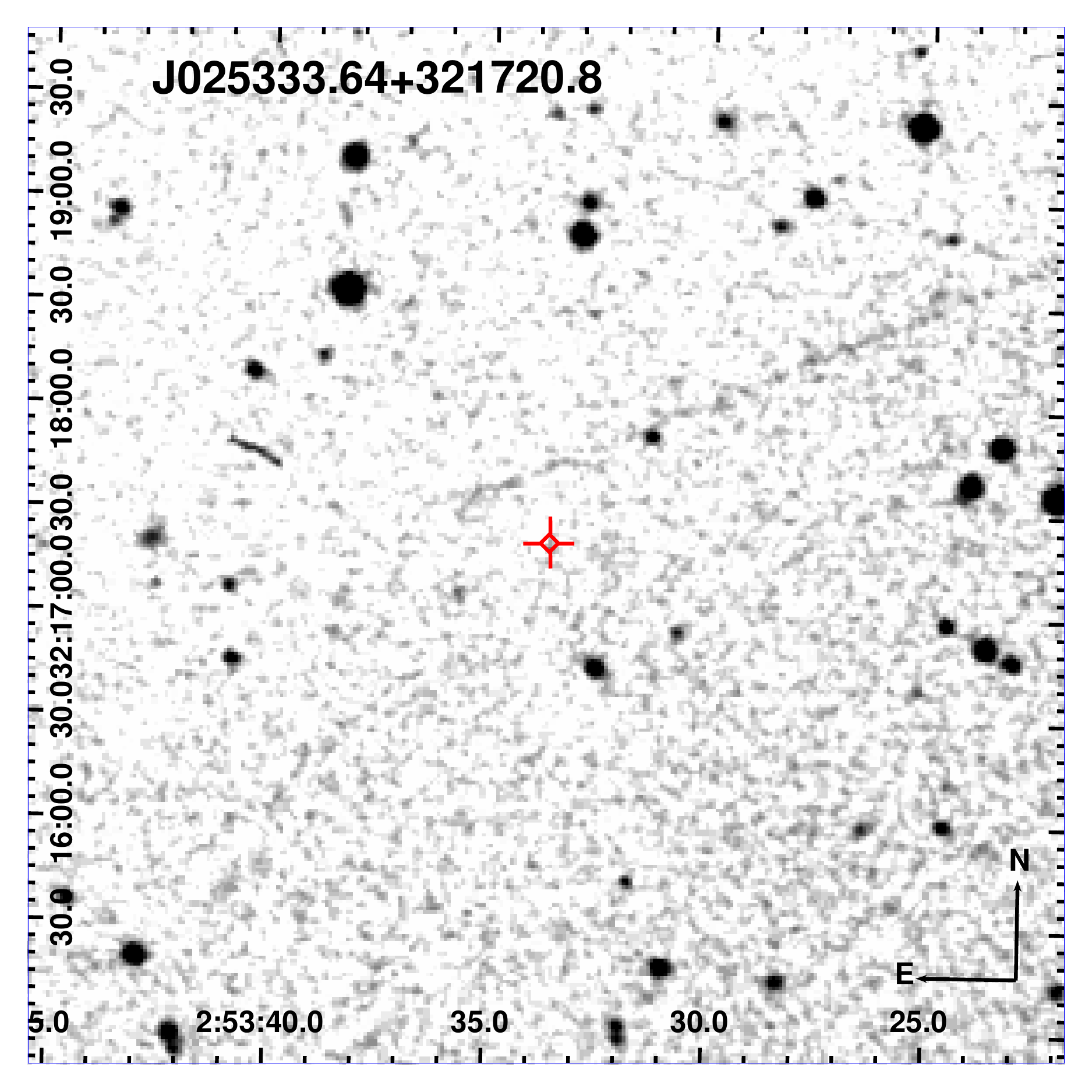}
	\includegraphics[width=0.25\paperwidth]{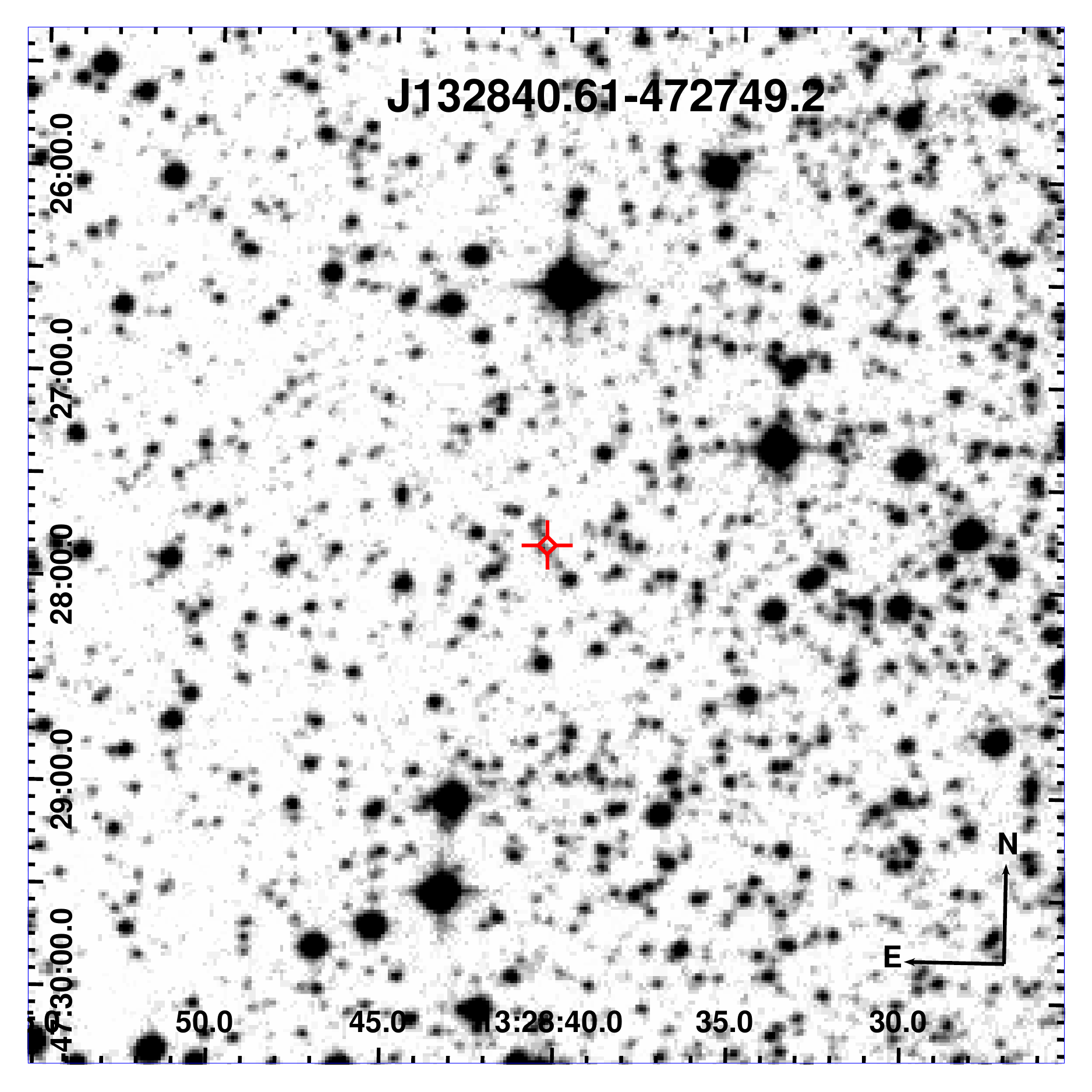}
	\includegraphics[width=0.25\paperwidth]{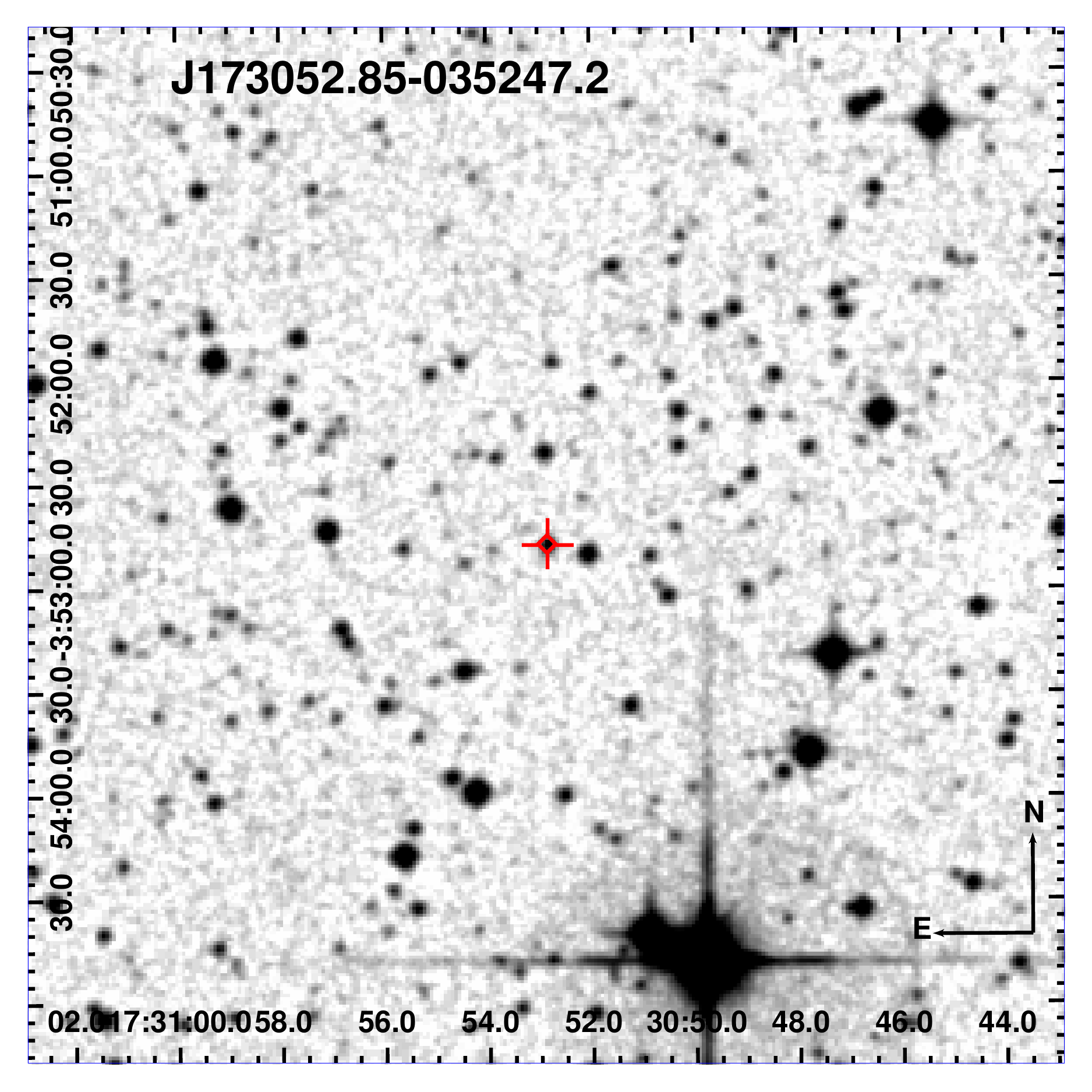}
	\includegraphics[width=0.25\paperwidth]{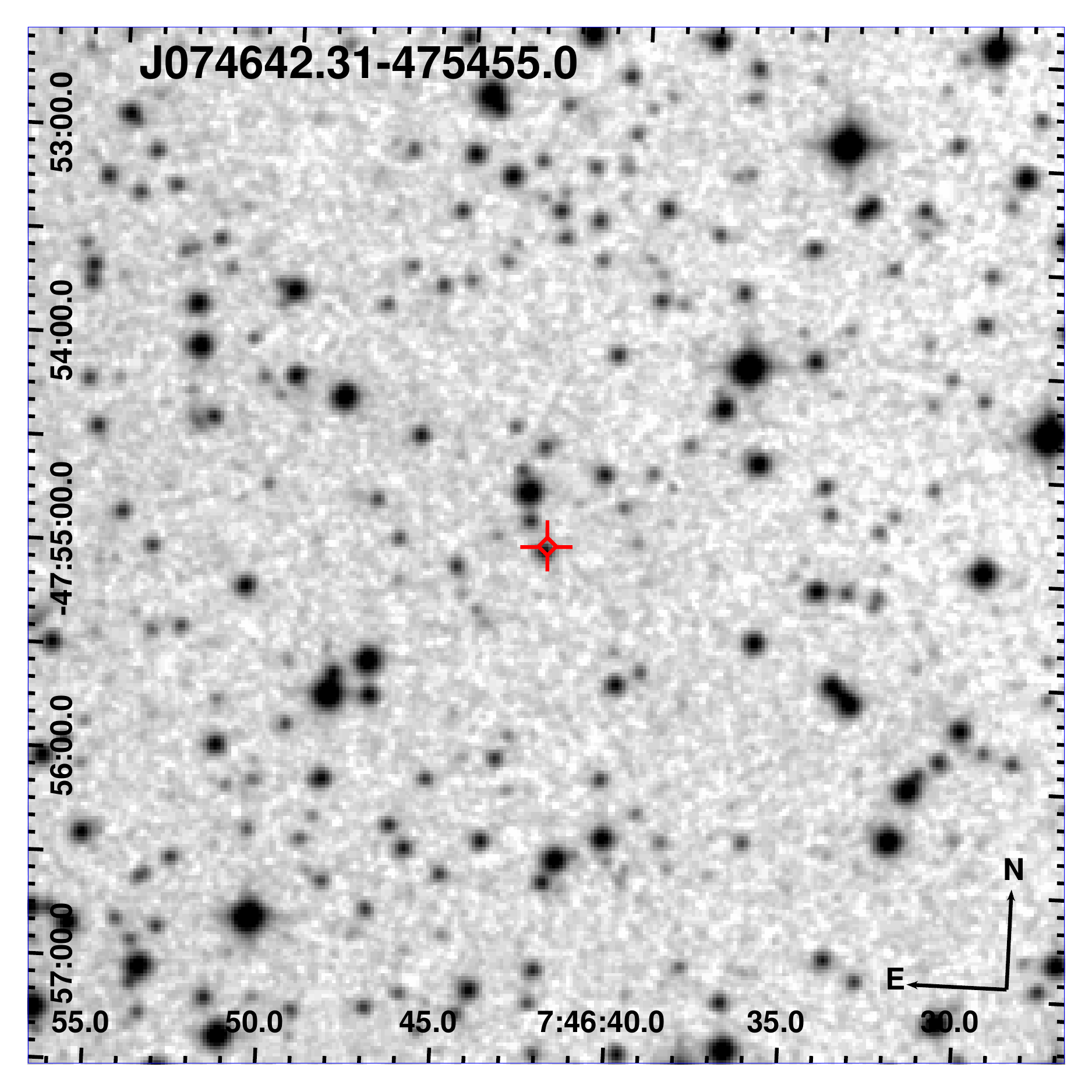}	
	\includegraphics[width=0.25\paperwidth]{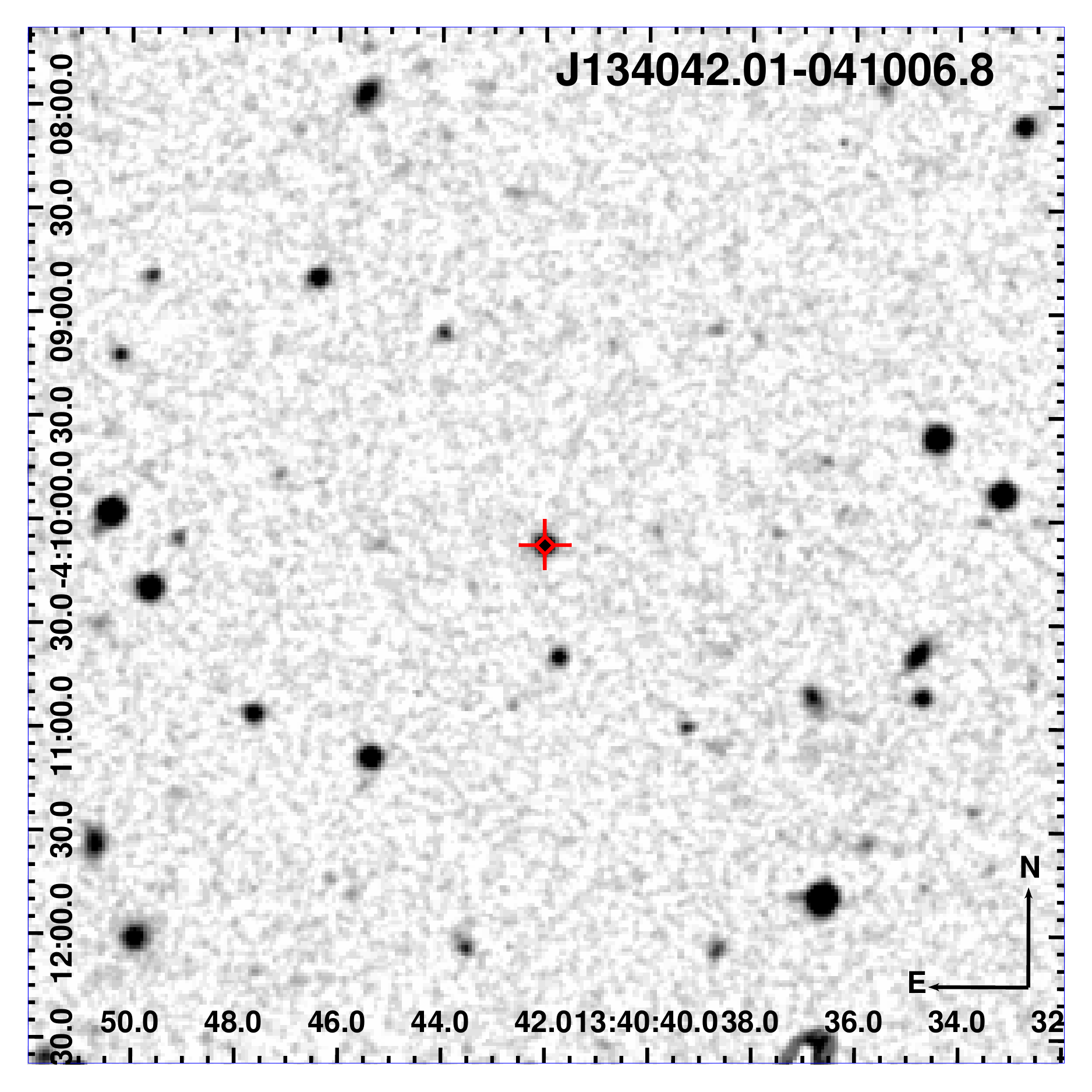}	
	\includegraphics[width=0.25\paperwidth]{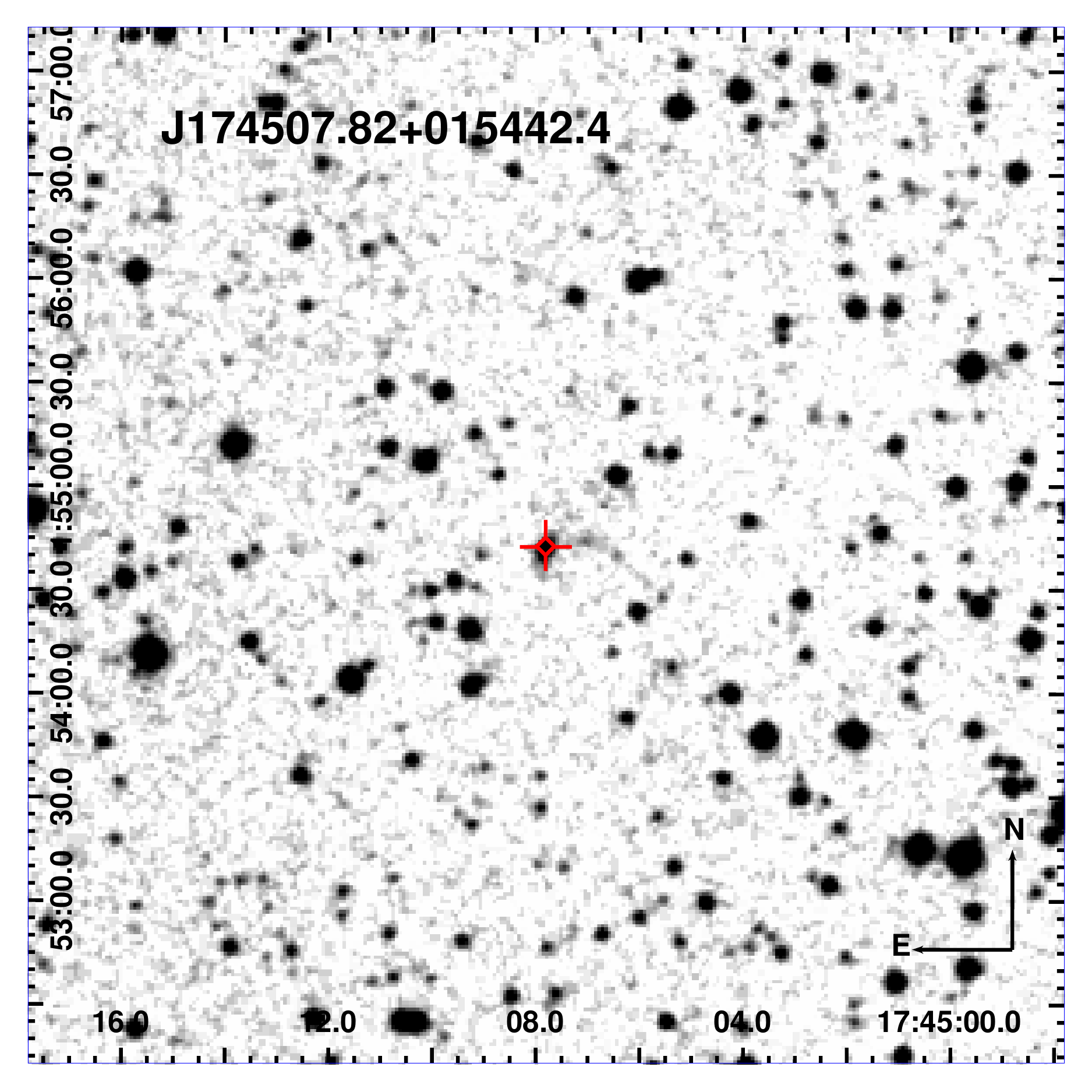}
\end{center}
	\caption{Optical images, 5$'$ on a side, of 9 of the \wse\ sources selected in this paper for 
	optical spectroscopic follow-up. The proposed optical counterparts are indicated with red marks and the fields are extracted from the DSS-II-Red survey.
	Object name, image scale and orientation are also reported in each panel.}
	\label{fig:fc1}
\end{figure*}

\begin{figure*}
	\includegraphics[width=0.25\paperwidth]{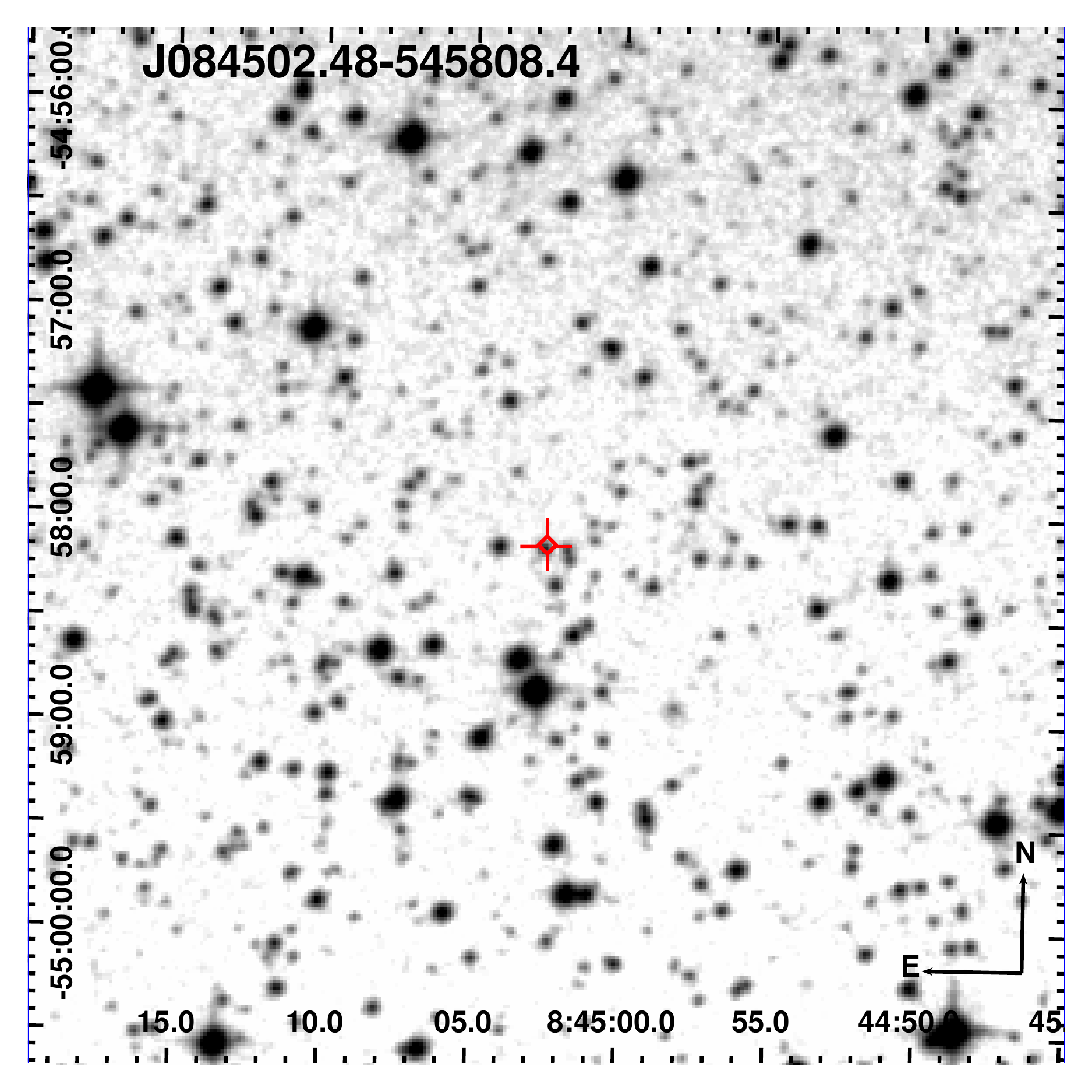}	
	\includegraphics[width=0.25\paperwidth]{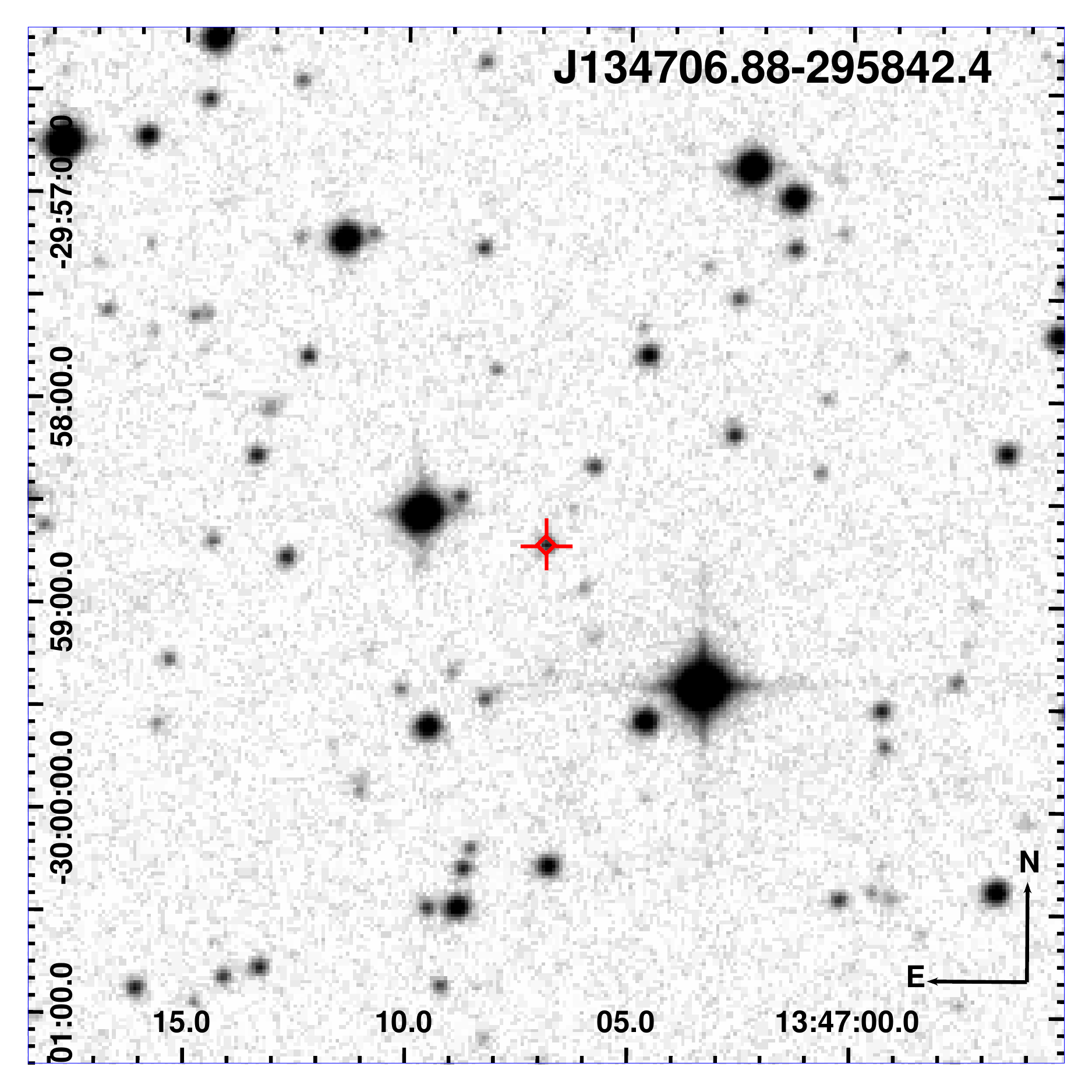}	
	\includegraphics[width=0.25\paperwidth]{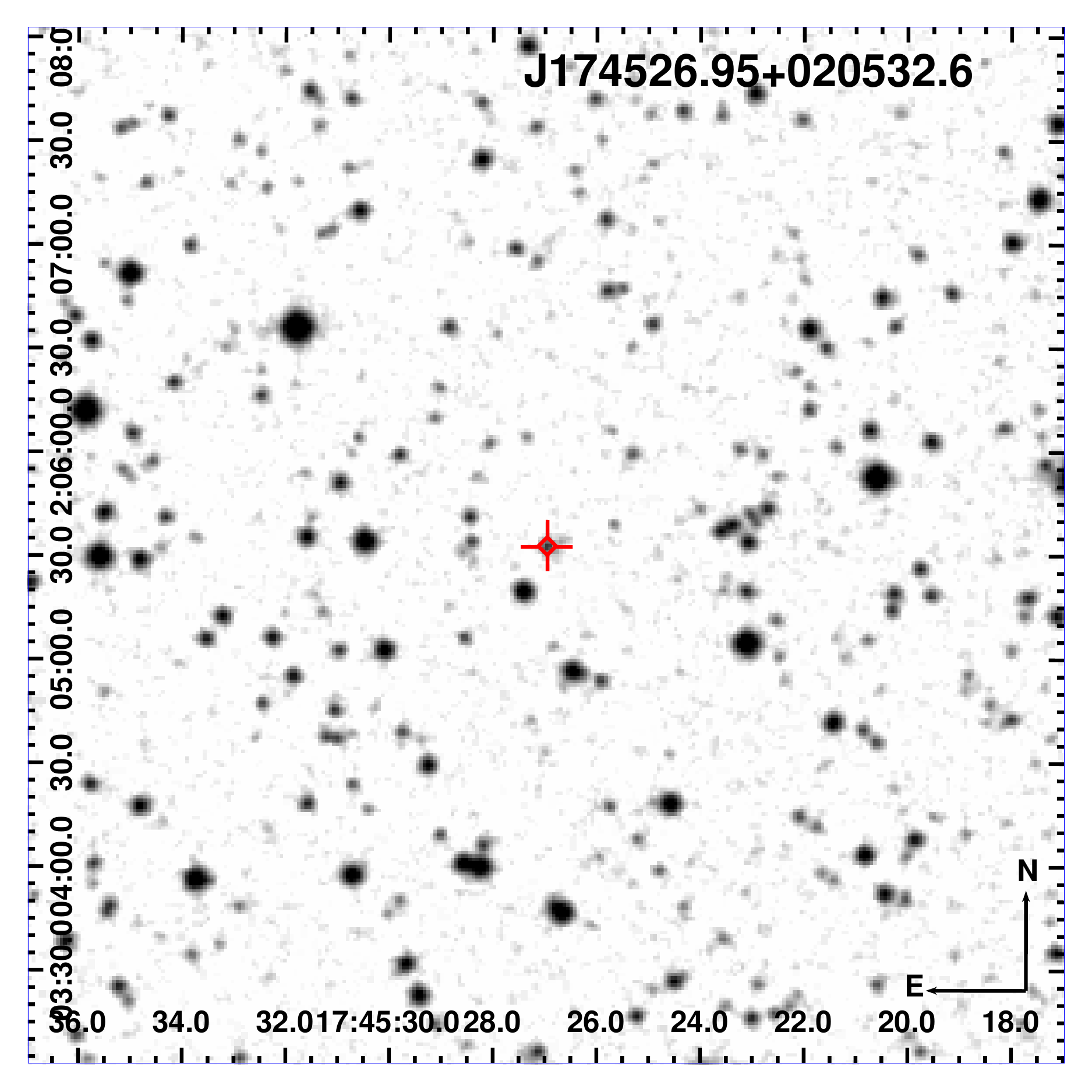}
	\includegraphics[width=0.25\paperwidth]{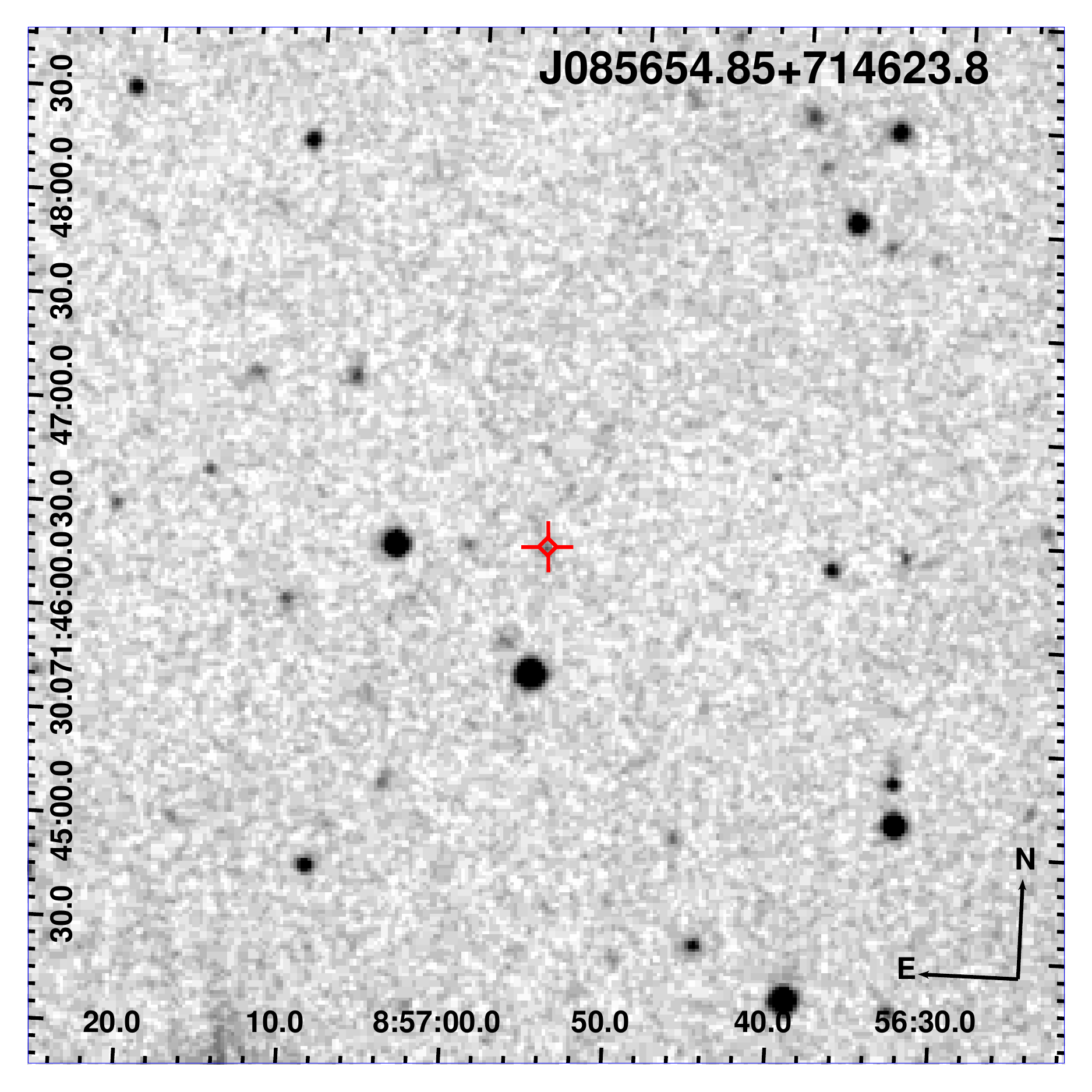}
	\includegraphics[width=0.25\paperwidth]{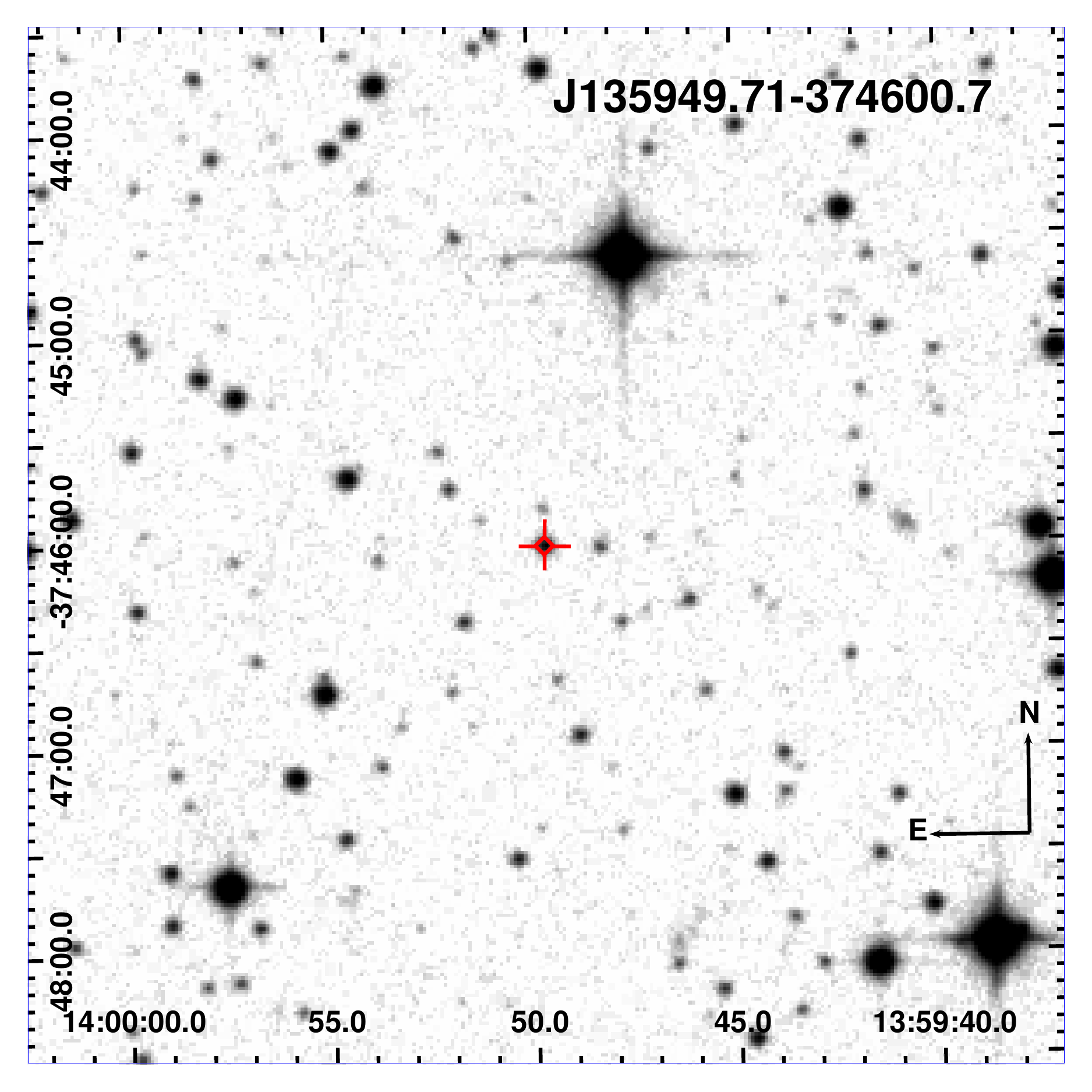}	
	\includegraphics[width=0.25\paperwidth]{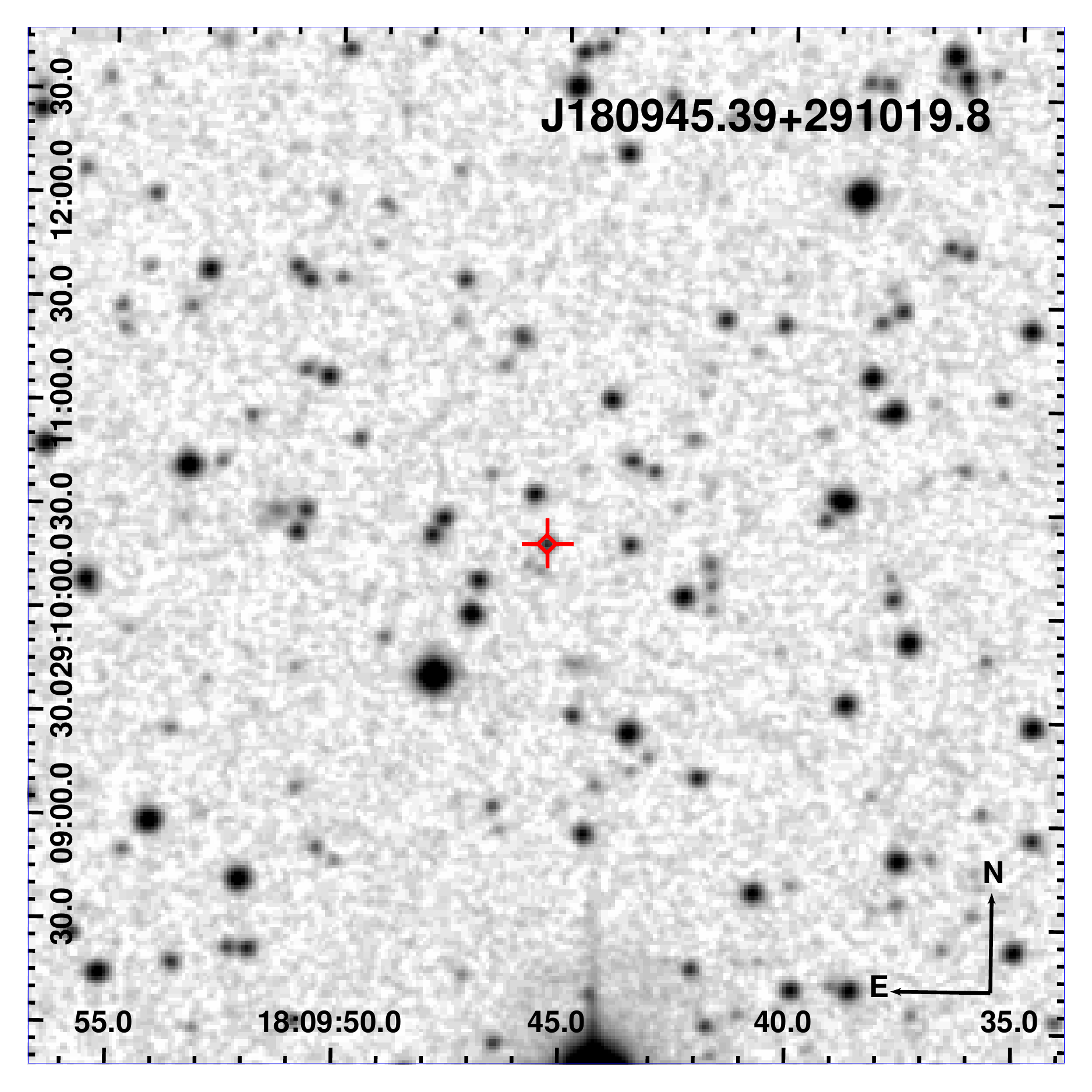}
	\includegraphics[width=0.25\paperwidth]{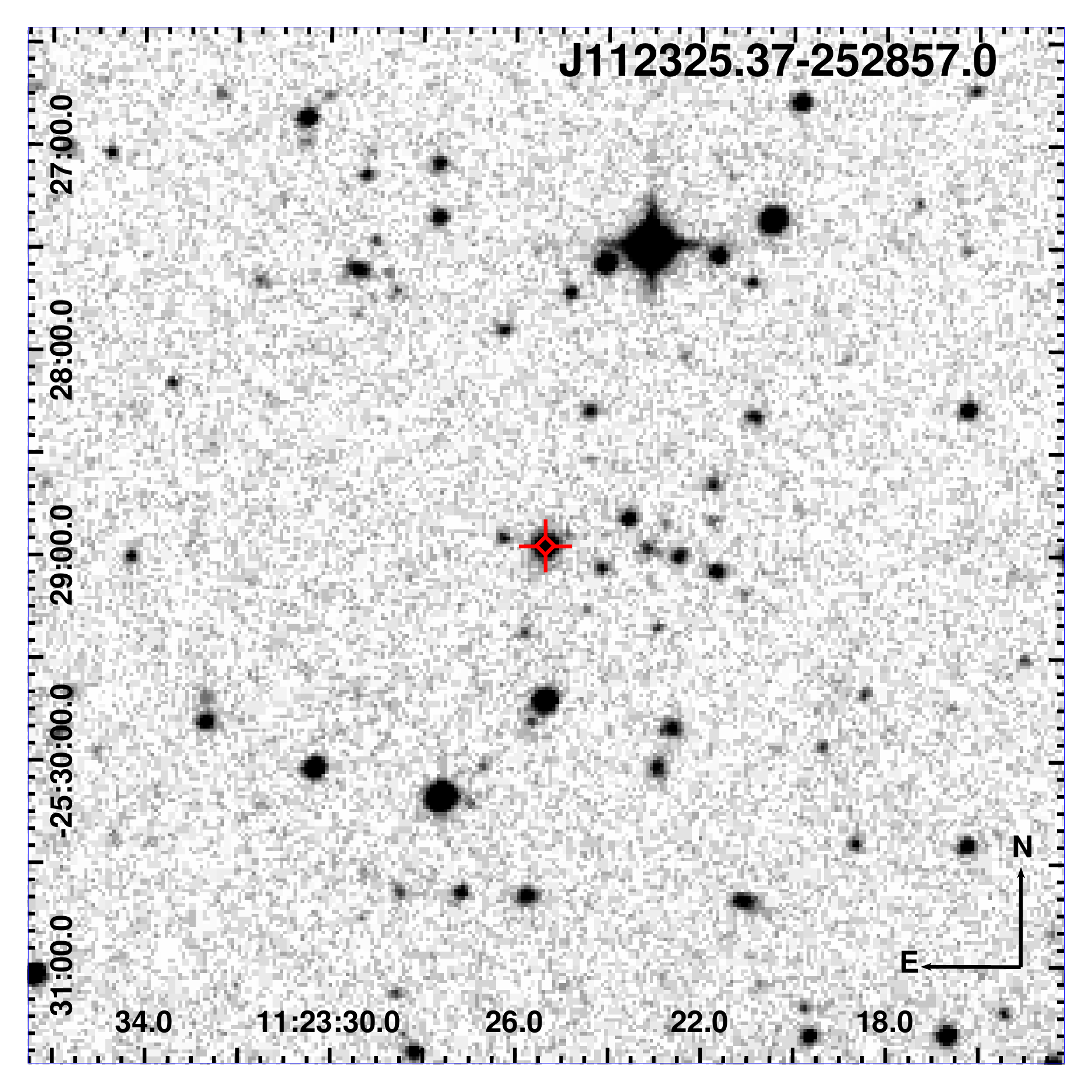}	
	\includegraphics[width=0.25\paperwidth]{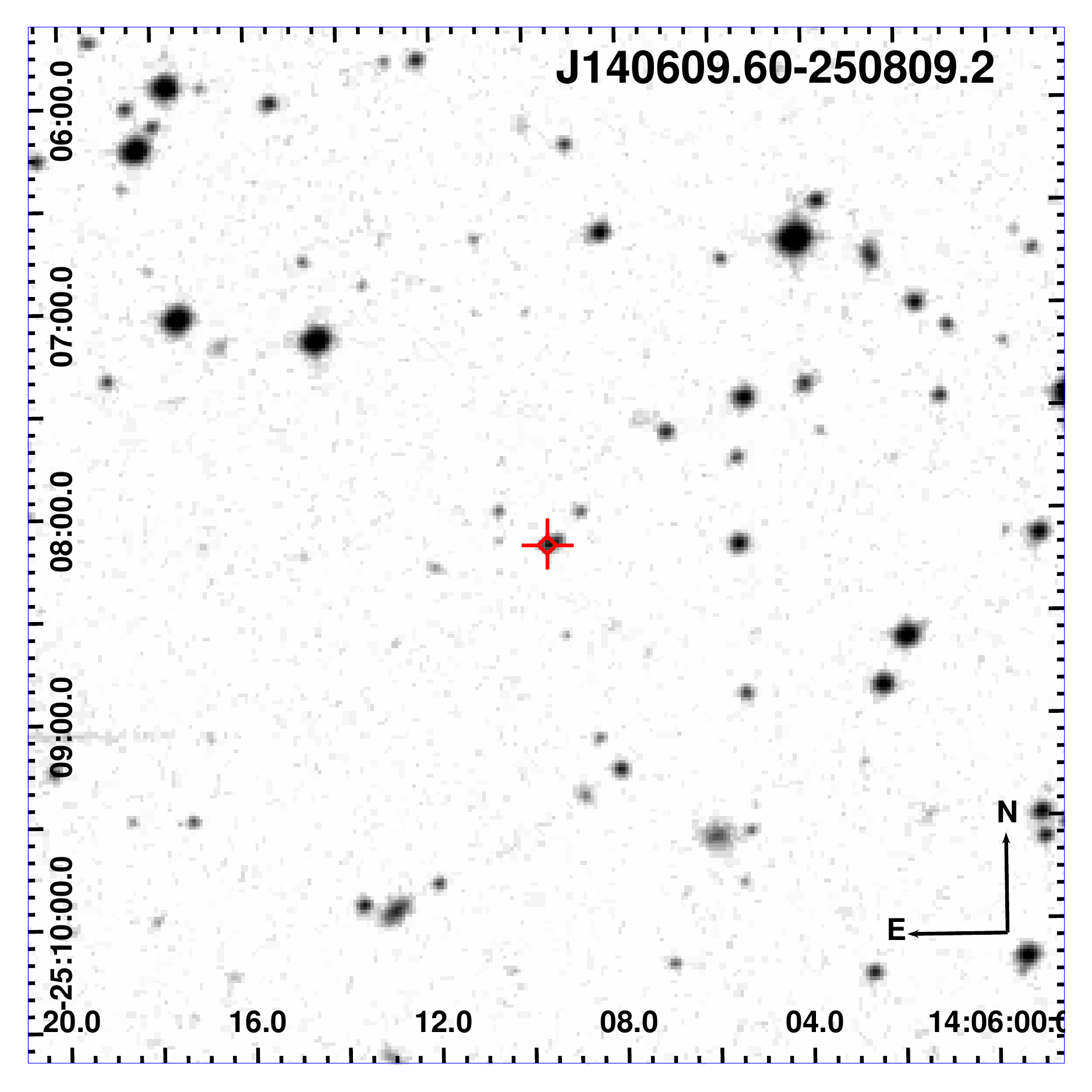}	
	\includegraphics[width=0.25\paperwidth]{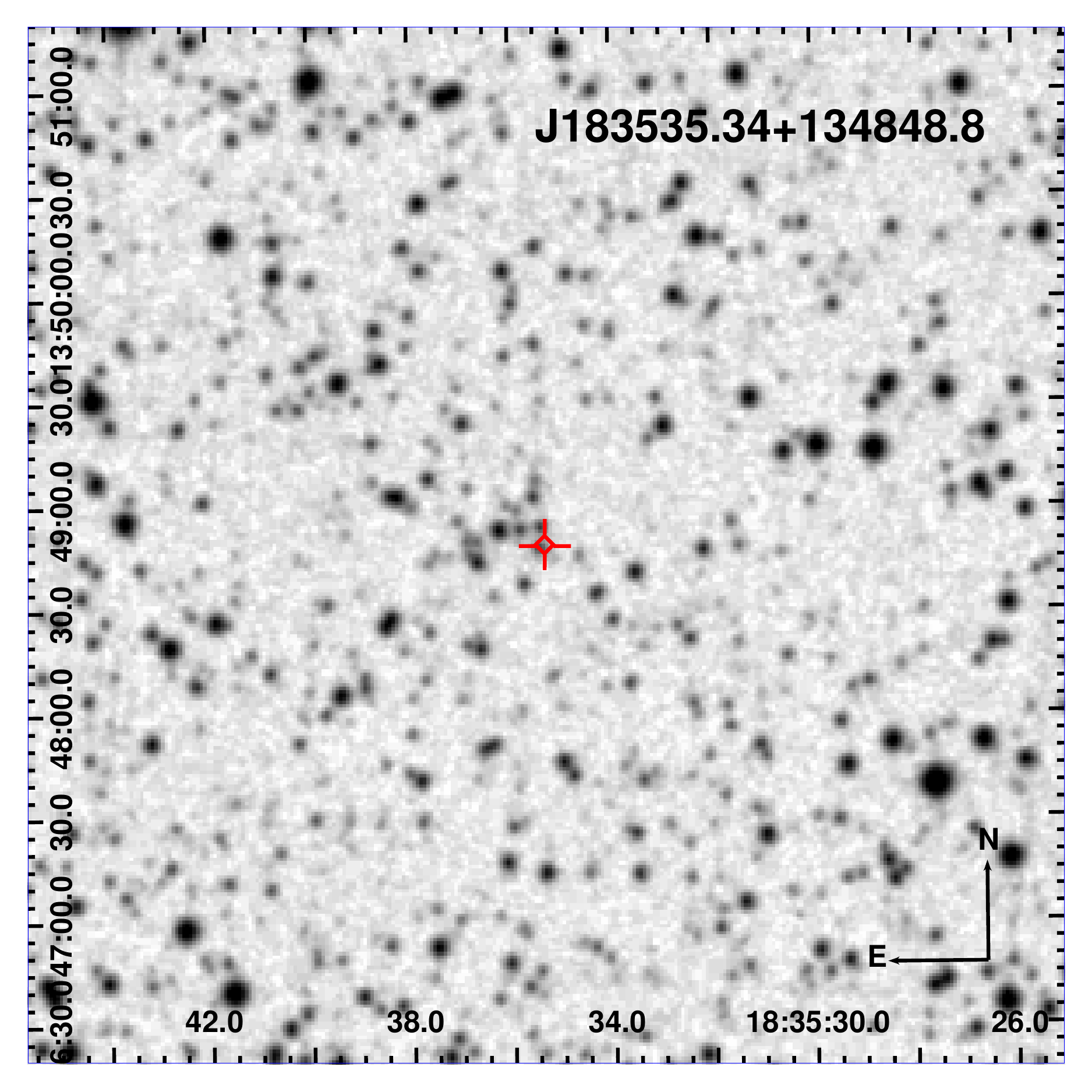}
	\caption{Same as Figure~\ref{fig:fc1}.}
	\label{fig:fc2}
\end{figure*}

\begin{figure*}
	\includegraphics[width=0.25\paperwidth]{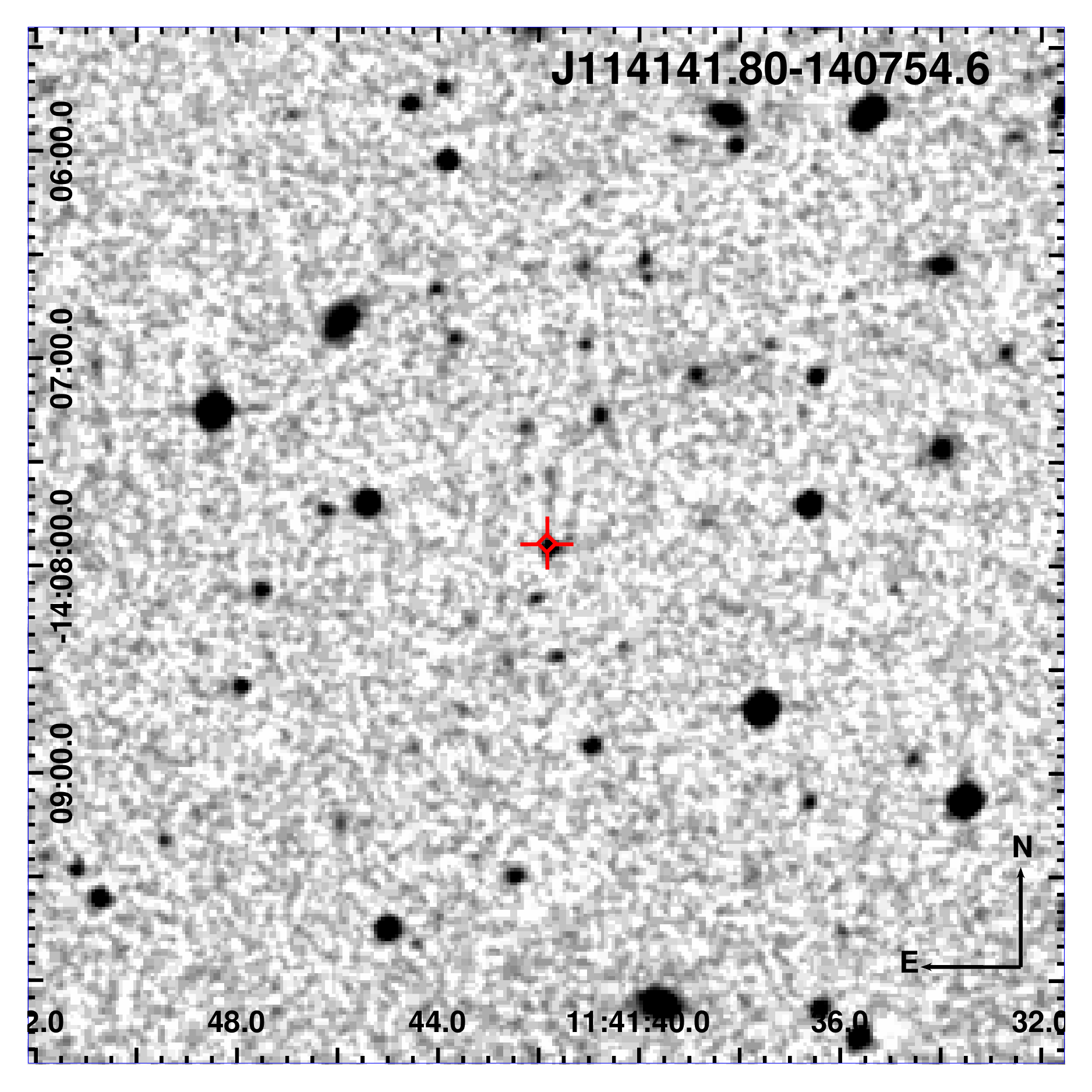}	
	\includegraphics[width=0.25\paperwidth]{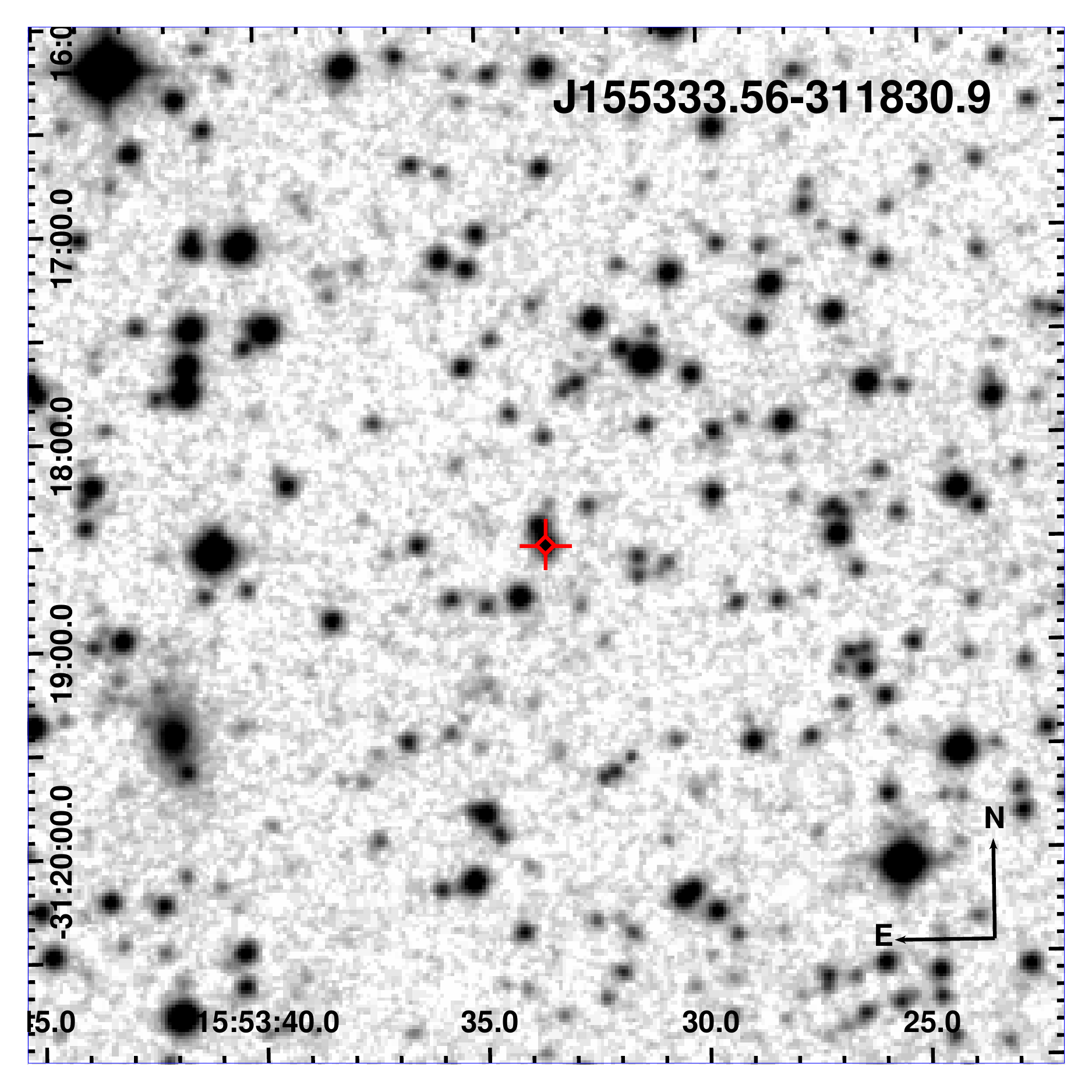}	
	\includegraphics[width=0.25\paperwidth]{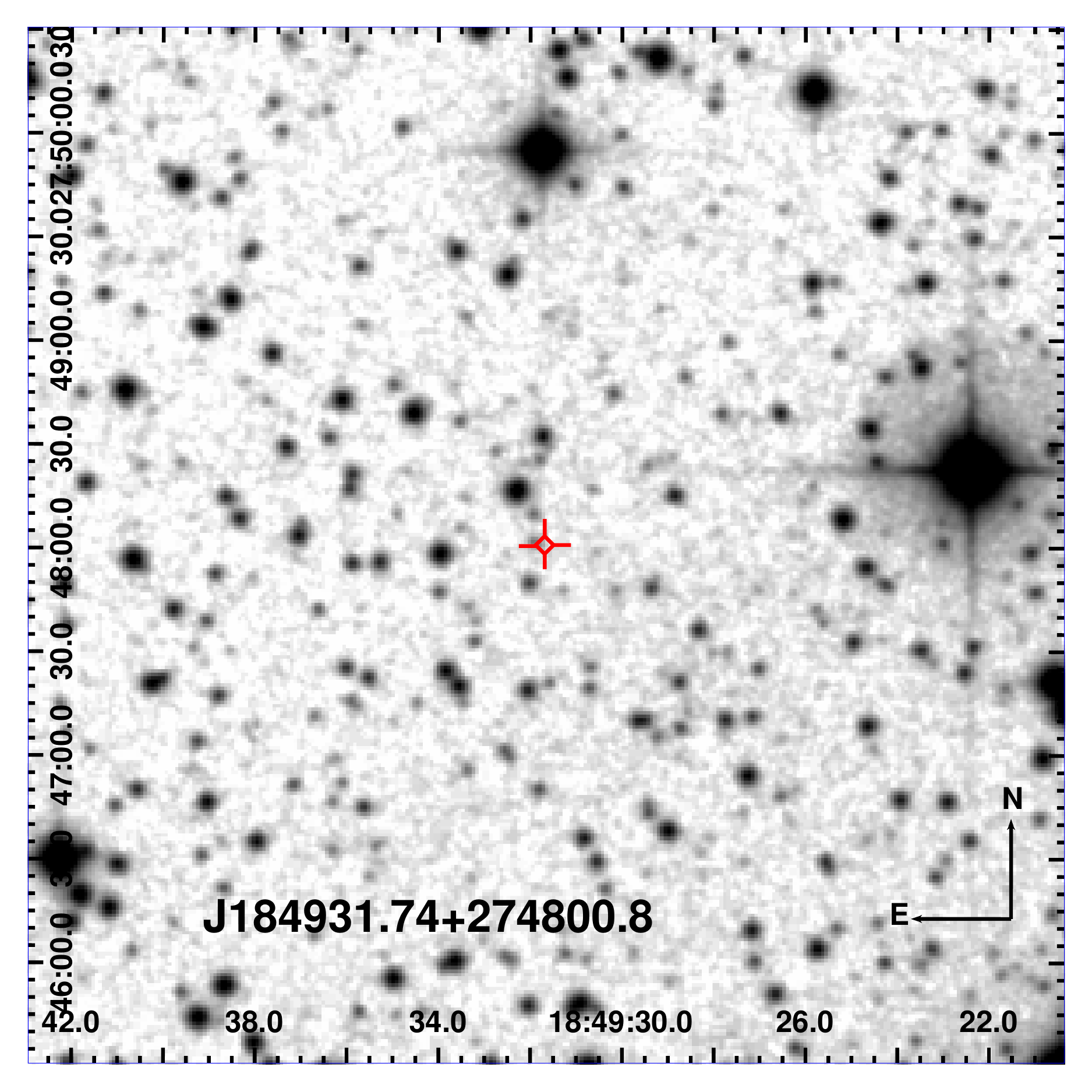}
	\includegraphics[width=0.25\paperwidth]{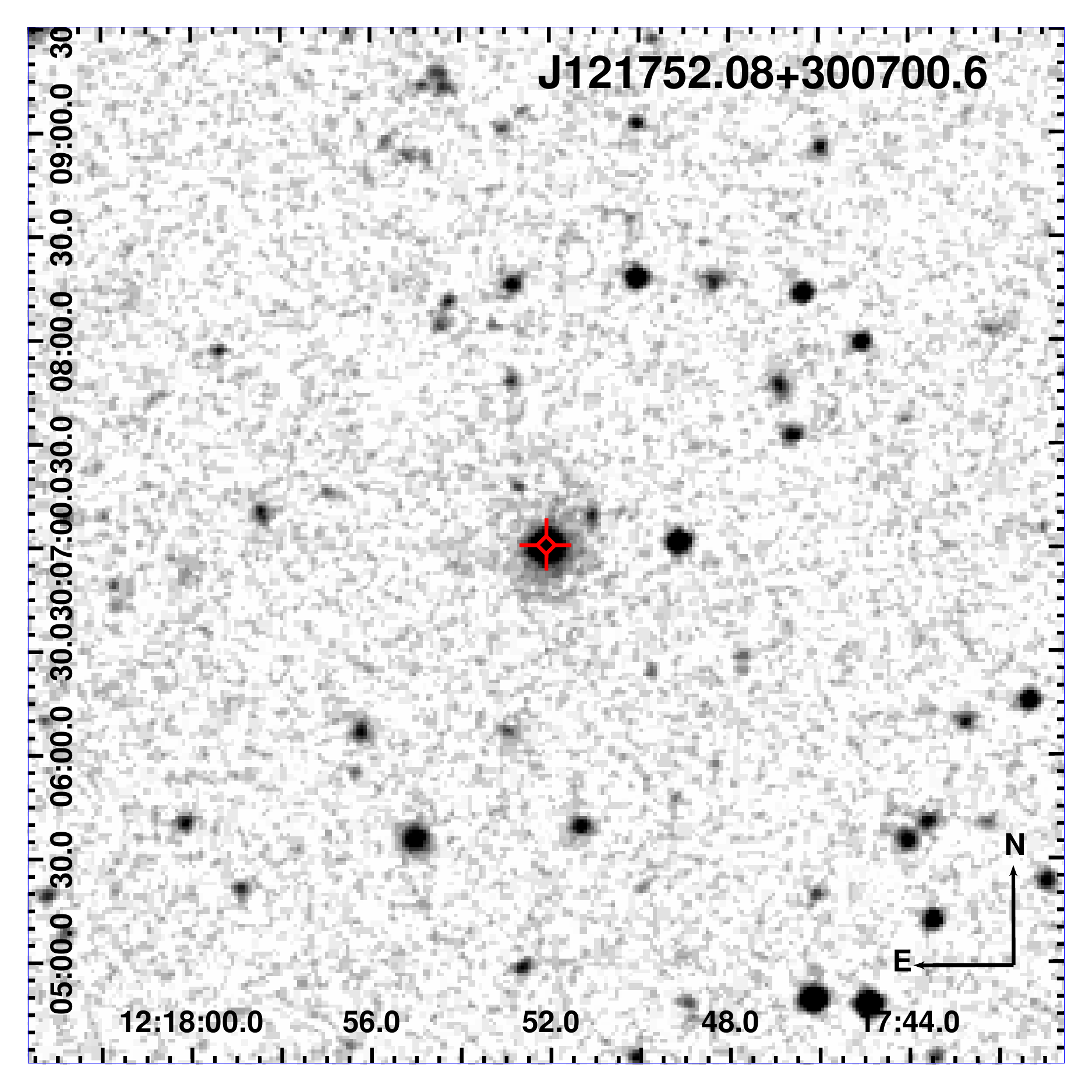}
	\includegraphics[width=0.25\paperwidth]{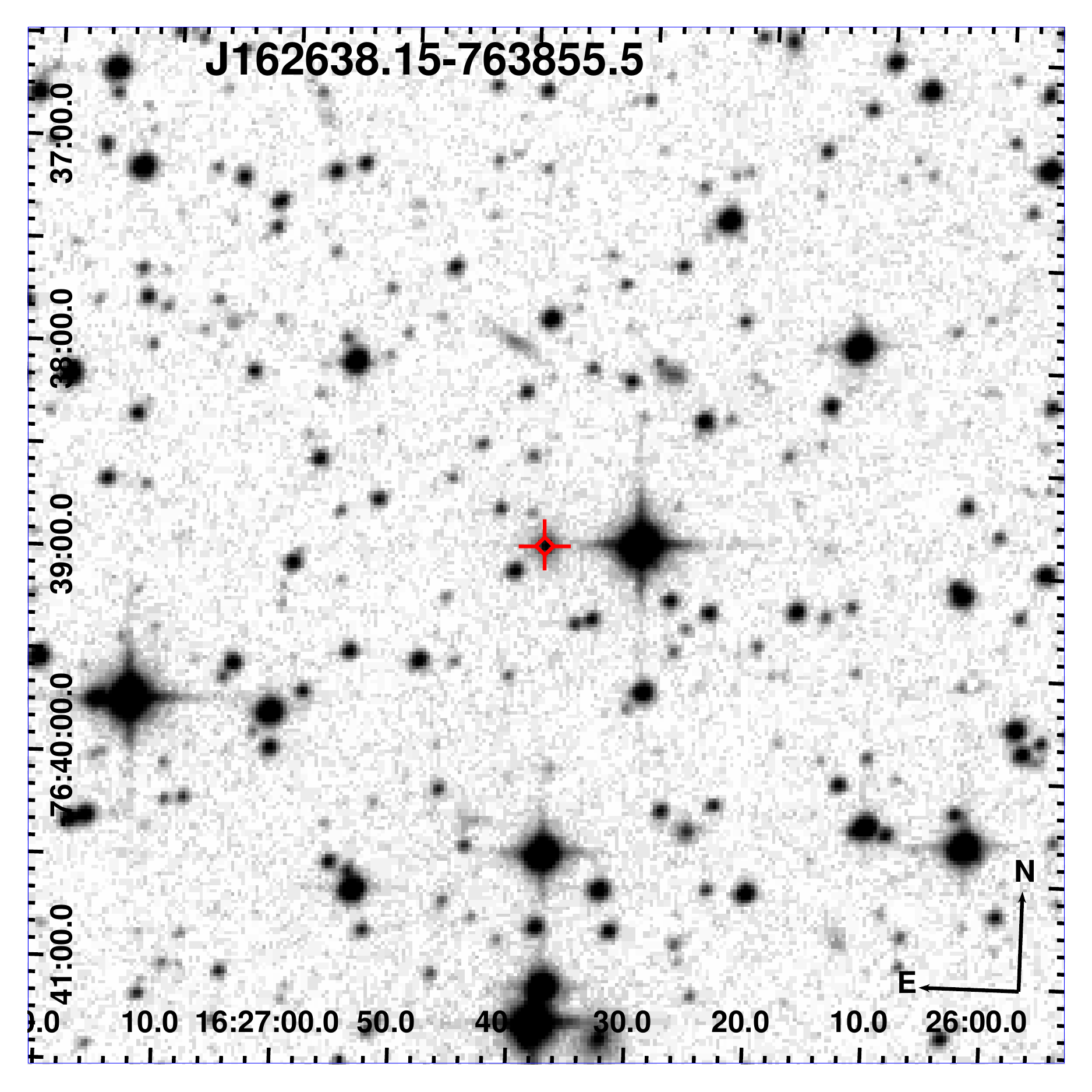}
	\includegraphics[width=0.25\paperwidth]{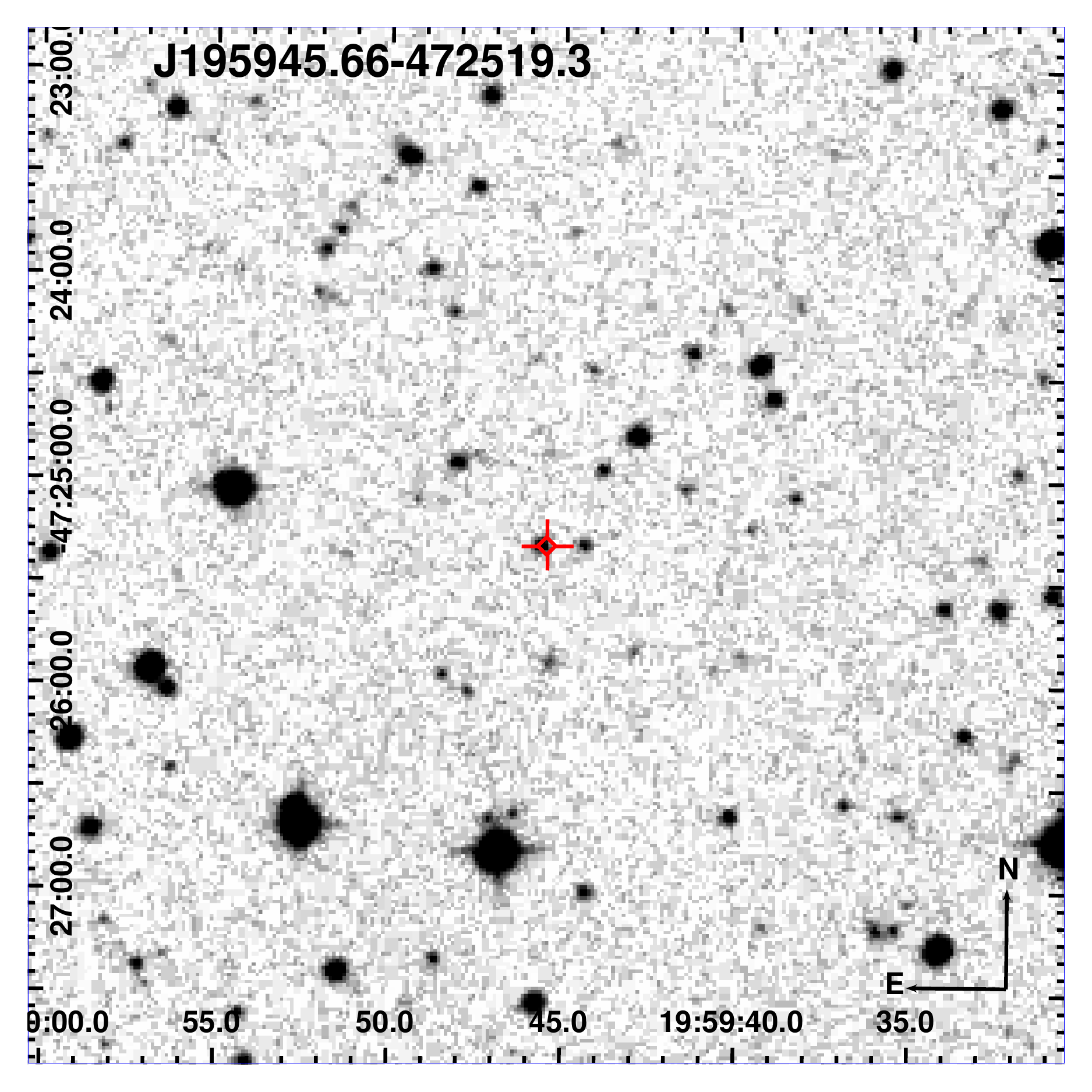}
	\includegraphics[width=0.25\paperwidth]{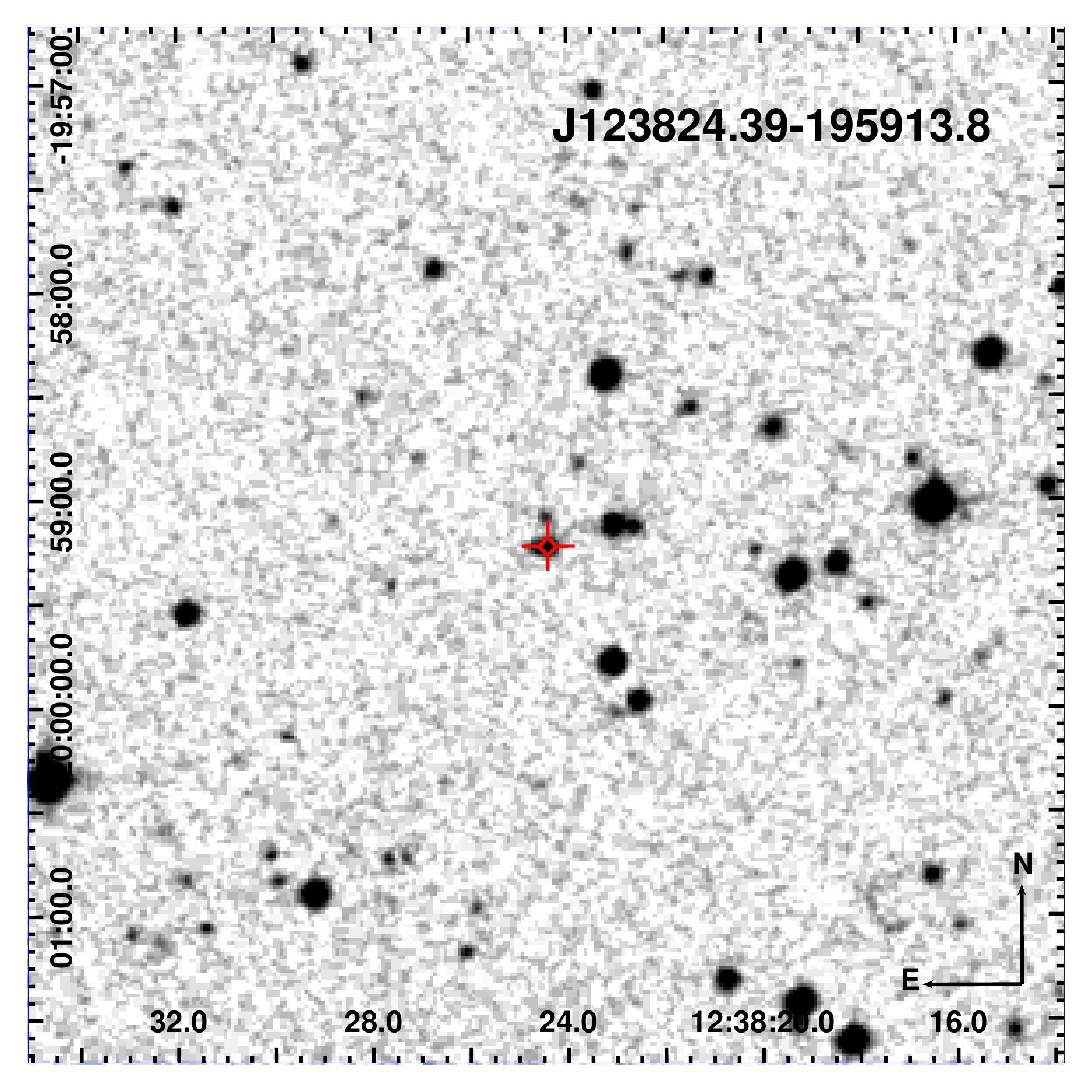}	
	\includegraphics[width=0.25\paperwidth]{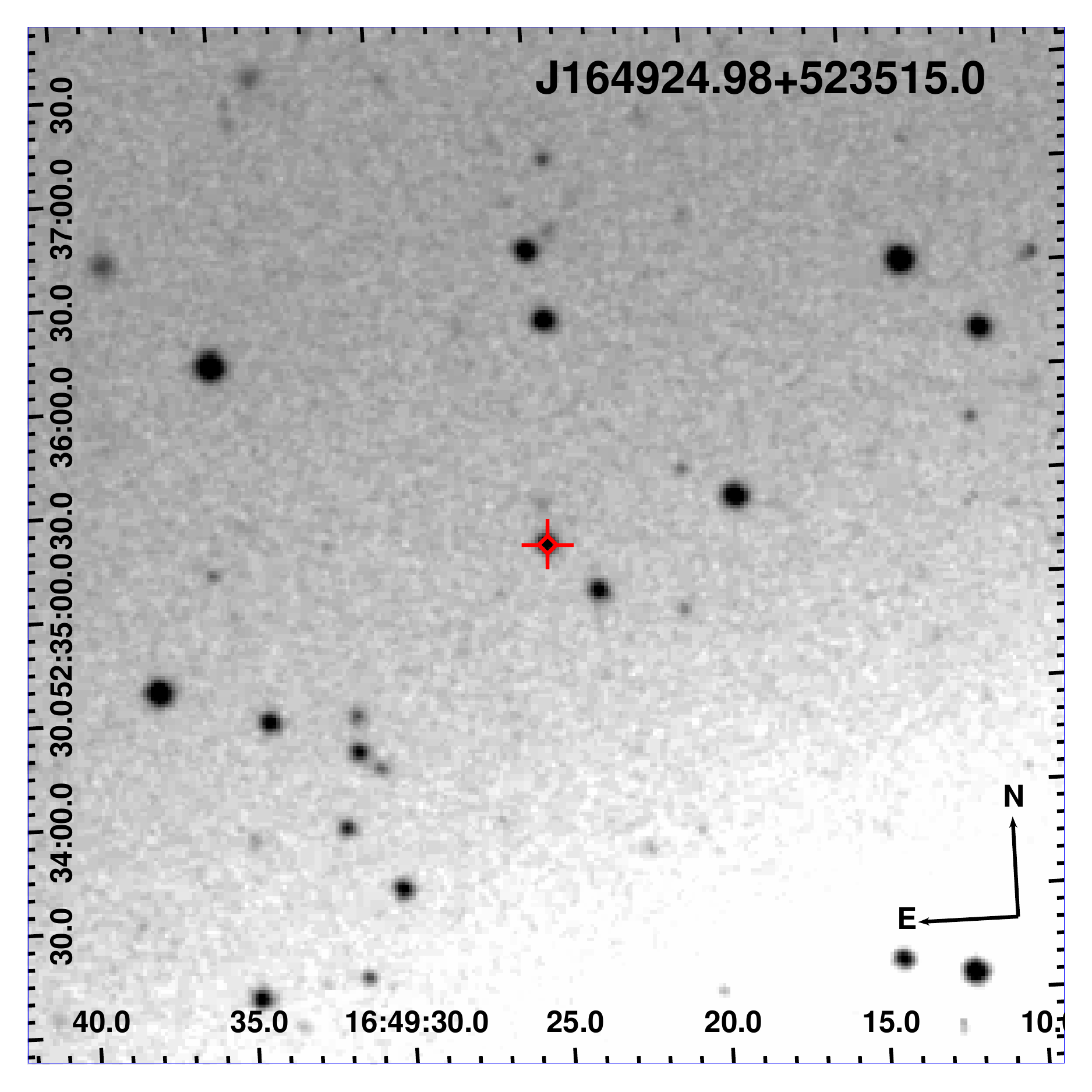}
	\includegraphics[width=0.25\paperwidth]{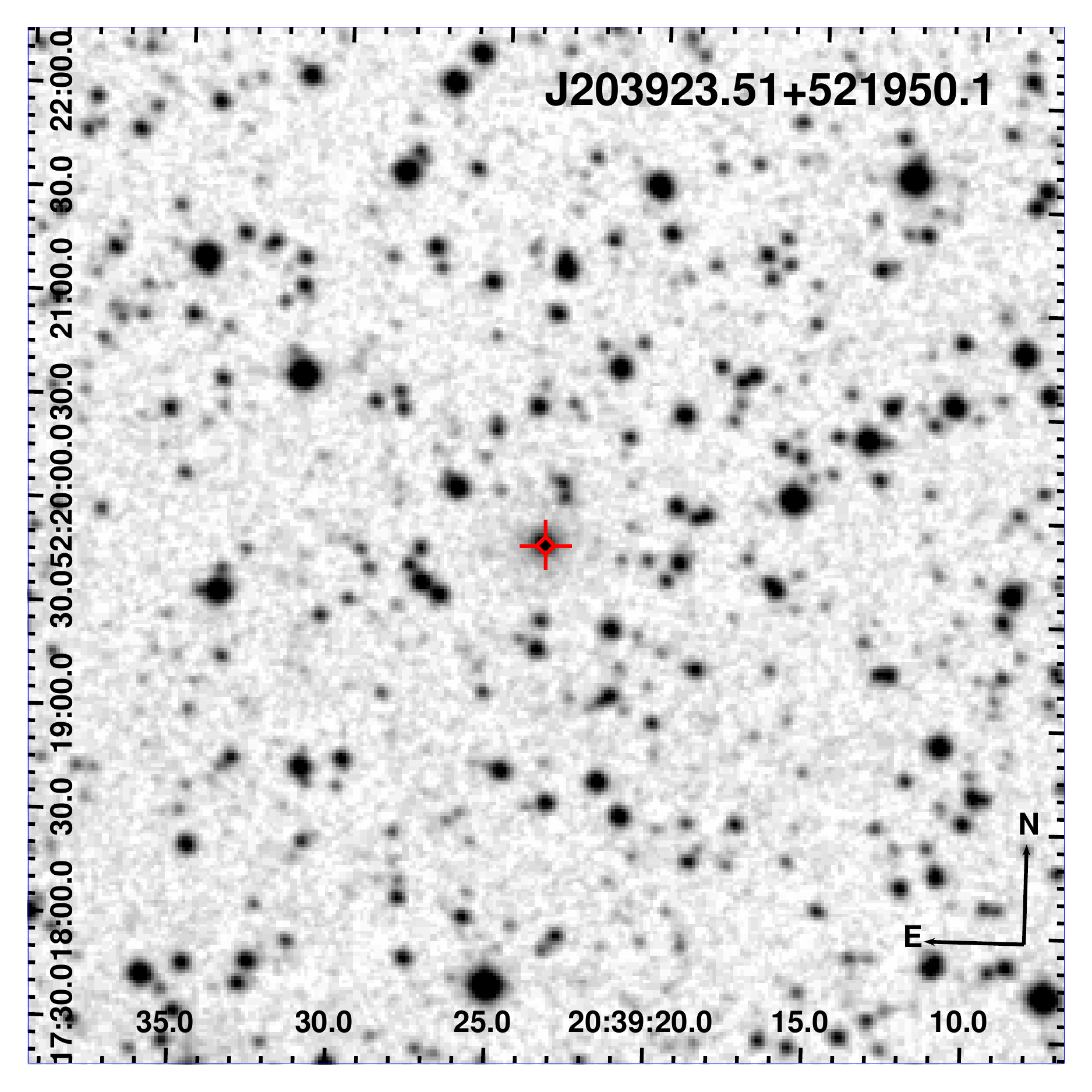}
	\caption{Same as Figure~\ref{fig:fc1}.}
	\label{fig:fc3}
\end{figure*}

We discuss our spectroscopic analysis in the following subsections while in
Table~\ref{tab:obslog} we summarize our results and report the multifrequency notes for each source
with the only exceptions of those listed in the \bzcat.
The finding charts for all the sources are shown in Figures~\ref{fig:fc1}-\ref{fig:fc3} using
the USNO-B1 Catalog \citep{monet03} and the Digitized Sky Survey\footnote{\scriptsize{\tt{http://archive.eso.org/dss/dss}}}.
We highlight that some of the sources observed during our campaign have also been observed at different observatories and groups. However we re-observed these targets for two main reasons: 1) when our observations were scheduled and performed
these spectra were not yet published; 2) due to the well-known BL Lac variability in the optical energy range, there is always the chance to
observe the source during a quiescent or low state and detect some emission and/or absorption features that could allow us to constrain their redshifts.

\section{Observations and Data Reduction}
\label{sec:obs}

Optical spectra of most of the sources accessible from the Northern hemisphere 
were acquired in remote observing mode with the KPNO Mayall 4-m telescope using the R-C spectrograph, while the sources in the Southern hemisphere
were observed remotely with the SOAR 4-m
telescope using the High Throughput Goodman spectrograph.
We also present the optical spectrum of the source \wse\ J024440.30-581954.5, 
observed on UT 2014 February 3 
in visitor mode at Las Campanas Observatory using
the Magellan 6.5-m telescope in combination with the IMACS instrument. Finally, we present optical
spectra of \wse\ J025333.64+321720.8 and \wse\ J085654.85+714623.8 obtained in visitor mode on UT 2014 February 22 with the Double Spectrograph (DBSP) at the Hale 200-inch Telescope at
Palomar Observatory.
Our scientific goal, i.e. the classification of the selected targets, is best achieved with broad spectral coverage. We thus adopted 
a slit width of 1.2$''$ (1$''$) and the
low resolution gratings yielding a dispersion of $\sim$3 (2) $\AA$ pixel$^{-1}$ for KPNO (SOAR). 
Our observations took place the nights
UT 2014 June 4 and 5
at KPNO and on UT 2014 April 24 at SOAR during grey time. 
The average seeing for both runs was
about 1$''$ and conditions were clear.
Additional observations were made on 2014 February 3 with the 6.5m
Baade Magellan telescope using the Inamori Magellan Areal Camera and
Spectrograph \citep[IMACS;][]{Bigelow98}. The f/2 camera was used in
combination with the 300 l mm$^{-1}$ grism (blaze angle 17.5 degrees)
and a 0.7$''$ slit to yield spectra with dispersion of 1.34 \AA\
pixel$^{-1}$ and FWHM resolution of $\sim$4 \AA. Conditions were
photometric and the seeing was generally $<1''$.
The whole set of spectroscopic data acquired at these telescopes 
was optimally extracted \citep{horne1986} and reduced following standard procedures with \textsc{IRAF} 
\citep{tody86}.
For each acquisition we performed 
bias subtraction, flat field correction and cosmic rays rejection. Since for
each target we secured two individual frames, we averaged them
according to their signal to noise ratios (SNR). 
For questionable spectral features, 
we have 
exploited the
availability of two individual exposures in order to better reject spurious 
signals. The wavelength calibration was achieved using the spectra
of He-Ne-Ar or Hg-Ar lamps that assure
coverage of the entire range. To take
into account drift and flexures of the instruments during the night, we took an arc frame before each target to guarantee a good wavelength solution
for the scientific spectra. The achieved accuracy is about $\sim$0.3 (0.4) $\AA$ rms for KPNO (SOAR).  
Although our program did not require accurate photometric
precision, we observed a spectrophotometric standard star to
perform relative flux calibration on each spectrum.
Finally, we corrected the whole sample for the Galactic absorption assuming $E_{B-V}$ values
computed using the Schlegel et al. (1998) 
and Cardelli et al. (1989)
relations.
To better detect faint spectral features and measure redshifts,
we normalised each spectrum to its continuum.
For the Palomar observations, we observed the candidates through a
1.5$''$ slit with the $\sim$ 5500 $\AA$ dichroic to split the light
across the blue and red arms of DBSP. Both sides have had resolving
powers $R \equiv \lambda / \Delta \lambda \sim 1000$, and the data
were reduced as above.

\begin{table*}
\caption{Selected sample and observation log. The sample is divided into three subsamples that are: 
potential counterparts of the UGSs, sources classified as AGUs, BL Lac candidates  and BL Lacs with uncertain redshift listed in the \bzcat . 
Among the \bzcat\ sources, those marked with an asterisk are detected by \fer\ \citep{nolan12}. Symbols used for the multifrequency notes are described in Sec. \ref{sec:sel}.  
\label{tab:obslog}}
\tiny
\resizebox{\textwidth}{!}{
\begin{tabular}{|lllllllll|}
\hline
Name & \fer\ & R.A.   & Dec.    & Obs.\,Date   & Exp. & notes & class & z\\
 & name & (J2000) & (J2000) & (yyyy-mm-dd) & (s) & & & \\ 
\hline
\hline
J112325.37$-$252857.0 & 2FGL J1123.3$-$2527 & 11:23:25.37 & $-$25:28:57.0 &  2014-04-23 & 300 & N,w,M,6,U,g -- ($z$=0.145784-QSO-Jones+09) & QSO & 0.148\\
J125949.80$-$374858.1 & 2FGL J1259.8$-$3749 & 12:59:49.80 & $-$37:48:58.2 & 2014-04-23 & 600 & S,N,w,U & BL Lac & ? \\ 
J132840.61$-$472749.2 & 2FGL J1328.5$-$4728 & 13:28:40.43 & $-$47:27:48.7 & 2014-04-23 & 600 & S,rf,w,M,U,g,u,x - SED in Takeuchi+13 & BL Lac & ? \\ 
J134042.01$-$041006.8 & 2FGL J1340.5$-$0412 & 13:40:42.01 & $-$04:10:07.2 & 2014-06-05 & 1200 & N,F,w,U,u,x - SED in Takeuchi+13 & BL Lac & ? \\ 
J134706.88$-$295842.4 & 2FGL J1347.0$-$2956 & 13:47:06.88 & $-$29:58:42.5 & 2014-04-23 & 900 & S,N,rf,w,M,U & BL Lac & ? \\ 
J173052.85$-$035247.2 & 2FGL J1730.6$-$0353 & 17:30:52.86 & $-$03:52:47.2 & 2014-06-05 & 900 & w,M & BL Lac & $\ge$0.776\\
J174507.82+015442.4   & 2FGL J1745.6+0203   & 17:45:07.82 & +01:54:42.6 & 2014-06-05 & 900 & N,w,M & QSO & 0.078 \\
J174526.95+020532.6   & 2FGL J1745.6+0203   & 17:45:26.96 & +02:05:32.8 & 2014-06-05 & 1200 & w,M & QSO & 0.335 \\ 
J183535.34+134848.8   & 2FGL J1835.4+1349   & 18:35:35.35 & +13:48:48.8 & 2014-06-05 & 600 & V,T,N,87,rf,w,M,U & BL Lac & ? \\
\hline                
\hline                
J025333.64+321720.8   & 2FGL J0253.4+3218   & 02:53:33.64 & +32:17:20.9 & 2014-02-22  & 600 & N,87,GB,c,rf,w & QSO &  0.859 \\
J074642.31$-$475455.0 & 2FGL J0746.5$-$4758 & 07:46:42.31 & $-$47:54:55.2 & 2014-04-23     & 600 & Pm,S,A,c,rf,w,M & BL Lac & ? \\ 
J084502.48$-$545808.4 & 2FGL J0844.8$-$5459 & 08:45:02.47 & $-$54:58:08.6 & 2014-04-23    & 600 & Pm,A,rf,U,X & BL Lac & ? \\ 
J085654.85+714623.8   & 2FGL J0856.0+7136   & 08:56:54.86 & +71:46:23.9 & 2014-02-22 & 600  & W,N,87,GB,c,rf,w,M,g,X & QSO & 0.542\\ 
J114141.80$-$140754.6 & 2FGL J1141.7$-$1404 & 11:41:41.84 & $-$14:07:53.5 & 2014-04-23    & 900 & L,Pm,N,c,w,M,U & BL Lac & ? \\
J123824.39$-$195913.8 & 2FGL J1238.1$-$1953 & 12:38:24.40 & $-$19:59:13.5 & 2014-04-23    & 900 & Pm,N,A,c,rf,w,M,g,X & BL Lac & ? \\
J140609.60$-$250809.2 & 2FGL J1406.2$-$2510 & 14:06:09.60 & $-$25:08:09.3 & 2014-04-23    & 600 & L,N,w,M,U,u,x - SED in Takeuchi+13 & BL Lac & ? \\
J162638.15$-$763855.5 & 2FGL J1626.0$-$7636 & 16:26:38.18 & $-$76:38:55.5 & 2014-04-23   & 300  & Pm,S,c,rf,w,M,g & BL Lac & 0.1050 \\
J184931.74+274800.8   & 2FGL J1849.5+2744   & 18:49:31.69 & +27:48:00.9 & 2014-06-04  &  1800  & W,N,87,c,rf,w,M & BL Lac & ? \\ 
J195945.66$-$472519.3 & 2FGL J1959.9$-$4727 & 19:59:45.48 & $-$47:25:19.3 & 2014-04-23 &  600  & S,w,M,U,g,u,X,x - SED in Takeuchi+13 & BL Lac & $\ge$0.519 \\ 
\hline                
\hline                
J024440.30$-$581954.5 & BZBJ0244$-$5819     & 02:44:40.31 & $-$58:19:54.6 & 2014-02-03 & 600 & BL Lac candidate at z=0.265 & BL Lac & $\ge$0.265 \\ 
J121752.08+300700.6   & BZBJ1217+3007$^*$   & 12:17:52.09 & +30:07:00.7 & 2014-06-04  &  300  & BL Lac at z=0.13? & BL Lac & ? \\ 
J135949.71$-$374600.7 & BZBJ1359$-$3746$^*$ & 13:59:49.72 & $-$37:46:00.8 & 2014-04-23   &  300 & BL Lac & BL Lac & ? \\
J155333.56$-$311830.9 & BZBJ1553$-$3118$^*$ & 15:53:33.56 & $-$31:18:31.0 & 2014-04-23 & 300 & BL Lac (z=0.132-BLLac-Masetti+13) & BL Lac & ?\\
J164924.98+523515.0   & BZBJ1649+5235$^*$   & 16:49:25.00 & +52:35:15.0 & 2014-06-05  &  900  & BL Lac candidate at z=? & BL Lac & ? \\
J170209.63+264314.7   & BZBJ1702+2643       & 17:02:09.64 & +26:43:14.8 & 2014-06-05   &  1200 & BL Lac candidate at z=? & BL Lac & ? \\
J180945.39+291019.8   & BZBJ1809+2910$^*$   & 18:09:45.39 & +29:10:20.0 & 2014-06-05   & 900  & BL Lac candidate at z=? & BL Lac & ? \\
J203923.51+521950.1   & BZBJ2039+5219$^*$   & 20:39:23.50 & +52:19:49.9 & 2014-06-04   &  600 & BL Lac candidate at z=0.053 & BL Lac & ? \\
\hline
\end{tabular}}
\end{table*}

\section{Results}
\label{sec:results}
\subsection{Unidentified Gamma-ray Sources}
\begin{figure}
	\includegraphics[width=0.3\paperwidth]{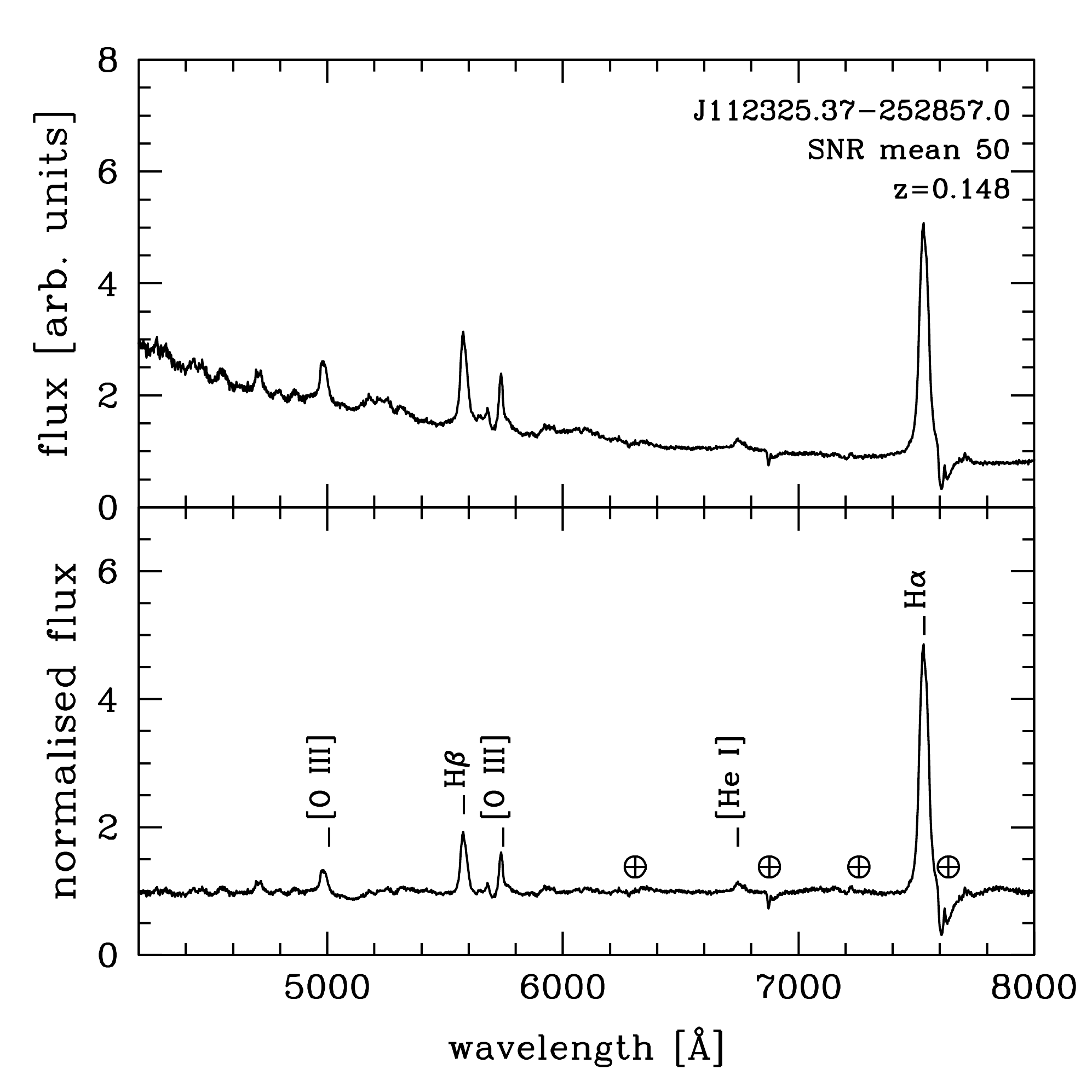}
	\caption{Upper panel: optical spectra observed at SOAR of 
		\wse\ J112325.37-252857.0, potential counterpart associated
		with 2FGL J1123.3-2527, classified as a QSO at z=0.148 on the basis of the 
		identification of the emission lines shown in the spectra. The average SNR is also reported
		in the figure. Lower panel: the normalised spectrum is shown. The symbol $\oplus$ indicates atmospheric telluric features.}
	\label{fig:ugs1}
\end{figure}

\begin{figure}
	\includegraphics[width=0.3\paperwidth]{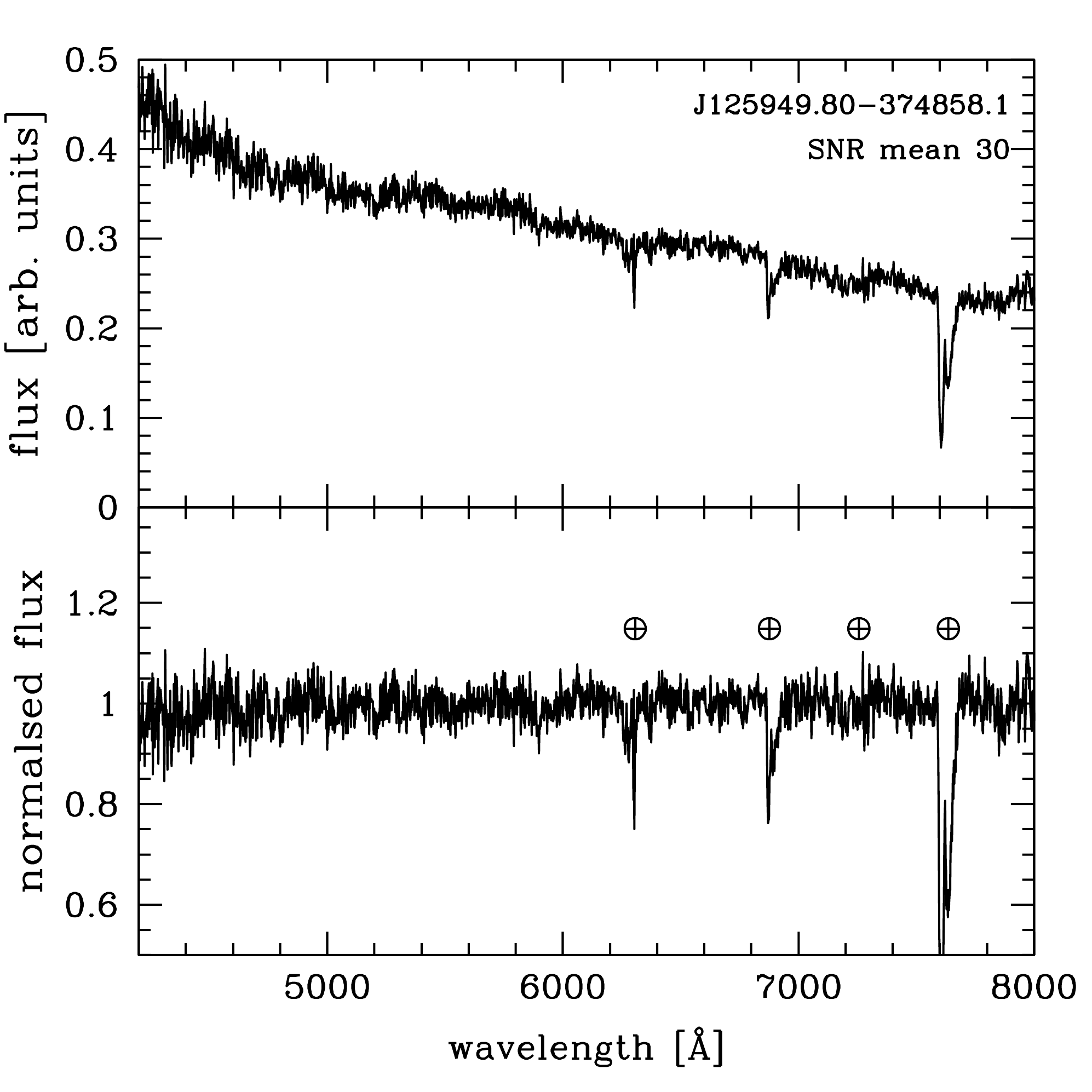}
	\caption{Upper panel: optical spectra observed at SOAR of 
		\wse\ J125949.80-374858.1, potential counterpart associated
		with FGL J1259.8-3749, classified as a BL Lac on the basis of its featureless continuum.
		Lower panel: as in Figure \ref{fig:ugs1}.}
	\label{fig:ugs2}
\end{figure}

\begin{figure}
	\includegraphics[width=0.3\paperwidth]{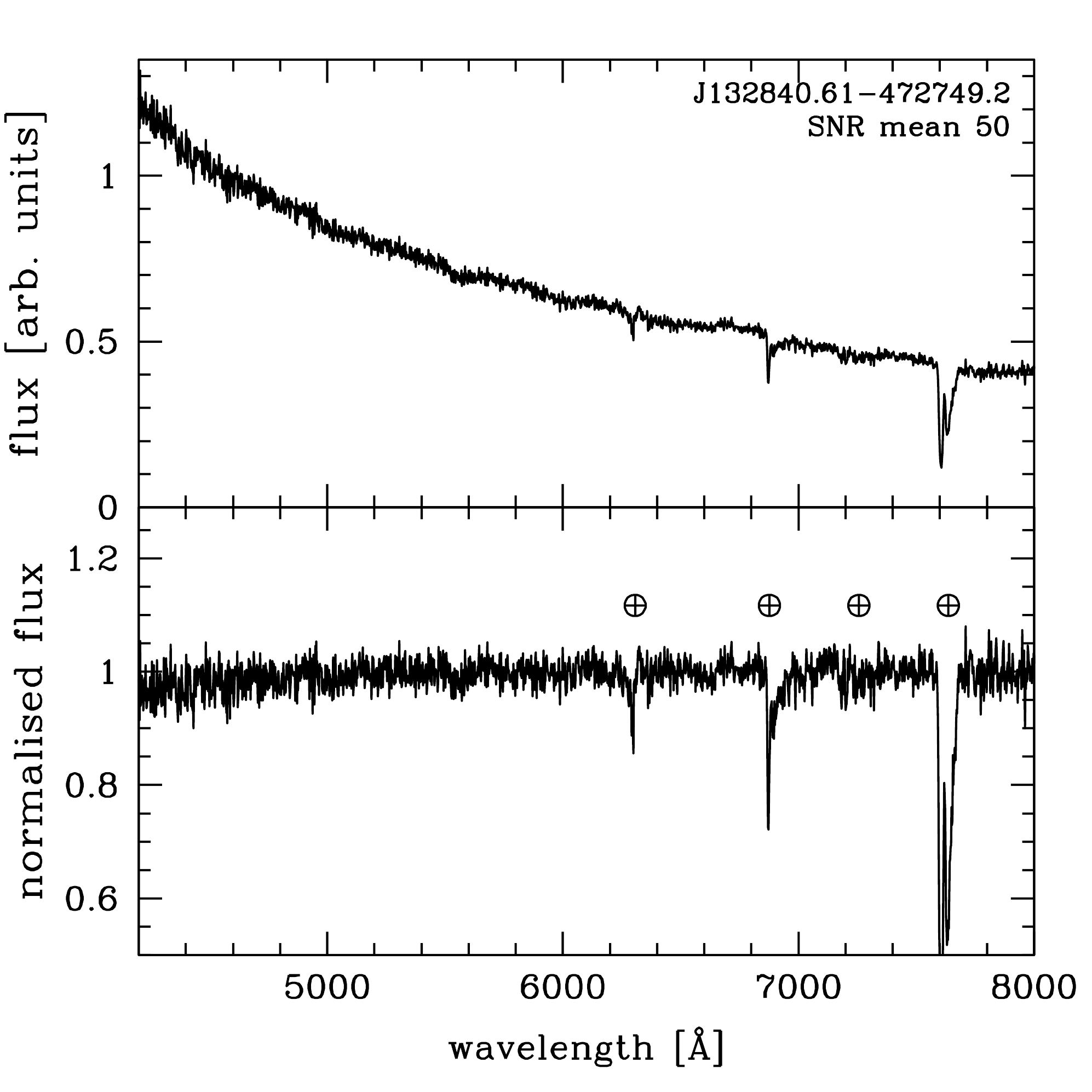}
	\caption{Upper panel: optical spectra observed at SOAR of 
		\wse\ J132840.61-472749.2, potential counterpart associated
		with 2FGL J1328.5-4728, classified as a BL Lac on the basis of its featureless continuum. 
		Lower panel: as in Figure \ref{fig:ugs1}.}
	\label{fig:ugs3}
\end{figure}

\begin{figure}
	\includegraphics[width=0.3\paperwidth]{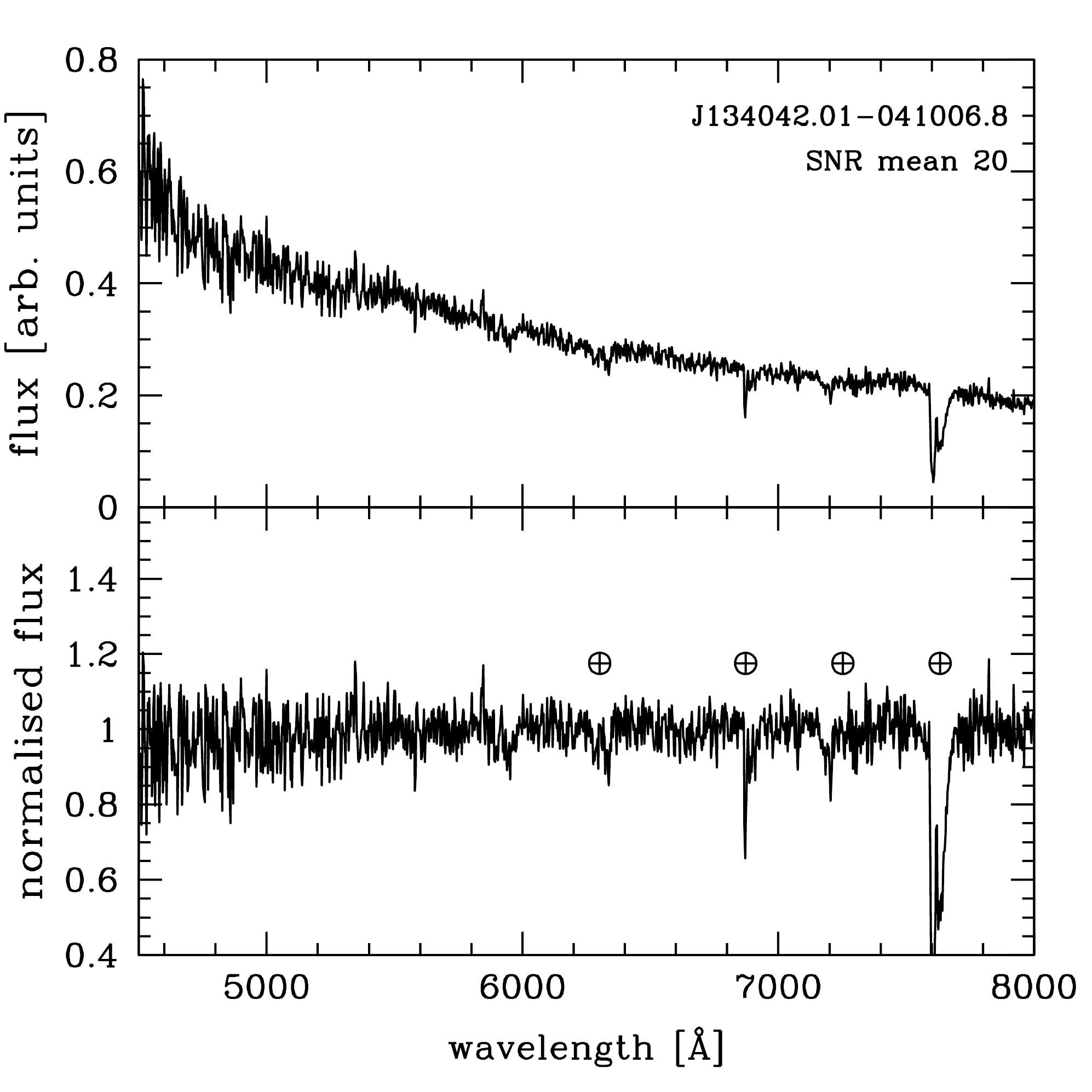}
	\caption{Upper panel: optical spectra observed at KPNO of 
		\wse\ J134042.01-041006.8, potential counterpart associated
		with 2FGL J1340.5-0412, classified as a BL Lac on the basis of its featureless continuum. 
		Lower panel: as in Figure \ref{fig:ugs1}.}
	\label{fig:ugs4}
\end{figure}

\begin{figure}
	\includegraphics[width=0.3\paperwidth]{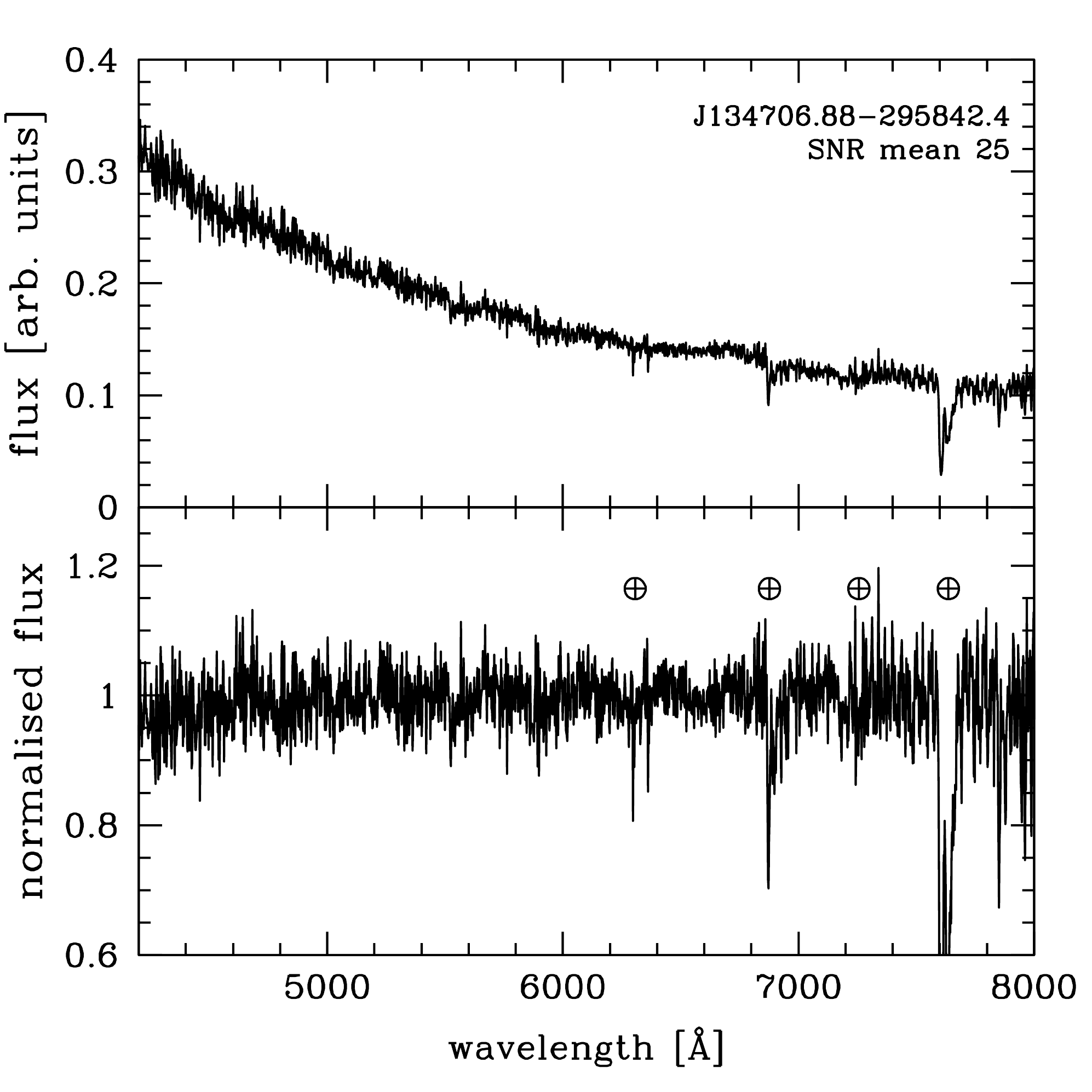}
	\caption{Upper panel: optical spectra observed at SOAR of 
		\wse\ J134706.88-295842.4, potential counterpart associated
		with 2FGL J1347.0-2956, classified as a BL Lac on the basis of its featureless continuum. 
		Lower panel: as in Figure \ref{fig:ugs1}.}
	\label{fig:ugs5}
\end{figure}

\begin{figure}
	\includegraphics[width=0.3\paperwidth]{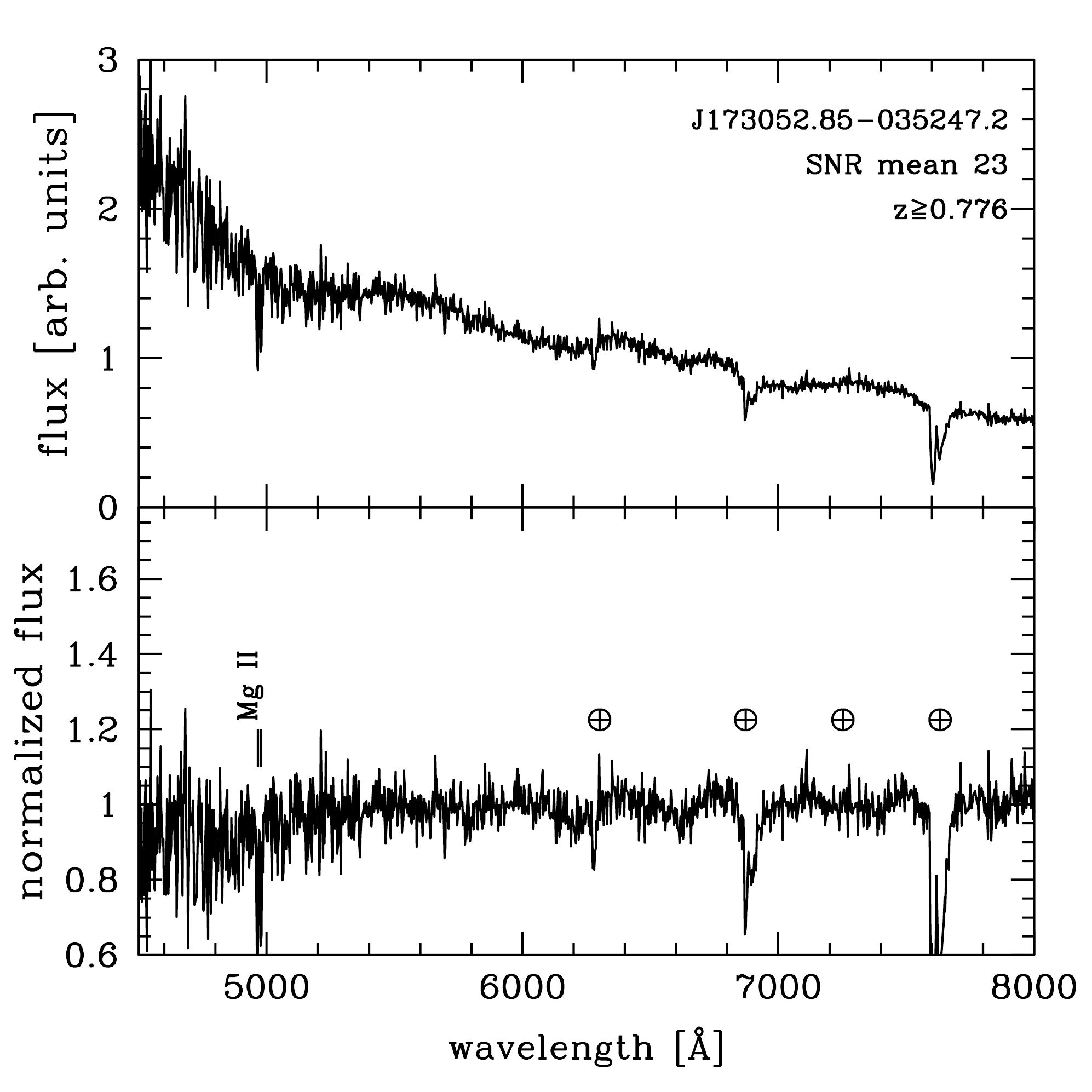}
	\caption{Upper panel: optical spectra observed at KPNO of 
		\wse\ J173052.85-035247.2, potential counterpart associated
		with  2FGL J1730.6-0353, classified as a BL Lac on the basis of its featureless continuum. 
		Lower panel: as in Figure \ref{fig:ugs1}.}
	\label{fig:ugs6}
\end{figure}

\begin{figure}
	\includegraphics[width=0.3\paperwidth]{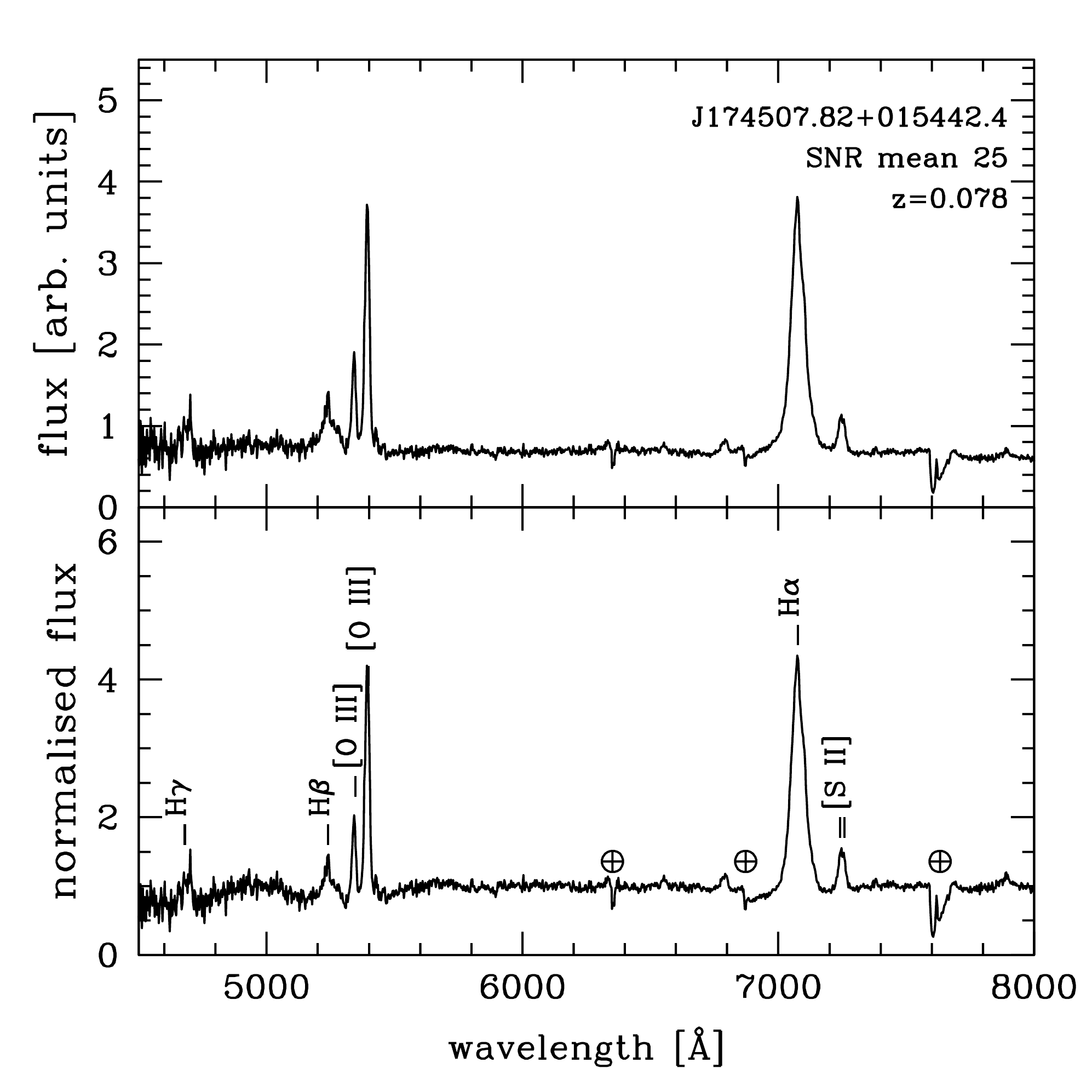}
	\caption{Upper panel: optical spectra observed at KPNO of 
		\wse\ J174507.82+015442.4, potential counterpart associated
		with 2FGL J1745.6+0203, classified as a QSO at z=0.078 on the basis of the identification
		of the emission lines visible in the spectra.
		Lower panel: as in Figure \ref{fig:ugs1}. }
	\label{fig:ugs7}
\end{figure}

\begin{figure}
	\includegraphics[width=0.3\paperwidth]{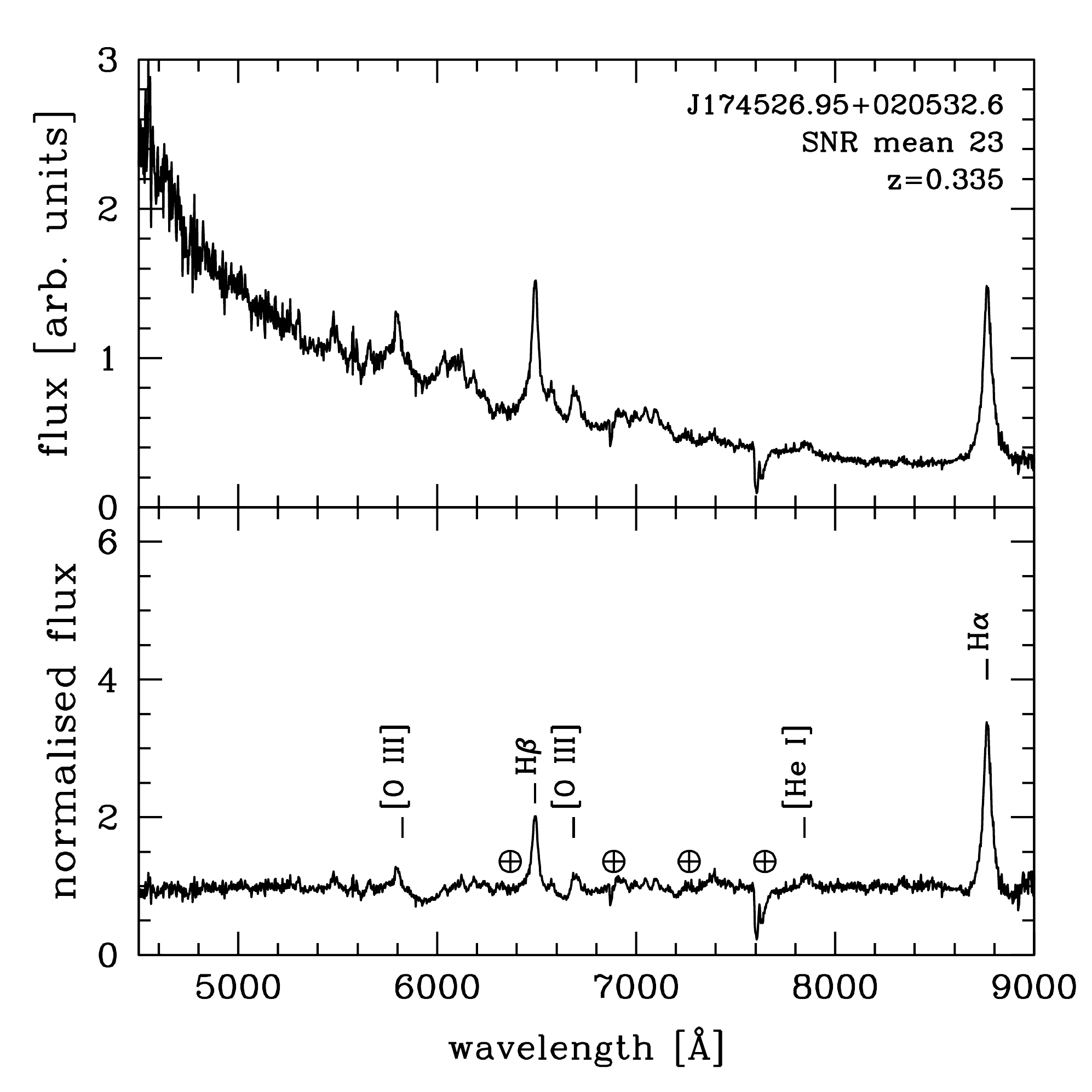}
	\caption{Upper panel: optical spectra observed at KPNO of 
		\wse\ J174526.95+020532.6, potential counterpart associated
		with 2FGL J1745.6+0203, classified as a QSO at z=0.335 on the basis of the identification
		of the emission lines visible in the spectra.
		Lower panel: as in Figure \ref{fig:ugs1}.}
	\label{fig:ugs8}
\end{figure}

\begin{figure}
	\includegraphics[width=0.3\paperwidth]{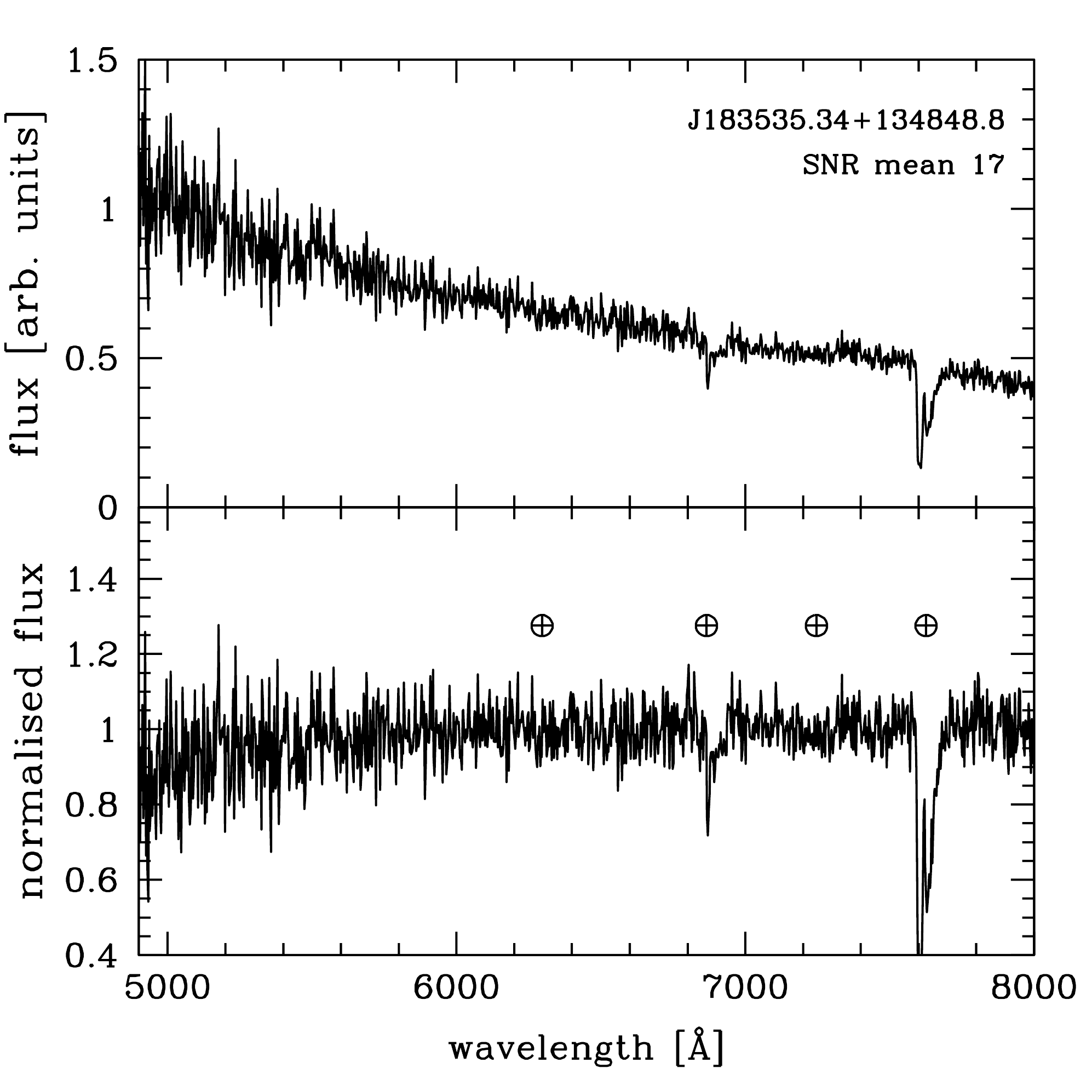}
	\caption{Upper panel: optical spectra observed at KPNO of 
		\wse\ J183535.34+134848.8, potential counterpart associated
		with 2FGL J1835.4+1349, classified as a BL Lac on the basis of its featureless continuum. Lower panel: as in Figure \ref{fig:ugs1}.}
	\label{fig:ugs9}
\end{figure}

Here we provide details for the low-energy counterparts of the 9 UGSs observed in our sample.
All these \wse-selected counterparts where found in the analysis performed by Massaro et al. (2013a)
to have IR \wse\ colors similar to known $\gamma$-ray blazars.
All of them were predicted to be BZB-like sources having the IR colors more consistent with those of the \fer\ BZBs
rather than the BZQs \citep[see][for more details]{ugs1}. 
The spectra of the whole UGSs listed in Table~\ref{tab:obslog} are shown in Figure~\ref{fig:ugs1}-\ref{fig:ugs9}. The blazar counterpart for one of these sources, \wse\ J173052.85-035247.2 (candidate counterpart
of 2FGL J1730.6-0353) appears intriguing. 
On the basis of our optical spectra, we classify the source as a BL Lac object and we put a 
lower limit on its redshift of 0.776 based on the detection of an intervening doublet system of Mg II ($\lambda\lambda_{obs}=4965$ - $4977$ $\AA$ with $EW_{obs}=3.4$ - $2.1$ $\AA$; see also Figure \ref{fig:ugs6}).
However, this source 
is not associated with any NVSS sources, as expected for BL Lac objects. 
We note that, even if deeper radio observations detect emission from the source, blazars are
traditionally defined as radio-loud sources on the basis of current radio data. All confirmed blazar in
\bzcat\ are indeed detected at 1.4 GHz with fluxes above a few mJy, and radio-quiet blazars are extremely
rare objects \citep{londish04,paggi13}. In particular, only 14 BZBs out of the 1220 present in the \bzcat\ have a radio flux density at 1.4GHz lower than 2.5 mJy
that is the average flux limit of the NVSS survey\footnote{\scriptsize{\tt{http://heasarc.gsfc.nasa.gov/W3Browse/radio-catalog/nvss.html}}}. \\
There are two sources, \wse\ J174507.82+015442.4 and \wse\ J174526.95+020532.6, potential counterparts to 
2FGL J1745.6+0203 found on the basis of the IR color selection method. 
According to our multifrequency investigation
we note that 
only \wse\ J174507.82+015442.4 
has a radio counterpart, thus we conclude that \wse\ J174526.95+020532.6 is a normal radio-quiet quasar contaminating our selection. 
However we cannot firmly establish the real blazar nature of \wse\ J174507.82+015442.4 since the lack of multi-wavelength radio observations 
did not allow us to verify the flatness of its radio spectrum.
For both 
objects we determined their redshifts based on
broad emission lines (H$\alpha$ and H$\beta$ are both present, see Figures~\ref{fig:ugs7}-\ref{fig:ugs8}). 
Their redshifts are reported in Table~\ref{tab:obslog}.

\subsection{Gamma-ray Active Galaxies of Uncertain type}
\begin{figure}
	\includegraphics[width=0.3\paperwidth]{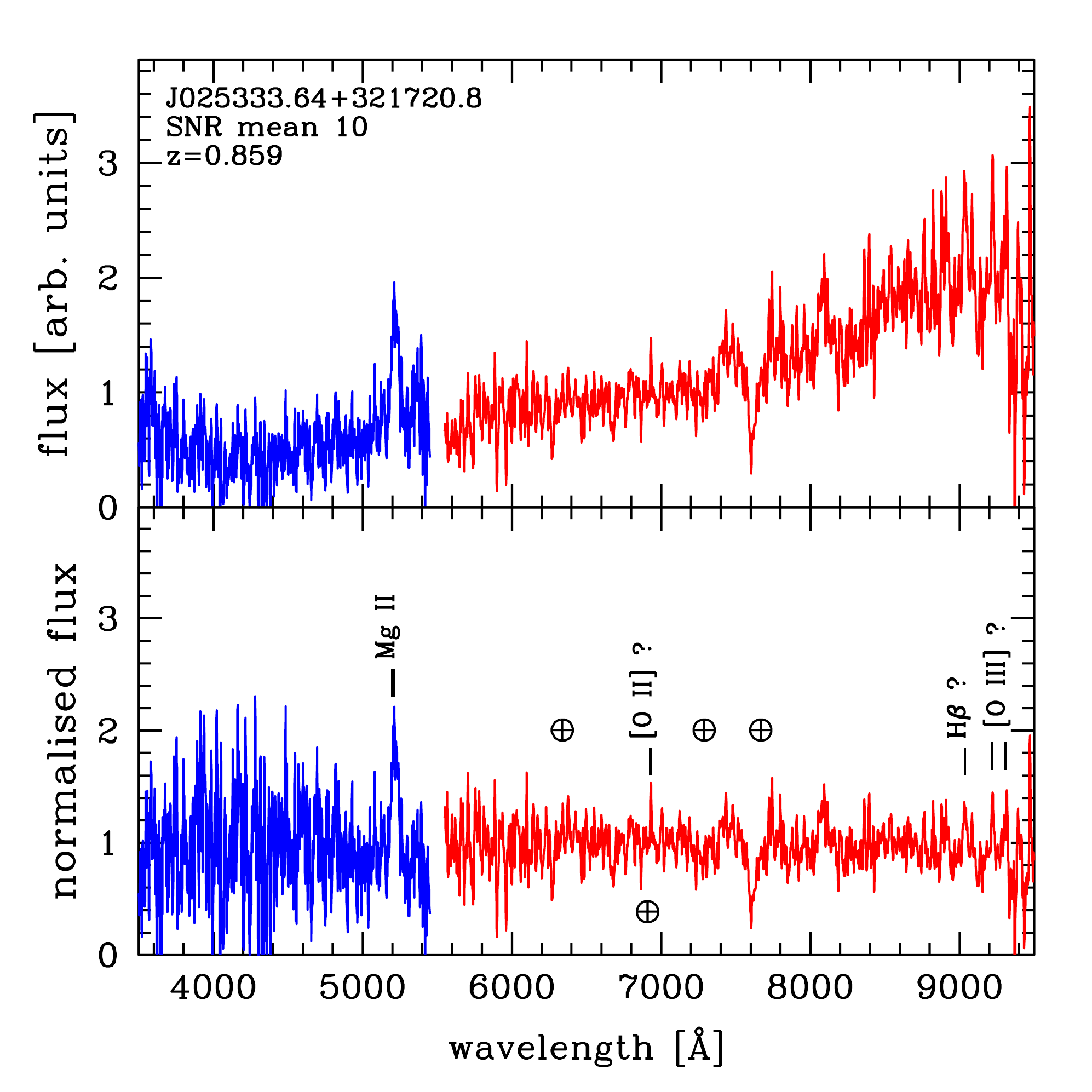}
	\caption{Upper panel: optical spectra observed at Palomar of 
		\wse\ J025333.64+321720.8, potential counterpart associated
		with 2FGL J0253.4+3218, classified as a QSO at z=0.859 on the basis of the identification of 
		the emission lines listed on the spectra. 
		Red and blue spectra 
		show the two sides of the dual-beam spectrograph.
		Lower panel: as in Figure \ref{fig:ugs1}. }
	\label{fig:agu1}
\end{figure}

\begin{figure}
	\includegraphics[width=0.3\paperwidth]{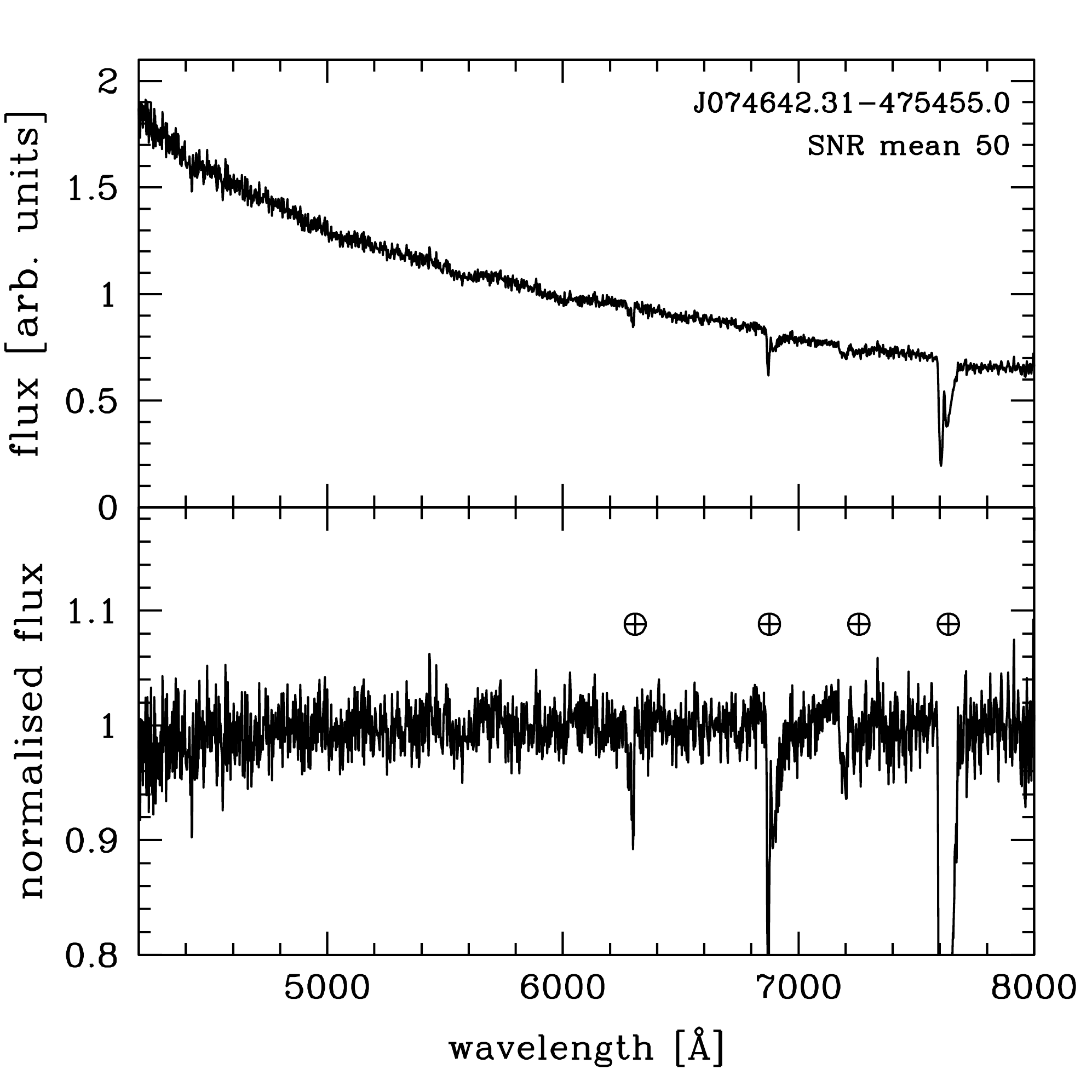}
	\caption{Upper panel: optical spectra observed at SOAR of 
		\wse\ J074642.31-475455.0, potential counterpart associated
		with 2FGL J0746.5-4758, classified as a BL Lac on the basis of its featureless
		continuum. Lower panel: as in Figure \ref{fig:ugs1}.}
	\label{fig:agu2}
\end{figure}

\begin{figure}
	\includegraphics[width=0.3\paperwidth]{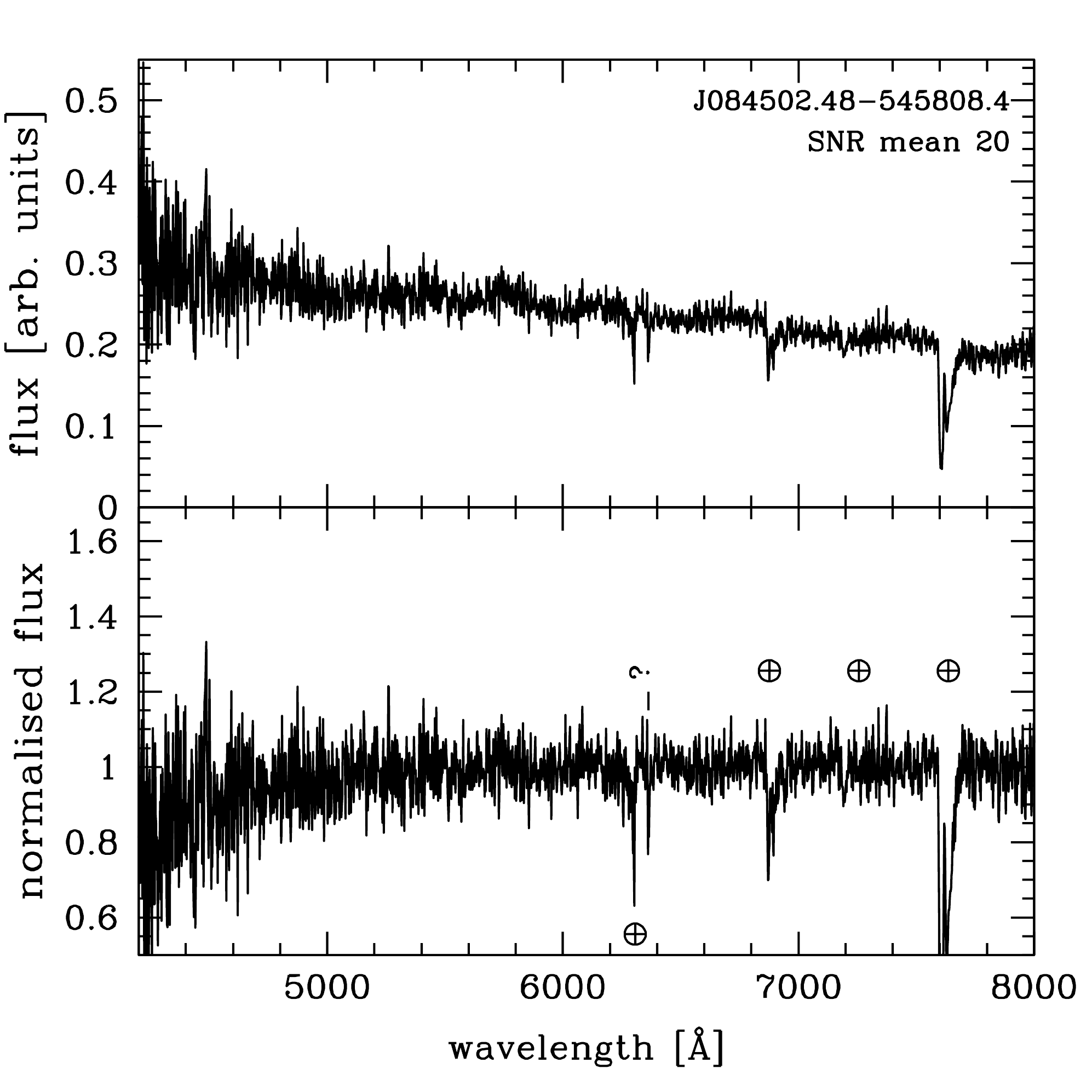}
	\caption{Upper panel: optical spectra observed at SOAR of 
		\wse\ J084502.48-545808.4, potential counterpart associated
		with 2FGL J0844.8-5459, classified as a BL Lac on the basis of its featureless
		continuum. 
		Lower panel: as in Figure \ref{fig:ugs1}.}
	\label{fig:agu3}
\end{figure}

\begin{figure}
	\includegraphics[width=0.3\paperwidth]{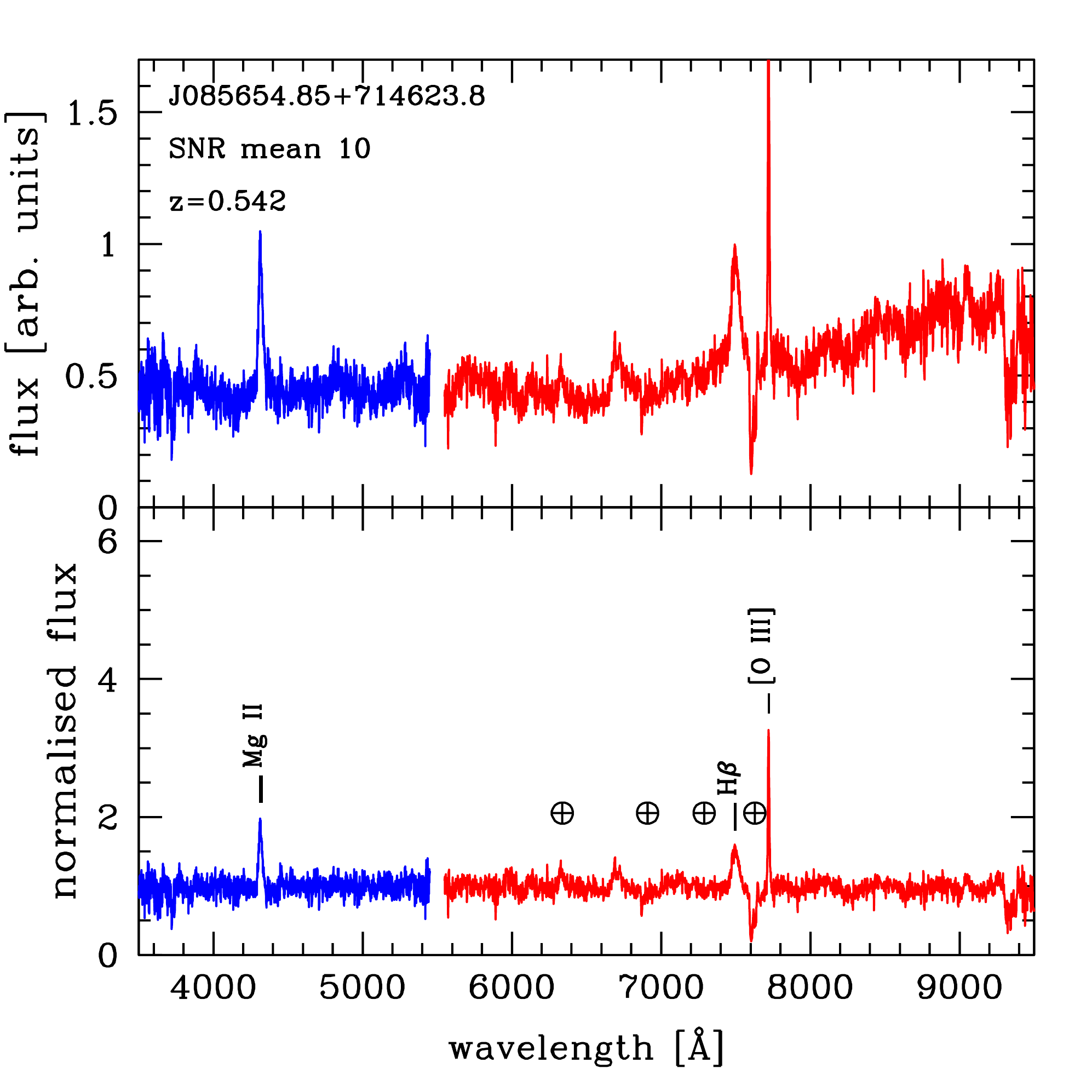}
	\caption{Upper panel: optical spectra observed at Palomar of 
		\wse\ J085654.85+714623.8, potential counterpart associated
		with 2FGL J0856.0+7136, classified as a QSO at z=0.542 on the basis of the identification
		of the emission lines visible in the spectra. Red and blue spectra 
		show the two sides of the dual-beam spectrograph.
		Lower panel: as in Figure \ref{fig:ugs1}.}
	\label{fig:agu4}
\end{figure}

\begin{figure}
	\includegraphics[width=0.3\paperwidth]{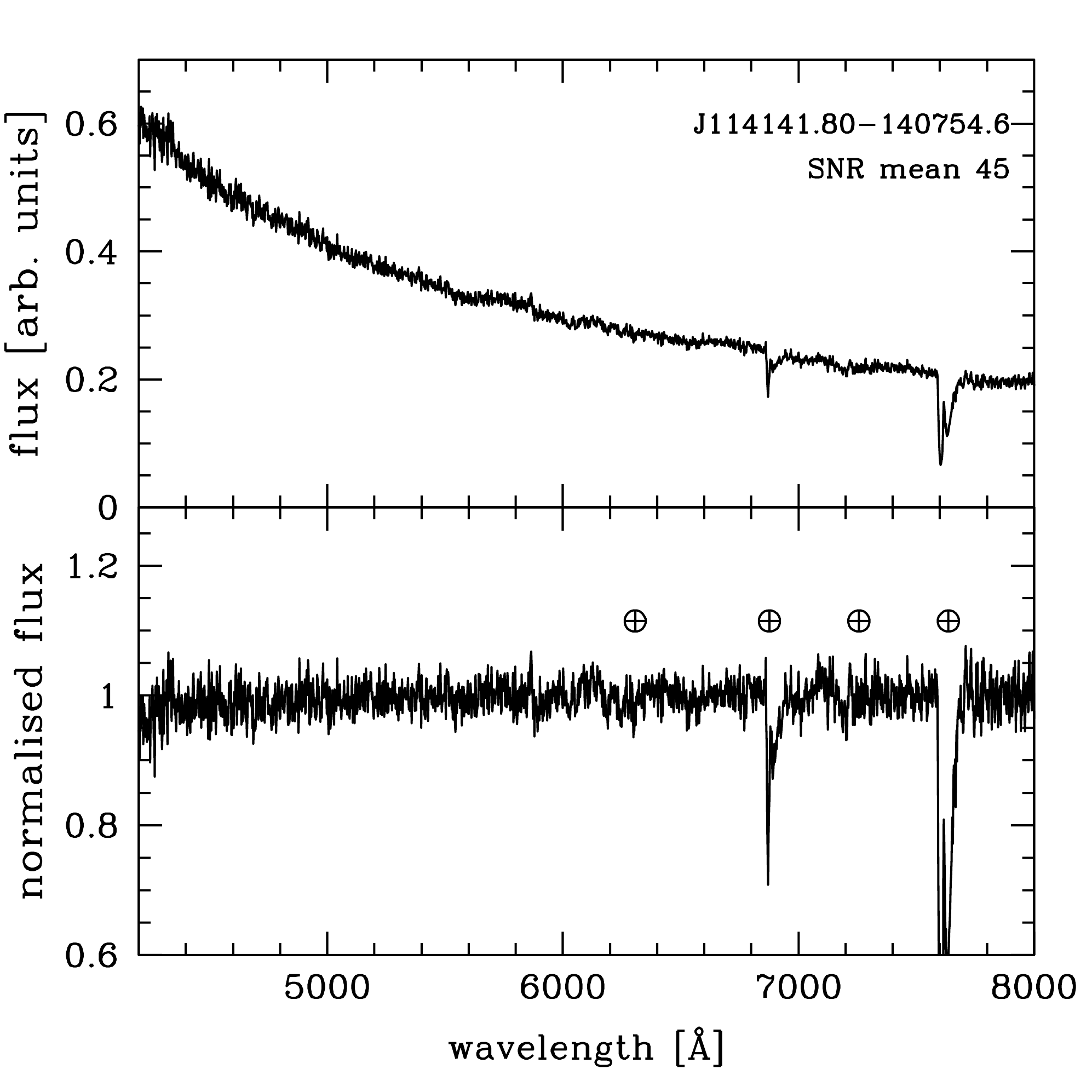}
	\caption{Upper panel: optical spectra observed at SOAR of 
		\wse\ J114141.80-140754.6, potential counterpart associated
		with 2FGL J1141.7-1404, classified as a BL Lac on the basis of its featureless continuum.
		Lower panel: as in Figure \ref{fig:ugs1}. }
	\label{fig:agu5}
\end{figure}

\begin{figure}
	\includegraphics[width=0.3\paperwidth]{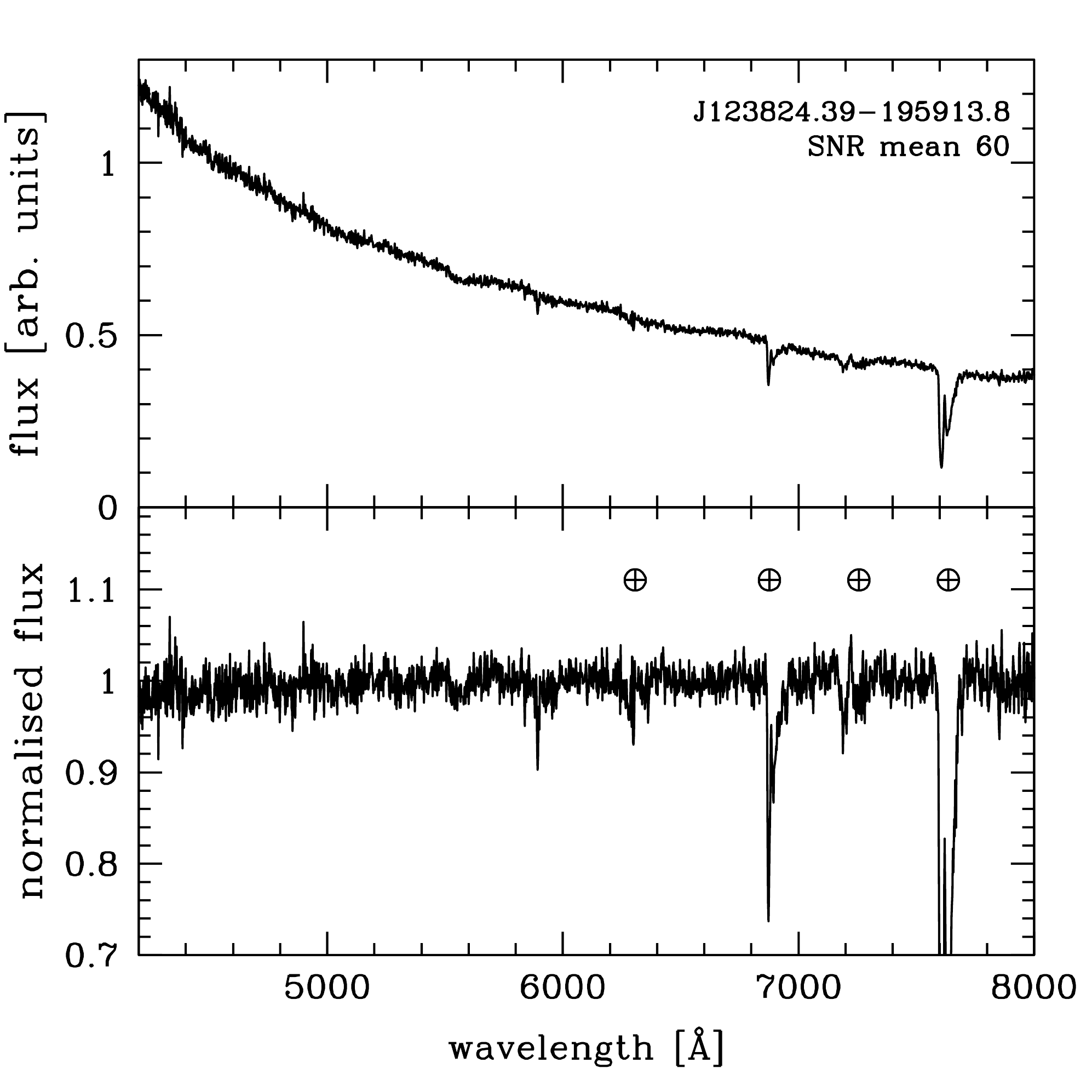}
	\caption{Upper panel: optical spectra observed at SOAR of 
		\wse\ J123824.39-195913.8, potential counterpart associated
		with 2FGL J1238.1-1953 classified as a BL Lac on the basis of its featureless continuum.
		Lower panel: as in Figure \ref{fig:ugs1}.}
	\label{fig:agu6}
\end{figure}

\begin{figure}
	\includegraphics[width=0.3\paperwidth]{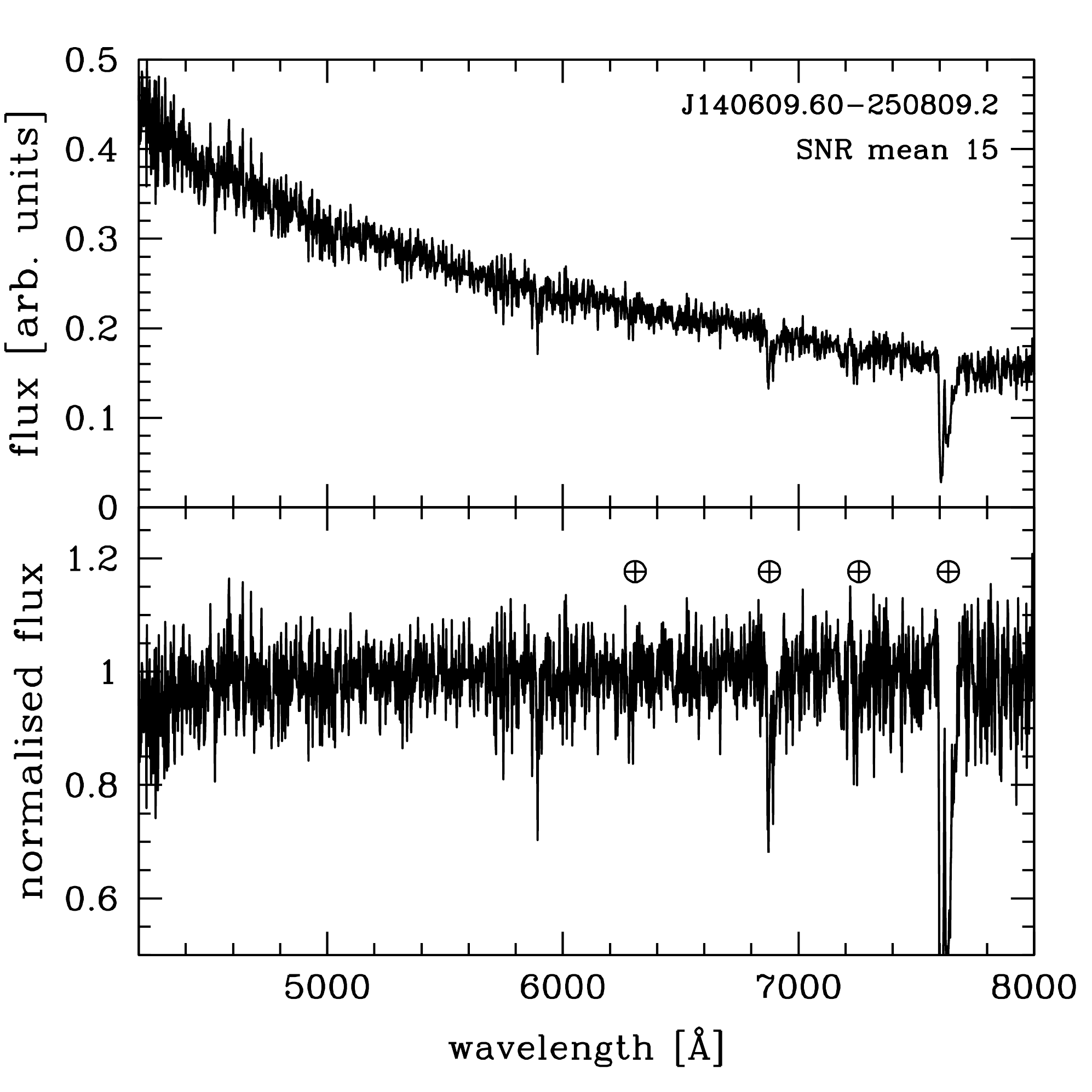}
	\caption{Upper panel: optical spectra observed at SOAR of 
		\wse J140609.60-250809.2, potential counterpart associated
		with 2FGL J1406.2-2510, classified as a BL Lac on the basis of its featureless continuum.
		Lower panel: as in Figure \ref{fig:ugs1}. }
	\label{fig:agu7}
\end{figure}

\begin{figure}
	\includegraphics[width=0.3\paperwidth]{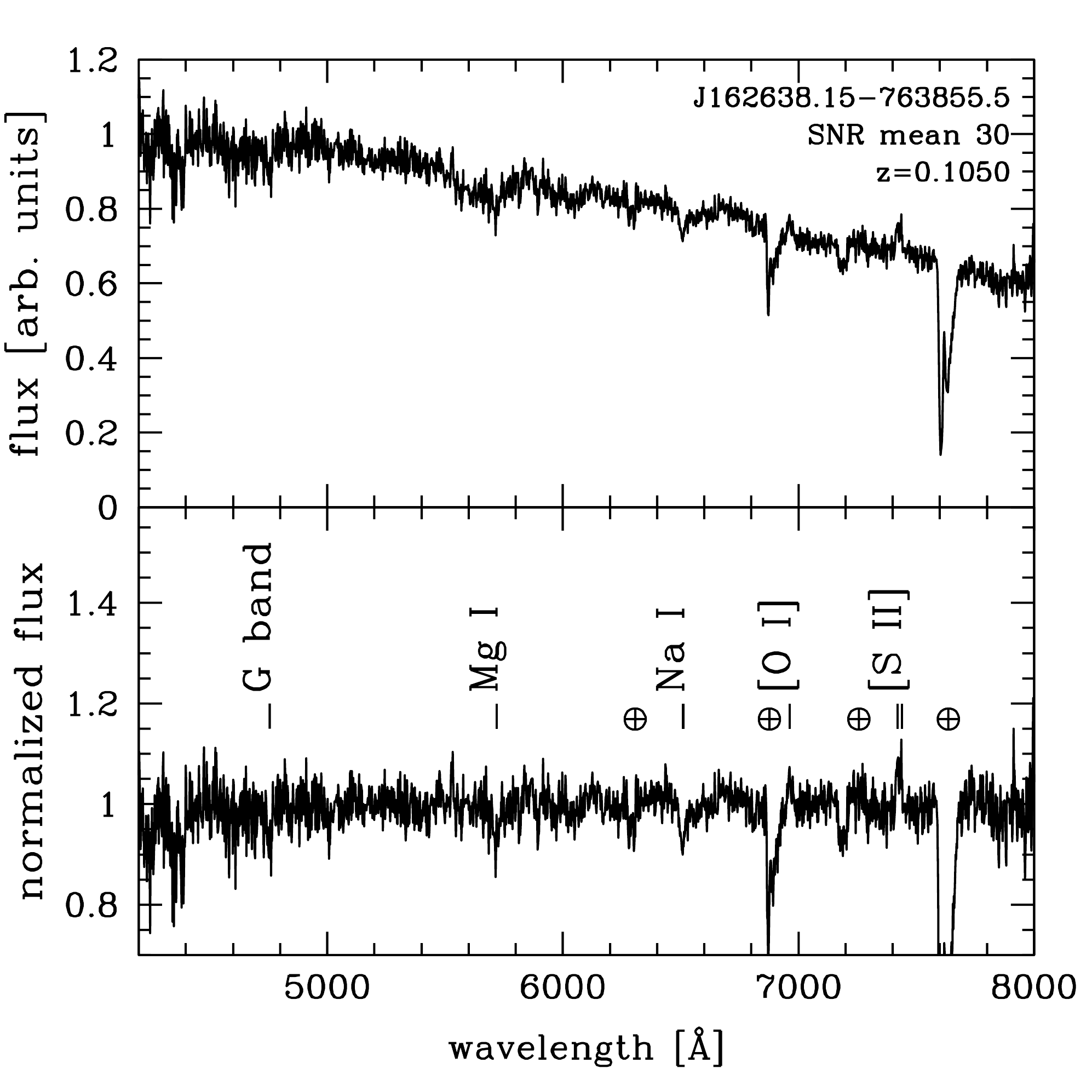}
	\caption{Upper panel: optical spectra observed at SOAR of 
		\wse\ J162638.15-763855.5, potential counterpart associated
		with 
		2FGL J1626.0-7636, classified as a BL Lac since its emission and absorption lines have EW < 5 $\AA$.
		The detection of emission ([O I] and the [S II] doublet) and absorption lines (G band, Mg I and Na I) has enabled us to estimate a redshift of 0.1050. 
		Lower panel: as in Figure \ref{fig:ugs1}. }
	\label{fig:agu8}
\end{figure}

\begin{figure}
	\includegraphics[width=0.3\paperwidth]{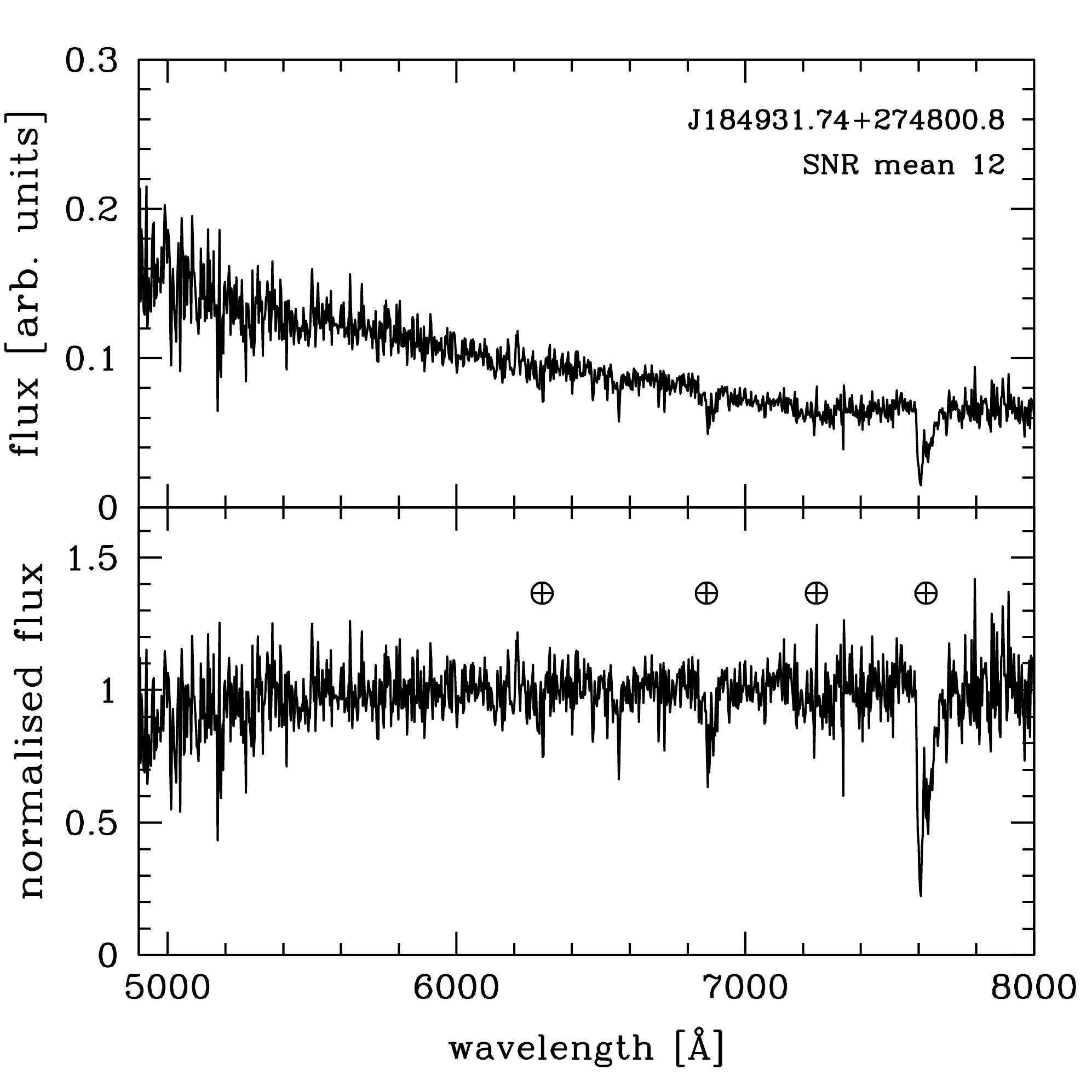}
	\caption{Upper panel: optical spectra observed at KPNO of 
		\wse\ J184931.74+274800.8, potential counterpart associated
		with 2FGL J1849.5+2744, classified as a BL Lac on the basis of its featureless
		continuum.
		Lower panel: as in Figure \ref{fig:ugs1}. }
	\label{fig:agu9}
\end{figure}

\begin{figure}
	\includegraphics[width=0.3\paperwidth]{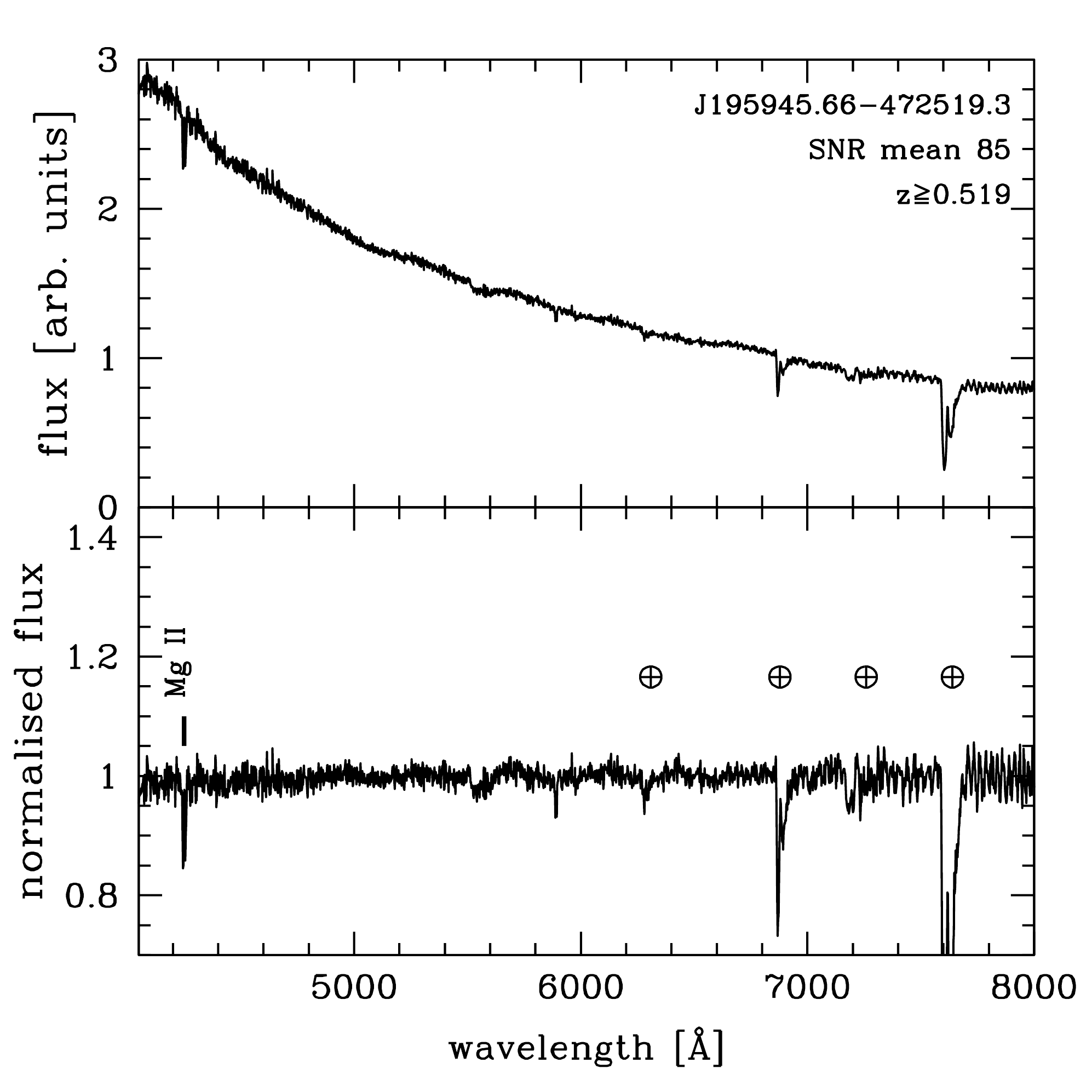}
	\caption{Upper panel: optical spectra observed at SOAR of 
		\wse\ J195945.66-472519.3, potential counterpart associated
		with 2FGL J1959.9-4727, classified as a BL Lac on the basis of its featureless
		continuum. Lower panel: as in Figure \ref{fig:ugs1}.}
	\label{fig:agu10}
\end{figure}

In this subsection we discuss the AGUs in our observed sample.
The multifrequency notes relative to each source are reported in Table~\ref{tab:obslog}, as previously done 
for the UGSs.
The spectra of the whole AGU sample are shown in Figure~\ref{fig:agu1}-\ref{fig:agu10}. 
Our spectroscopic observations confirm that 8 out of 10 sources are BL Lac objects while the remaining two, having quasar-like optical spectra and flat radio spectra, are indeed BZQs.
The optical spectra of these two BZQs, 
\wse\ J025333.64+321720.8 and \wse\ J085654.85+714623.8, 
contain broad emission lines identified as Mg II and H$\beta$ (see Figures~\ref{fig:agu1} and \ref{fig:agu4} for more details),
and permit us to measure 
their redshifts of 0.859 and 0.542, respectively. 
In particular, the identifications of [OII], H$\beta$ and 
[OIII] doublet in the optical spectrum 
of \wse\ J025333.64+321720.8 are uncertain due to low SNR.
The optical spectrum of \wse\ J084502.48-545808.4,
candidate counterpart of 2FGL J0844.8-5459, show an absorption
feature ($\lambda_{obs}$ = 6364 $\AA$ with $EW_{obs}$ = 2.2 $\AA$) 
that we are not able to clearly identify (see Figure \ref{fig:agu3}).
We classify the \wse\ J162638.15-763855.5
counterpart to 2FGL J1626.0-7636 as a BL Lac since its emission lines have EW $<$ 5 $\AA$. The detection of 
emission ([O I] with $EW_{obs}=2.3$ $\AA$, the [S II] doublet $\lambda \lambda_{obs}=7421$ - $7438$ $\AA$ with $EW_{obs}=2.5$ - $2.4$ $\AA$) and absorption lines (G band, Mg I with $EW_{obs}=4.9$ $\AA$ and Na I with $EW_{obs}=3.3$ $\AA$), enable us to measure a redshift of 0.1050 (see Figure~\ref{fig:agu8}). 
The optical spectra of the AGUs associated with the $\gamma$-ray source 2FGL J1849.5+2744 
are also published in Shaw et al. (2013a).
The source is listed as a BZB in the 2LAC catalog, although the optical spectrum was not yet available.  
The optical spectrum collected by us for this object 
(\wse\ J184931.74+274800.8) is nearly featureless (see Figure~\ref{fig:agu9}), 
so we are not able to confirm the lower limit $z$ estimate (i.e., 1.466) of this source proposed by Shaw et al. (2013a)  
on the basis 
of Mg II absorption doublet. 
Thanks to our optical spectrum of the \wse\ J195945.66-472519.3, candidate counterpart to 2FGL J1959.9-4727, 
we can put a 
lower limit on its redshift of 0.519 based on the detection of an intervening doublet system of Mg II ($\lambda \lambda_{obs}=4246$ - $4256$ $\AA$ with $EW_{obs}=1.4$ - $0.9$ $\AA$; see also Figure \ref{fig:agu10}).

\subsection{BL Lac objects}
\begin{figure}
	\includegraphics[width=0.3\paperwidth]{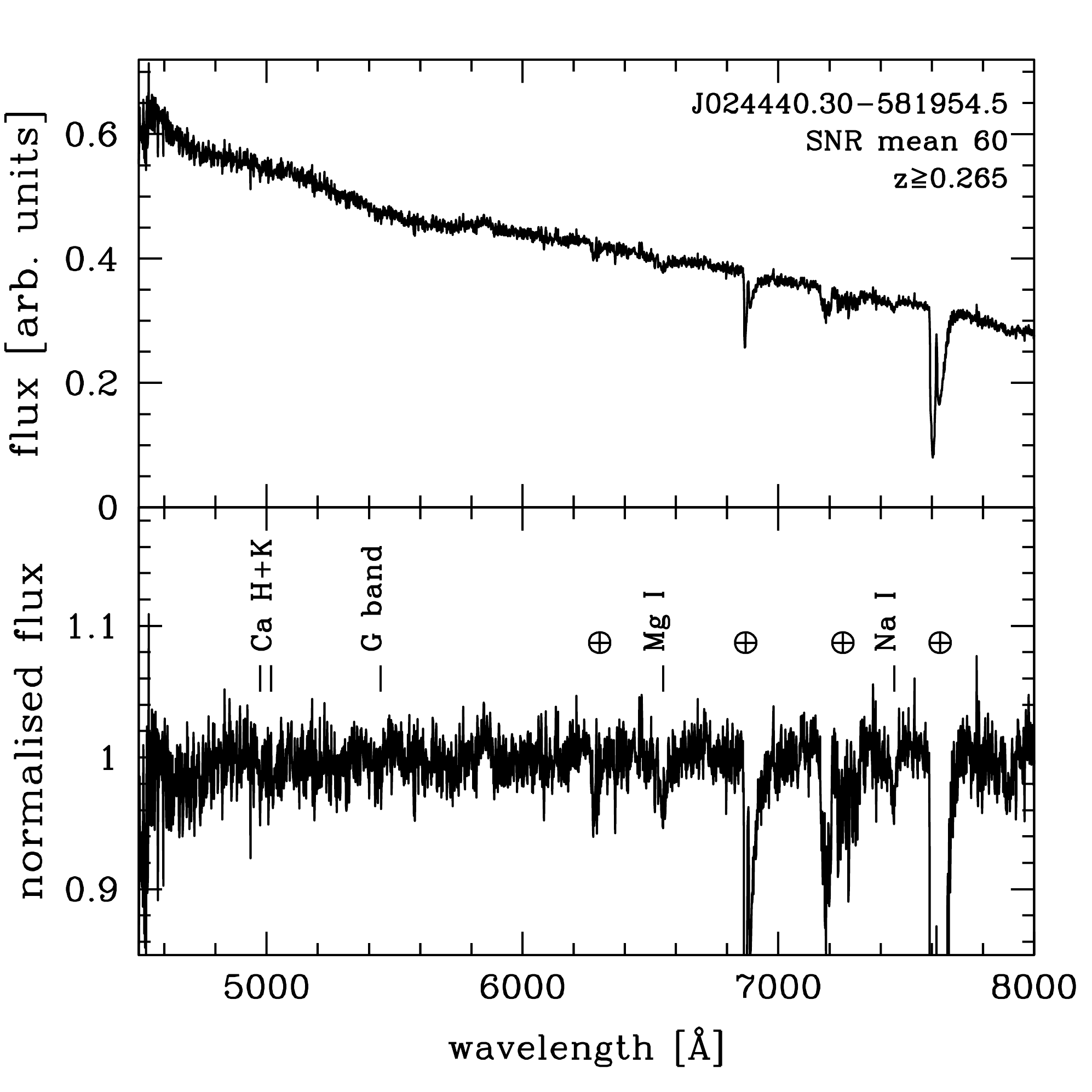}
	\caption{Upper panel: optical spectra observed at Magellan of 
		\wse\ J024440.30-581954.5, potential counterpart associated
		with BZB J0244-5819, classified as a BL Lac since its absorption lines have EW < 5 $\AA$. The detection of absorption lines (Ca H+K, G band, Mg I and Na I) has enabled us to estimate a lower limit on its redshift of 0.265.
		Lower panel: as in Figure \ref{fig:ugs1}.}
	\label{fig:bzb1}
\end{figure}

\begin{figure}
	\includegraphics[width=0.3\paperwidth]{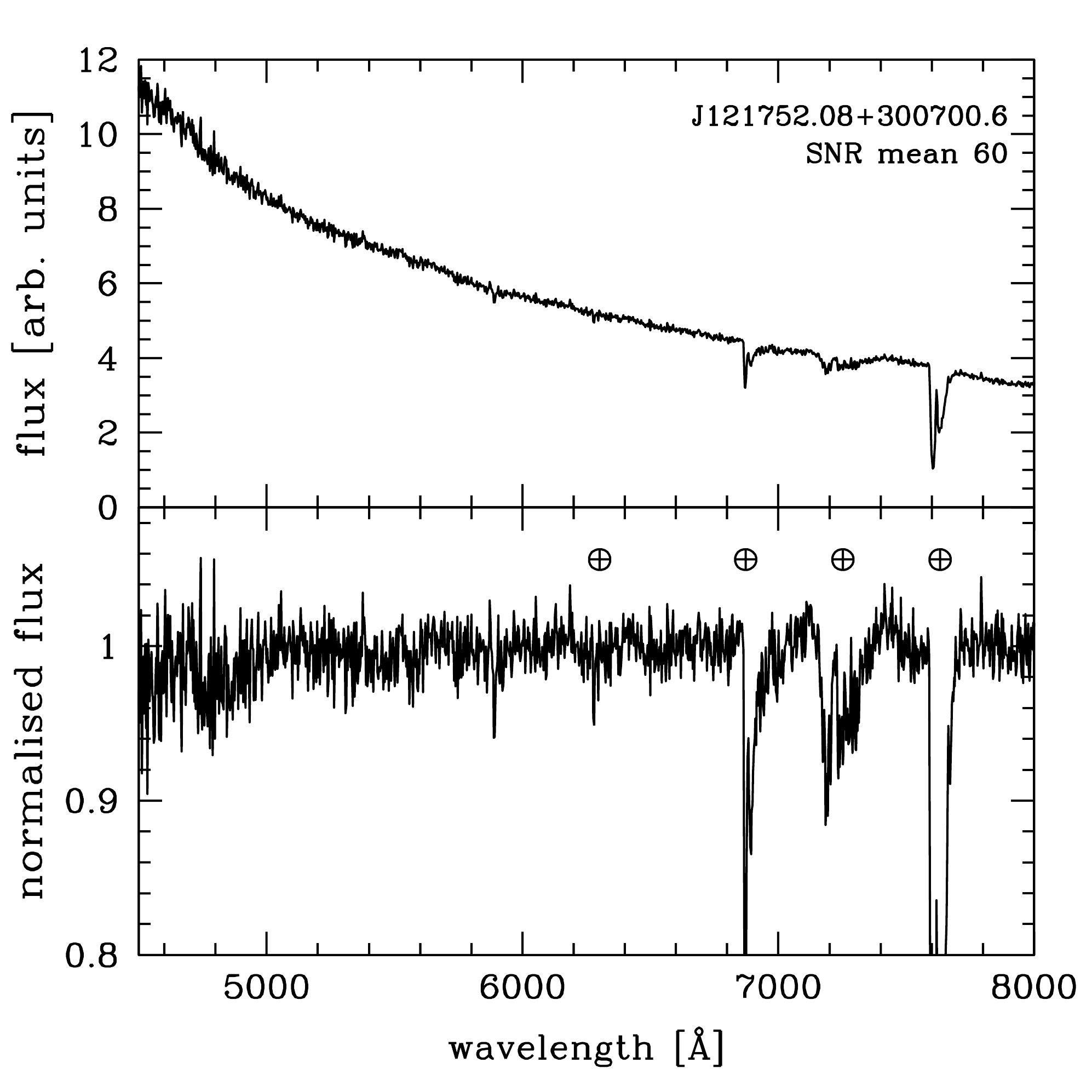}
	\caption{Upper panel: optical spectra observed at KPNO of 
		\wse\ J121752.08+300700.6, potential counterpart associated
		with BZB J1217+3007, classified as a BL Lac on the basis of its featureless 
		continuum. 
		Lower panel: as in Figure \ref{fig:ugs1}. }
	\label{fig:bzb2}
\end{figure}

\begin{figure}
	\includegraphics[width=0.3\paperwidth]{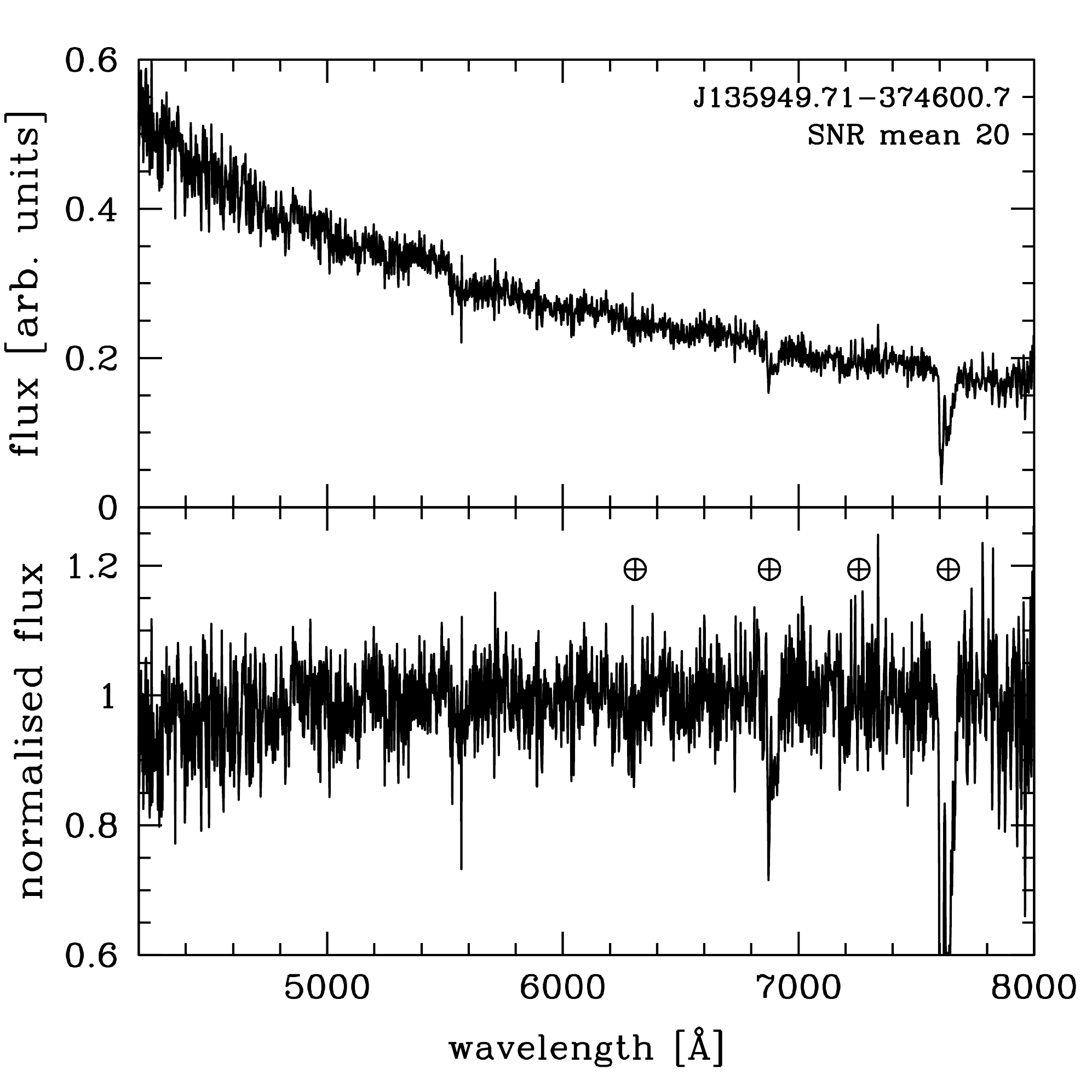}
	\caption{Upper panel: optical spectra observed at SOAR of 
		\wse\ J135949.71-374600.7, potential counterpart associated
		with BZB J1359-3746, classified as a BL Lac on the basis of its featureless 
		continuum. 
		Lower panel: as in Figure \ref{fig:ugs1}. }
	\label{fig:bzb3}
\end{figure}

\begin{figure}
	\includegraphics[width=0.3\paperwidth]{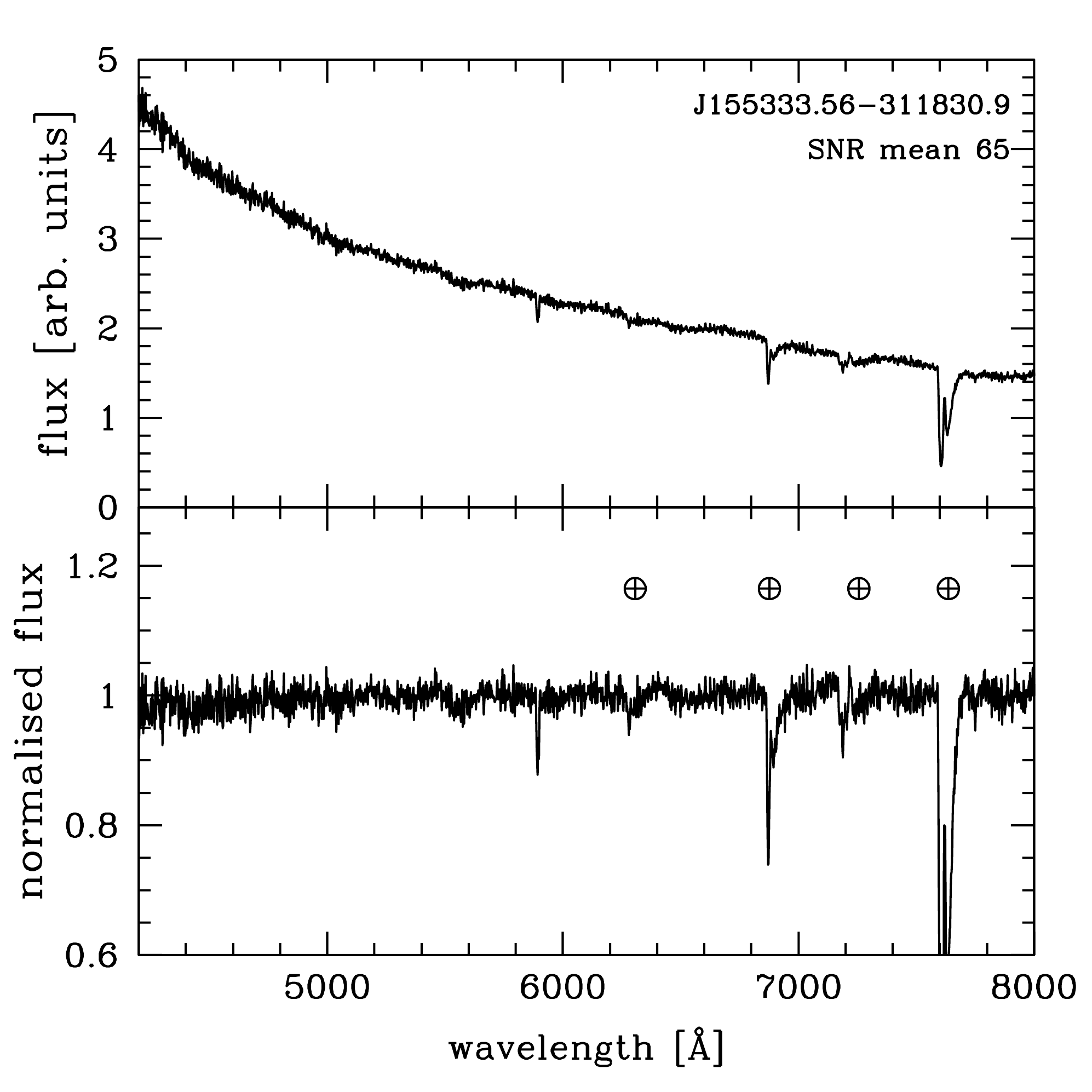}
	\caption{Upper panel: optical spectra observed at SOAR of 
		\wse\ J155333.56-311830.9, potential counterpart associated
		with BZB J1553-3118, classified as a BL Lac on the basis of its featureless 
		continuum. 
		Lower panel: as in Figure \ref{fig:ugs1}.}
	\label{fig:bzb4}
\end{figure}

\begin{figure}
	\includegraphics[width=0.3\paperwidth]{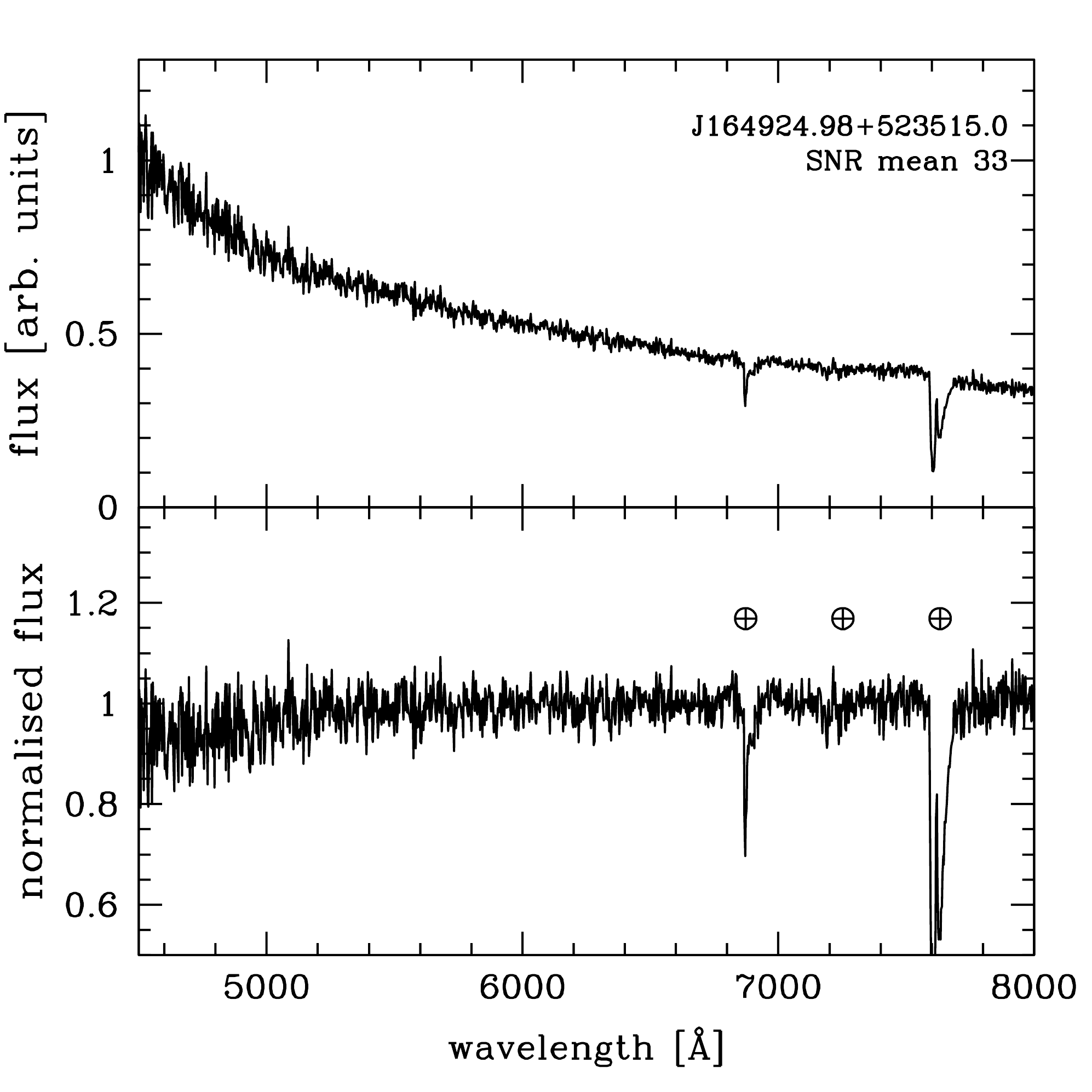}
	\caption{Upper panel: optical spectra observed at KPNO of 
		\wse\ J164924.98+523515.0, potential counterpart associated
		with BZB J1649+5235, classified as a BL Lac on the basis of its featureless 
		continuum. 
		Lower panel: as in Figure \ref{fig:ugs1}.}
	\label{fig:bzb5}
\end{figure}

\begin{figure}
	\includegraphics[width=0.3\paperwidth]{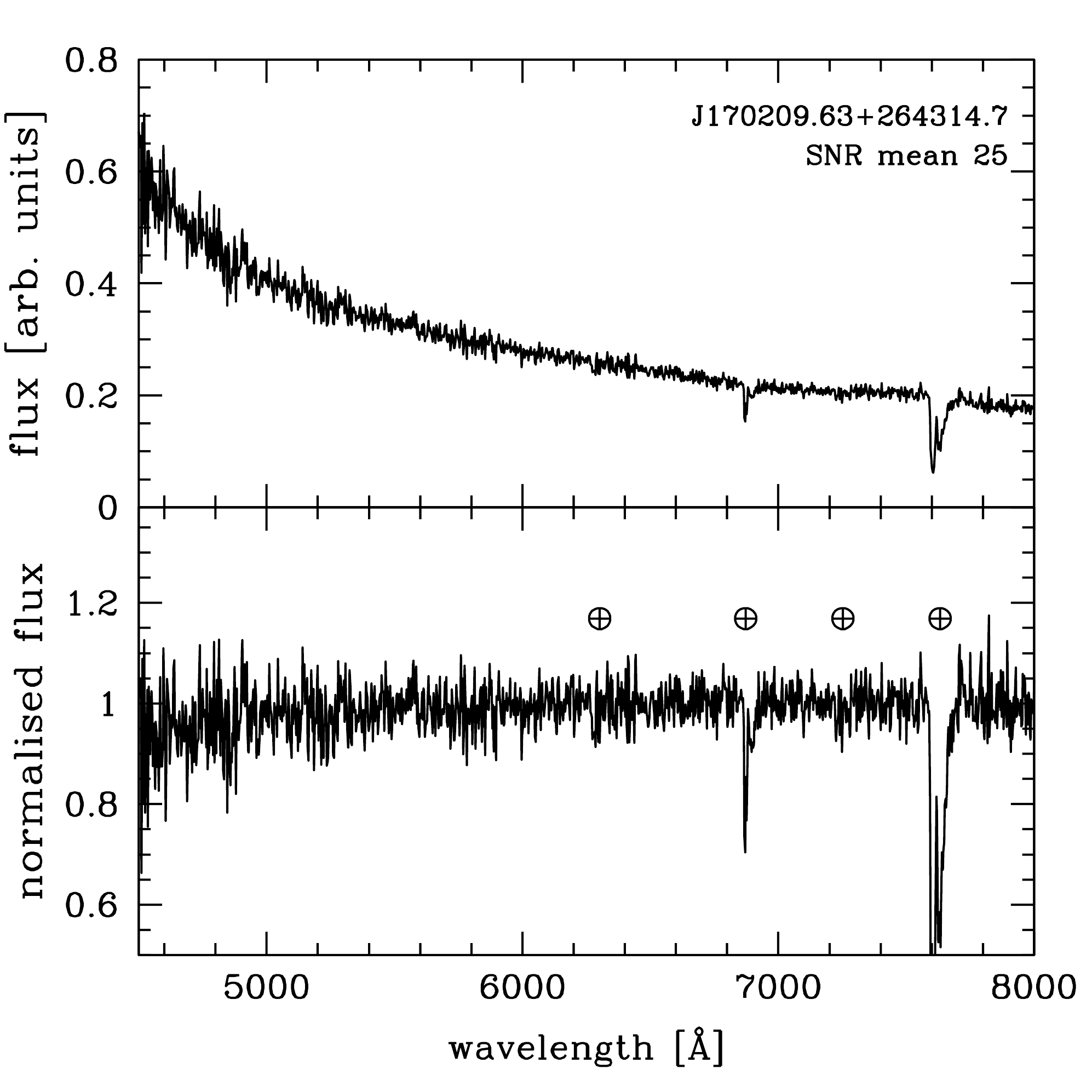}
	\caption{Upper panel: optical spectra observed at KPNO of 
		\wse\ J170209.63+264314.7, potential counterpart associated
		with BZB J1702+2643, classified as a BL Lac on the basis of its featureless 
		continuum. 
		Lower panel: as in Figure \ref{fig:ugs1}.}
	\label{fig:bzb6}
\end{figure}

\begin{figure}
	\includegraphics[width=0.3\paperwidth]{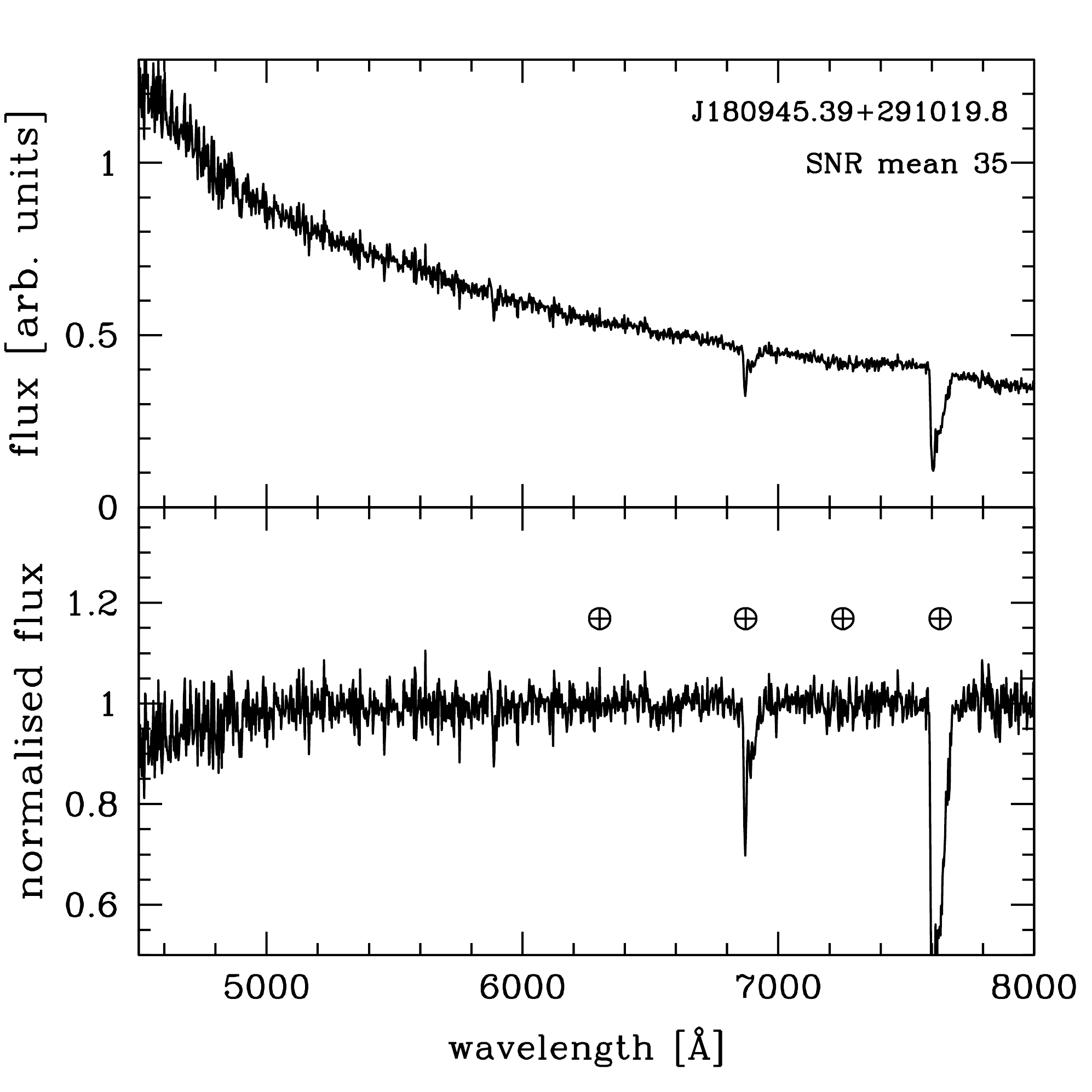}
	\caption{Upper panel: optical spectra observed at KPNO of 
		\wse\ J180945.39+291019.8, potential counterpart associated
		with BZB J1809+2910, classified as a BL Lac on the basis of its featureless 
		continuum. 
		Lower panel: as in Figure \ref{fig:ugs1}.}
	\label{fig:bzb7}
\end{figure}

\begin{figure}
	\includegraphics[width=0.3\paperwidth]{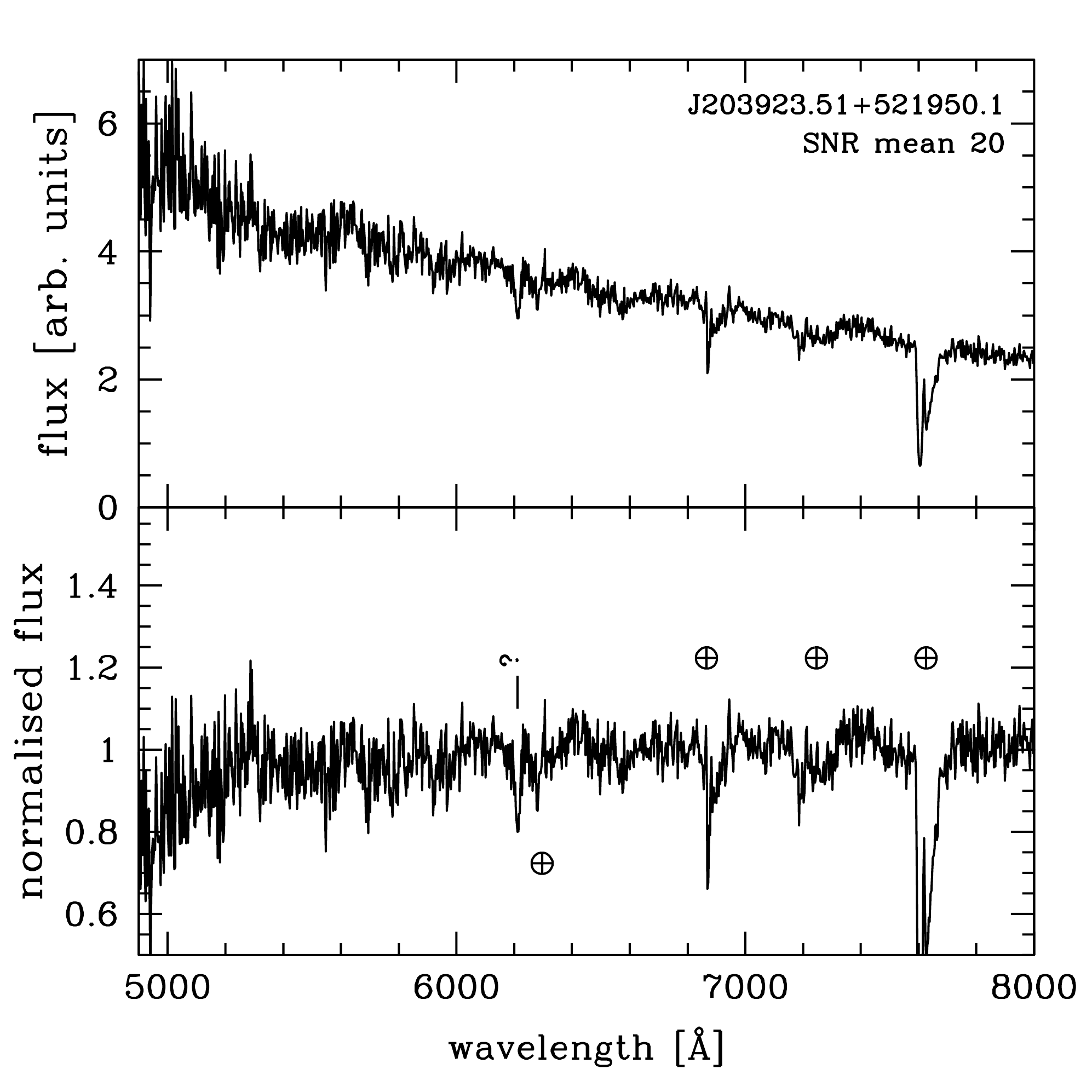}
	\caption{Upper panel: optical spectra observed at KPNO of 
		\wse\ J203923.51+521950.1, potential counterpart associated
		with BZB J2039+5219, classified as a BL Lac on the basis of its featureless 
		continuum. 
		Lower panel: as in Figure \ref{fig:ugs1}.}
	\label{fig:bzb8}
\end{figure}

Details for the BL Lacs in our sample are listed below.
Table~\ref{tab:obslog} reports their \bzcat\ name and the name of the \wse\ counterpart with the coordinates.
Multifrequency notes are not presented in this table since they are already extensively discussed in the \bzcat.
All the BZBs that are also detected by \fer\ belong to the sample named: {\it locus} \citep[i.e. the three-dimensional
region in the WISE color space occupied by the $\gamma$-ray-emitting blazars, well separated from the regions occupied by other classes,][]{ugs1}, with the only exception being BZB J2039+5219. 
They were used in D'Abrusco et al. (2013)
to build the method to search for blazar-like sources within the positional uncertainty region of the \fer\ UGSs.
Thus all their \wse\ counterparts have the IR colors consistent with the \fer\ blazar population.
There are five objects listed in Table~\ref{tab:obslog} as BL Lac candidates for which no optical spectra are present in literature
that allowed a firm classification. 
Our spectroscopic observations confirmed that all the BL Lac candidates have featureless optical spectra
as shown in Figures~\ref{fig:bzb1}-\ref{fig:bzb8}.
In addition, the remaining three objects, namely BZB J1217+3007, BZB J1359-3746 and BZB J1553-3118, that were indeed 
classified BZBs with an uncertain redshift estimate when the \bzcat\ was released. For one BZB, BZB J0244-5819, we have been able to 
estimate a lower limit on its redshift, z$\ge$0.265, on the basis of absorption features in its optical spectrum. 
Our observation of 
\wse\ J024440.30-581954.5 
show
the doublet Ca H+K ($\lambda \lambda_{obs}=4975$ - $5017$ $\AA$ with $EW_{obs}=0.7$ - $1$ $\AA$), G band, Mg I ($EW_{obs}=1.5$ $\AA$) and Na I ($EW_{obs}=1.1$ $\AA$)
absorption features that could be due to the host galaxy 
and/or to intervening systems (see Figure \ref{fig:bzb1}). 
We note that BZB J1359-3746, BZB J1649+5235 and BZB J1809+2910 also have published optical spectra for their candidate counterparts 
released in Shaw et al. (2013b).  
In addition, Shaw et al. (2013b) 
reported the spectrum of BZB J1359-3746, with the detection of the 
Calcium break that allowed them to determine a redshift of 0.334. We cannot confirm this result since our optical spectrum 
of \wse\ J135949.71-374600.7 shows a featureless continuum (see Figure~\ref{fig:bzb3}). 
We also present the spectrum of the low-energy counterpart of BZB J1553-3118, which was also published
in Masetti et al. (2013)
with a redshift of 0.132. 
In our lower signal to noise observation, its IR counterpart \wse\ J135949.71-374600.7 appears 
featureless, as shown in Figure~\ref{fig:bzb4}. 
The optical spectrum of \wse\ J203923.51+521950.1, 
candidate counterpart of BZB J2039+5219, show an absorption feature at
$\lambda_{obs}=6212$ $\AA$ with $EW_{obs}=4.8$ $\AA$. We do not have a 
clear identification for this feature (see also Figure \ref{fig:bzb8}).

\section{Summary and conclusions}
\label{sec:conclusions}
We present here the results of our 2014 spectroscopic campaign 
carried out in the Northern hemisphere with the KPNO and Palomar telescopes, and
in the Southern hemisphere 
with the SOAR and Magellan telescopes.
The main goal of our campaign is to confirm
the nature of sources selected for having IR colors or low radio frequency spectra (i.e., below $\sim$1GHz) similar to  known
\fer-detected blazars and lying within the positional uncertainty
regions of the UGSs through optical spectroscopic observations. 
Given the positional uncertainty of the UGSs, there could be a possible contamination by
different source classes in these selection procedures of $\gamma$-ray blazar candidates \citep[e.g.,][]{stern13}
and spectroscopic observations are the only way to remove 
such degeneracy \citep[e.g.,][]{masetti13,shaw13a,shaw13b,paggi14,sdss} and discover new gamma-ray blazars.
The confirmation of the blazar nature of these selected objects will improve/refine future
associations for the \fer\ catalog and will also yield to understand the efficiency and completeness of the association method based on the
\wse\ colors once our campaign is completed.
Our spectroscopic observations  could potentially allow us to obtain redshift estimates for the UGS candidate counterparts.
During our campaign we also observed several AGUs, 
as defined according to the \fer\ catalogs \citep[see e.g.,][]{ackermann11a,nolan12}, to verify if
they are indeed blazars. In addition we observed several sources that already belong to the \bzcat\ 
but were classified as BL Lac
candidates due to the lack of optical spectra available in literature,
or were BL Lac objects with uncertain redshift estimates, when the \bzcat\ v4.1 was released.

We observed a total of 27 targets.
The results of this spectroscopic campaign are reported as follows:

\begin{itemize}
\item In the sample of potential counterparts for the UGS,
selected based on their IR colors \citep{paper1,
ugs1,ugs2} and
on the basis of their flat radio spectra below $\sim$ 1 GHz 
\citep{ugs3,ugs6,lorcat}, we confirm the blazar nature of 
8 of 9 UGSs. 
Among them, six are clearly BL Lacs presenting
featureless optical spectra. The remaining two are QSOs.
One potential counterparts for 2FGL
J1745.6+0203 found on the basis of the IR color selection
method, namely \wse\ J174526.95+020532.6, 
is a QSO that probably contaminates our selection method.

\item We classify \wse\ J173052.85-035247.2 as a BL Lac although 
this source is not associated with any NVSS
counterpart, as expected for BL Lac objects.
The detection of an intervening doublet system of Mg II
enabled us to set a lower limit on its redshift of 0.776.

\item All the sources that belong to our AGU subsample have a blazar nature. Two of them are QSOs, while for one of them, 
namely \wse\ J162638.15-763855.5, we have been able to detect some features in the optical spectrum, 
both in emission and in absorption, 
leading to a redshift measurement of 0.1050.
The AGU associated with the \wse\ J184931.74+274800.8 has been 
classified as a BZB in the 2LAC paper but at that time there was no optical spectra. 
Our observations confirm its BL Lac nature. 
We have been also able to set a lower limit for the AGU associated 
with \wse\ J195945.66-472519.3 of 0.519 thanks to the detection in its optical spectrum of a Mg II intervening system.

\item Within the \bzcat\ sources we found five BL Lac candidates
that thanks to the collected optical spectra are all confirmed
BL Lacs.
For one of them, BZB J0244-5819, we
have been able to set a lower limit on their redshifts on the basis
of absorption systems (Ca H+K, G band, Mg I and Na I) that could be due to the host
galaxy and/or to intervening systems.
For the remaining four BZBs listed in the \bzcat\ with
uncertain redshift estimates, we were not able to obtain any
redshift values. 
\end{itemize}

\acknowledgements
We are grateful to Dr. D. Hammer and Dr. S. Points for their help to schedule, 
prepare and perform the KPNO and the SOAR observations, respectively.
We are grateful to F. La Franca for the fruitful discussions that significantly improved the paper.
This investigation is supported by the NASA grants NNX12AO97G and NNX13AP20G.
F. Ricci acknowledges the grants MIUR PRIN 2010-2011 and INAF-PRIN 2011.
H. A. Smith acknowledges partial support from NASA/JPL grant RSA 1369566.
The work by G. Tosti is supported by the ASI/INAF contract I/005/12/0.
Part of this work is based on archival data, software or on-line services provided by the ASI Science Data Center.
This research has made use of data obtained from the high-energy Astrophysics Science Archive
Research Center (HEASARC) provided by NASA's Goddard Space Flight Center; 
the SIMBAD database operated at CDS,
Strasbourg, France; the NASA/IPAC Extragalactic Database
(NED) operated by the Jet Propulsion Laboratory, California
Institute of Technology, under contract with the NASA.
Part of this work is based on the NVSS (NRAO VLA Sky Survey):
the National Radio Astronomy Observatory is operated by Associated Universities,
Inc., under contract with the National Science Foundation and on the VLA low-frequency Sky Survey (VLSS).
The Molonglo Observatory site manager, D. Campbell-Wilson, and the staff, J. Webb, M. White and J. Barry, 
are responsible for the smooth operation of Molonglo Observatory Synthesis Telescope (MOST) and the day-to-day observing programme of SUMSS. 
The SUMSS survey is dedicated to M. Large whose expertise and vision made the project possible. 
The MOST is operated by the School of Physics with the support of the Australian Research Council and the Science Foundation for Physics within the University of Sydney.
This publication makes use of data products from the {\it Wide-field Infrared Survey Explorer}, 
which is a joint project of the University of California, Los Angeles, and 
the Jet Propulsion Laboratory/California Institute of Technology, 
funded by NASA.
This publication makes use of data products from the Two Micron All Sky Survey, which is a joint project of the University of 
Massachusetts and the Infrared Processing and Analysis Center/California Institute of Technology, funded by NASA and NSF.
This research has made use of the USNOFS Image and Catalogue Archive
operated by the United States Naval Observatory, Flagstaff Station\footnote{\scriptsize{\tt{http://www.nofs.navy.mil/data/fchpix/}}}.
Funding for the SDSS and SDSS-II has been provided by the Alfred P. Sloan Foundation, 
the Participating Institutions, the NSF, the U.S. Department of Energy, 
NASA, the Japanese Monbukagakusho, 
the Max Planck Society, and the Higher Education Funding Council for England. 
The SDSS Web Site is http://www.sdss.org/.
The SDSS is managed by the Astrophysical Research Consortium for the Participating Institutions. 
The Participating Institutions are the American Museum of Natural History, 
Astrophysical Institute Potsdam, University of Basel, University of Cambridge, 
Case Western Reserve University, University of Chicago, Drexel University, 
Fermilab, the Institute for Advanced Study, the Japan Participation Group, 
Johns Hopkins University, the Joint Institute for Nuclear Astrophysics, 
the Kavli Institute for Particle Astrophysics and Cosmology, the Korean Scientist Group, 
the Chinese Academy of Sciences (LAMOST), Los Alamos National Laboratory, 
the Max-Planck-Institute for Astronomy (MPIA), the Max-Planck-Institute for Astrophysics (MPA), 
New Mexico State University, Ohio State University, University of Pittsburgh, 
University of Portsmouth, Princeton University, the United States Naval Observatory, 
and the University of Washington.
The WENSS project was a collaboration between the Netherlands Foundation 
for Research in Astronomy and the Leiden Observatory. 
We acknowledge the WENSS team consisted of G. de Bruyn, Y. Tang, 
R. Rengelink, G. Miley, H. R\"ottgering, M. Bremer, 
M. Bremer, W. Brouw, E. Raimond and D. Fullagar 
for the extensive work aimed at producing the WENSS catalog.
TOPCAT\footnote{\scriptsize{\tt{http://www.star.bris.ac.uk/$\sim$mbt/topcat/}}} 
\citep{taylor05} for the preparation and manipulation of the tabular data and the images.
The Aladin Java applet\footnote{\scriptsize{\tt{http://aladin.u-strasbg.fr/aladin.gml}}}
was used to create the finding charts reported in this paper \citep{bonnarel00}. 
It can be started from the CDS (Strasbourg, France), from the CFA (Harvard, USA), from the ADAC (Tokyo, Japan), 
from the IUCAA (Pune, India), from the UKADC (Cambridge, UK), or from the CADC (Victoria, Canada).

\end{document}